\documentclass[preprint]{aastex}

\usepackage{hyperref}
\usepackage{breakurl}

\slugcomment{}

\shorttitle{}
\shortauthors{Pearson et al.}

\begin{document}

\title{HERUS: A CO Atlas from SPIRE Spectroscopy of local ULIRGs}
\author{Chris Pearson\altaffilmark{1,2,3}}
\author{Dimitra Rigopoulou\altaffilmark{1,3}}
\author{Peter Hurley\altaffilmark{4}}
\author{Duncan Farrah\altaffilmark{5}}
\author{Jose Afonso\altaffilmark{6,7}}
\author{Jeronimo Bernard-Salas\altaffilmark{8}}
\author{Colin Borys\altaffilmark{9}}
\author{David L. Clements\altaffilmark{10}}
\author{Diane Cormier\altaffilmark{11}}
\author{Andreas Efstathiou\altaffilmark{12}}
\author{Eduardo Gonzalez-Alfonso\altaffilmark{13}}
\author{Vianney Lebouteiller\altaffilmark{14,15}}
\author{Henrik Spoon\altaffilmark{15}}

\altaffiltext{1}{RAL Space, CCLRC Rutherford Appleton Laboratory, Chilton, Didcot, Oxfordshire OX11 0QX, United Kingdom} 
\altaffiltext{2}{Department of Physical Sciences, The Open University, Milton Keynes, MK7 6AA, UK} 
\altaffiltext{3}{Oxford Astrophysics, Denys Wilkinson Building, University of Oxford, Keble Rd, Oxford OX1 3RH, UK} 
\altaffiltext{4}{Department of Physics and Astronomy, University of Sussex, Falmer, Brighton BN1 9QH, UK} 
\altaffiltext{5}{Department of Physics, Virginia Tech, Blacksburg, VA  24061, USA} 
\altaffiltext{6}{Instituto de Astrof\'{i}sica e Ci\^{e}ncias do Espa\c co, Universidade de Lisboa, OAL, Tapada da Ajuda, PT1349-018 Lisboa, Portugal}
\altaffiltext{7}{Departamento de F\'{i}sica, Faculdade de Ci\^{e}ncias, Universidade de Lisboa, Edif\'{i}cio C8, Campo Grande, PT1749-016 Lisbon, Portugal}
\altaffiltext{8}{The Open University, MK7 6, AA Milton Keynes, UK}
\altaffiltext{9}{Riot Games, 12333 W Olympic Blvd, Los Angeles, CA, 90064} 
\altaffiltext{10}{Astrophysics Group, Imperial College London, Blackett Laboratory, Prince Consort Road, London SW7 2AZ, UK}
\altaffiltext{11}{Institut fur theoretische Astrophysik, Zentrum fur Astronomie der Universitat Heidelberg, Albert-Ueberle Str. 2, D-69120 Heidelberg, Germany} 
\altaffiltext{12}{School of Sciences, European University Cyprus, Diogenes Street, Engomi, 1516 Nicosia, Cyprus}
\altaffiltext{13}{Universidad de Alcala de Henares, Departamento de Fisica, Campus Universitario, E-28871 Alcala de Henares, Madrid, Spain}
\altaffiltext{14}{CEA-Saclay, DSM/IRFU/SAp, F-91191 Gif-sur-Yvette, France} 
\altaffiltext{15}{Cornell University, CRSR, Space Sciences Building, Ithaca, NY 14853, USA}

\begin{abstract}
We present the Herschel SPIRE Fourier Transform Spectroscopy (FTS) atlas for a complete flux limited sample of local Ultra-Luminous Infra-Red Galaxies as part of the HERschel ULIRG Survey (HERUS). The data reduction is described in detail and was optimized for faint FTS sources with particular care being taken with the subtraction of the background which dominates the continuum shape of the spectra. Special treatment in the data reduction has been given to any observation suffering from artefacts in the data caused by anomalous instrumental effects to improve the final spectra. Complete spectra are shown covering $200 - 671\mu$m with photometry in the SPIRE bands at 250$\mu$m, 350$\mu$m and 500$\mu$m. The spectra include near complete CO ladders for over half of our sample, as well as fine structure lines from [CI] 370 $\mu$m, [CI] 609 $\mu$m, and [NII] 205 $\mu$m. We also detect H$_{2}$O lines in several objects. We construct CO Spectral Line Energy Distributions (SLEDs) for the sample, and compare their slopes with the far-infrared colours and luminosities. We show that the CO SLEDs of ULIRGs can be broadly grouped into three classes based on their excitation. We find that the mid-J (5$<$J$<$8) lines are better correlated with the total far-infrared luminosity, suggesting that the warm gas component is closely linked to recent star-formation. The higher J transitions do not linearly correlate with the far-infrared luminosity, consistent with them originating in hotter, denser gas unconnected to the current star-formation. {\bf We conclude that in most cases more than one temperature components are required to model the CO SLEDs.}
\end{abstract}


\section{Introduction}\label{sec:introduction}
Among the most luminous objects in the low-redshift ($z\leq 0.2$) Universe are the Ultra Luminous InfraRed Galaxies (ULIRGs, L(8-1000)$\, \mu$m$ >$10$^{12}$L$_{\odot}$, \citealt{soifer86,sanders88,lfs06}) with star-formation rates of up to several hundred M$_{\odot}$yr$^{-1}$. Although low-redshift ULIRGs are rare, with a space density similar to QSOs (0.001 per deg$^{-2}$), their contribution to the co-moving star formation rate density increases dramatically with redshift, by approximately three orders of magnitude by $z=1$ (\citealt{clements96,kim98,lefloch05,murphy11}), making them an important population for understanding the cosmic history of stellar and Super-Massive Black Hole (SMBH) mass assembly.

Recent studies, especially from ESA's {\it Herschel} Space Observatory ~\citep{pilbratt10}, have shown that the low-redshift ULIRGs may not be direct analogues of their high redshift counterparts. In the local Universe, ULIRGs are invariably mergers hosting compact ($<$kpc) star-forming regions and AGN (e.g. \citealt{clements96,melnick90,rigo99,soifer00,farrah01}), but at higher redshifts ULIRGs may host more extended star-forming regions, and be associated both with interacting and isolated galaxies (e.g. \citealt{pope06}, \citealt{farrah08}, \citealt{magdis11}, \citealt{symeonidis13}, \citealt{bethermin14}). The high redshift ULIRGs thus lie on the galaxy mass-specific star-formation rate density (SSFRD) relation (the so called 'main sequence'), making them typical rather than extreme at these redshifts (e.g. \citealt{elbaz11}).

An invaluable tool in understanding why this change in the properties and importance of ULIRGs with redshift occurs is infrared spectroscopy, since it probes ionization conditions of the interstellar medium and star-forming regions. In particular, infrared spectroscopy can probe the large reservoirs of carbon monoxide (CO) in ULIRGs, which trace the total molecular gas reservoir, by constructing their CO Spectral Line Energy Distributions (SLEDs, \citealt{sanders91,downes93,wolfire10}). The lower rotational transitions, CO(1-0) up to about CO(3-2), trace the total cold dense gas (e.g. \citealt{solomon97}, \citealt{sanders91}), while the higher J transitions trace the warmer gas associated with the PDR and XDR regions. Therefore, to trace the different temperature components within the gas, observations covering many CO transitions are required. However, since only the low-J lines (J$<$6) are accessible from the ground, observations with {\it Herschel} over the CO(5-4) up to CO(13-12) range are necessary to discriminate the contributions from UV and X-ray excitation via star-formation and AGN, respectively (e.g. \citealt{meijerink05}). Observations of the CO SLED in low-redshift ULIRGs can also be used as templates for comparison for intermediate to high redshift (e.g. \citealt{greve14}). In addition, far-infrared spectroscopy gives access to several important fine-structure lines, which provide complementary diagnostics of ISM conditions. For example, low-redshift ULIRGs show a deficit in the strength of their [CII] fine structure line at 158$\, \mu$m relative to the far-infrared (FIR) dust continuum compared to lower luminosity galaxies (e.g. \citealt{luhman98}, \citealt{luhman03}, \citealt{farrah13}), but  this deficit is not apparent at higher redshift (\citealt{haileydunsheath10b}, \citealt{stacey10}, \citealt{magdis14}).

In this paper, we report on a comprehensive {\it Herschel} survey of a flux-limited sample of low-redshift ULIRGs, producing a CO atlas including all ULIRGs in the Universe out to $z=0.2$. An extensive CO spectral atlas of extragalactic objects has also been published in \citet{kamenet16}. In Section \ref{sec:sampleobservations}, we introduce the source sample and describe the  {\it Herschel} observations. In Section \ref{sec:Reduction} we describe the data reduction steps using standard processing pipelines and also the post-processing steps required to produce the final spectra. In Section \ref{sec:LineFitting}, we describe our line fitting procedure, providing identifications and line fluxes for our sample. Our results are presented in Section \ref{sec:results}. Discussion and conclusions are given in Sections \ref{sec:discussion} and \ref{sec:conclusions}. A detailed analysis of the CO SLED fits will be presented in Hurley et al, in preparation. We assume a Hubble constant of  $H_0=70$\,km\,s$^{-1}$\,Mpc$^{-1}$ and density parameters of $\Omega_{\rm M}=0.3$ and $\Omega_\Lambda=0.7$.

\section{Observations}\label{sec:sampleobservations}
The {\it Herschel} ULIRG Survey (HERUS, PI Farrah) was the largest extragalactic Open Time survey (OT1, 250 hrs) carried out by the {\it Herschel} Space Observatory. HERUS is a flux limited sample of low-redshift ULIRGs comprising the 43 ULIRGs from the IRAS PSC-z survey \citep{saunders00} with 60$\mu$m fluxes greater than 1.8Jy (Table \ref{tab:observations}). 

The sample was observed by {\it Herschel} using the Spectral and Photometric Imaging REceiver (SPIRE, \citealt{griffin10}) in both photometry (250, 350, 500$\mu$m) and spectroscopy using the SPIRE Fourier Transform Spectrometer, (FTS, \citealt{swinyard10}) from 194--671$\mu$m (except 3C273 which only has photometric data). The SPIRE photometry and spectroscopy observations were carried out between 26th July 2011 ({\it Herschel} Operational Day, OD 804) to 19th October 2012 (OD 1178). The observations are summarised in Table \ref{tab:observations}. The SPIRE photometer observations were carried out in Small Map mode ({\it SpirePhotoSmallScan}, POF10,  \citealt{dowell10}) with  fixed, 3 repetition (445s), cross-linked 1$\times$1 scan legs covering a field of 4$\arcmin$ radius. Images are taken simultaneously in the SPIRE  250$\mu$m (PSW), 350$\mu$m (PMW), and 500$\mu$m (PLW) bands.  Note that 3C273  only has photometric data and the photometric data for IRAS 06035-7102 was in fact extracted from the Open Time Key Programme: KPOT$\_$mmeixner observations of the Large Magellanic Cloud, Level 2.5  {\it Herschel} data product. 

\begin{figure} 
\begin{center}
\includegraphics[width=0.45\columnwidth,angle=0]{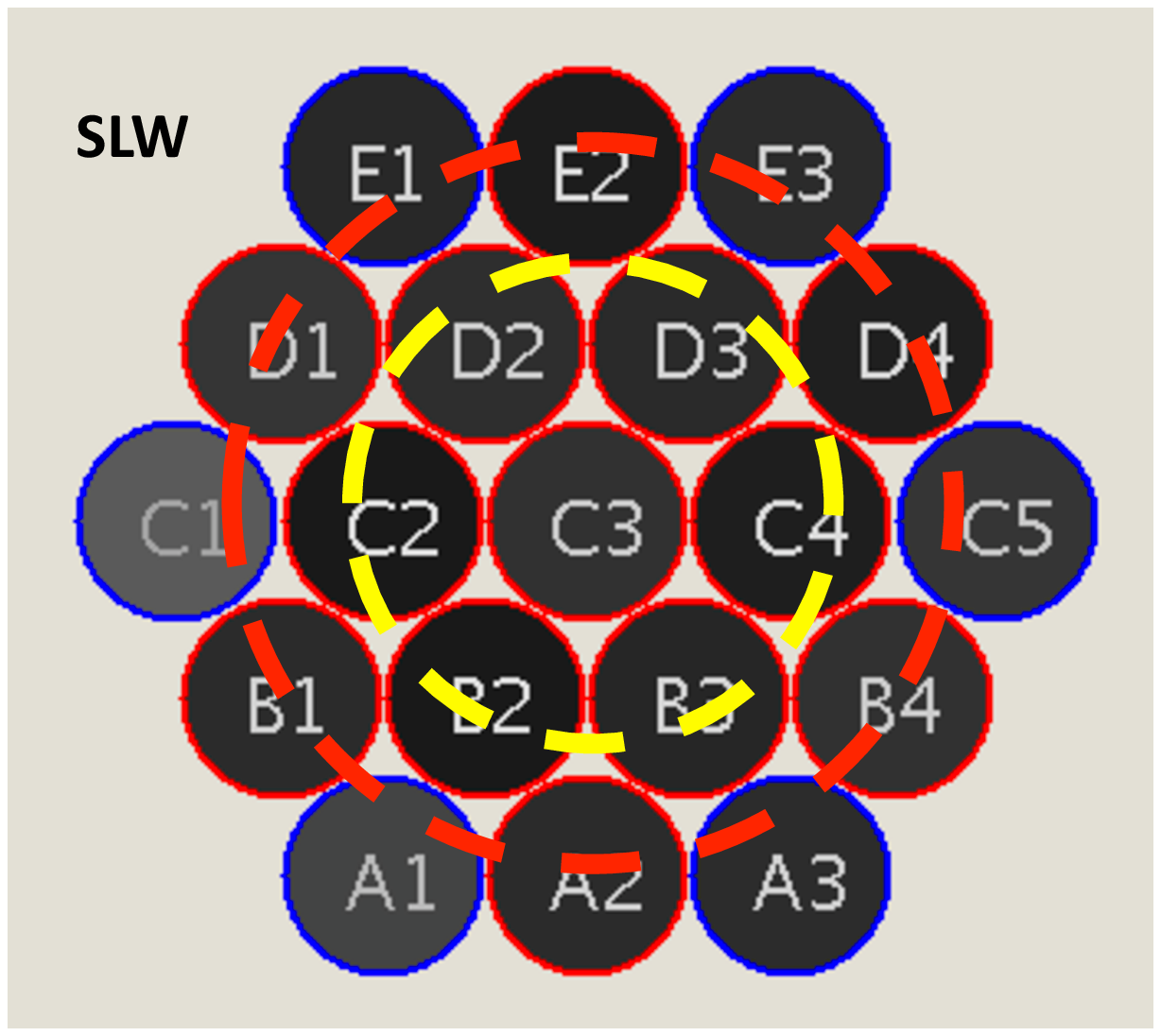}
\includegraphics[width=0.45\columnwidth,angle=0]{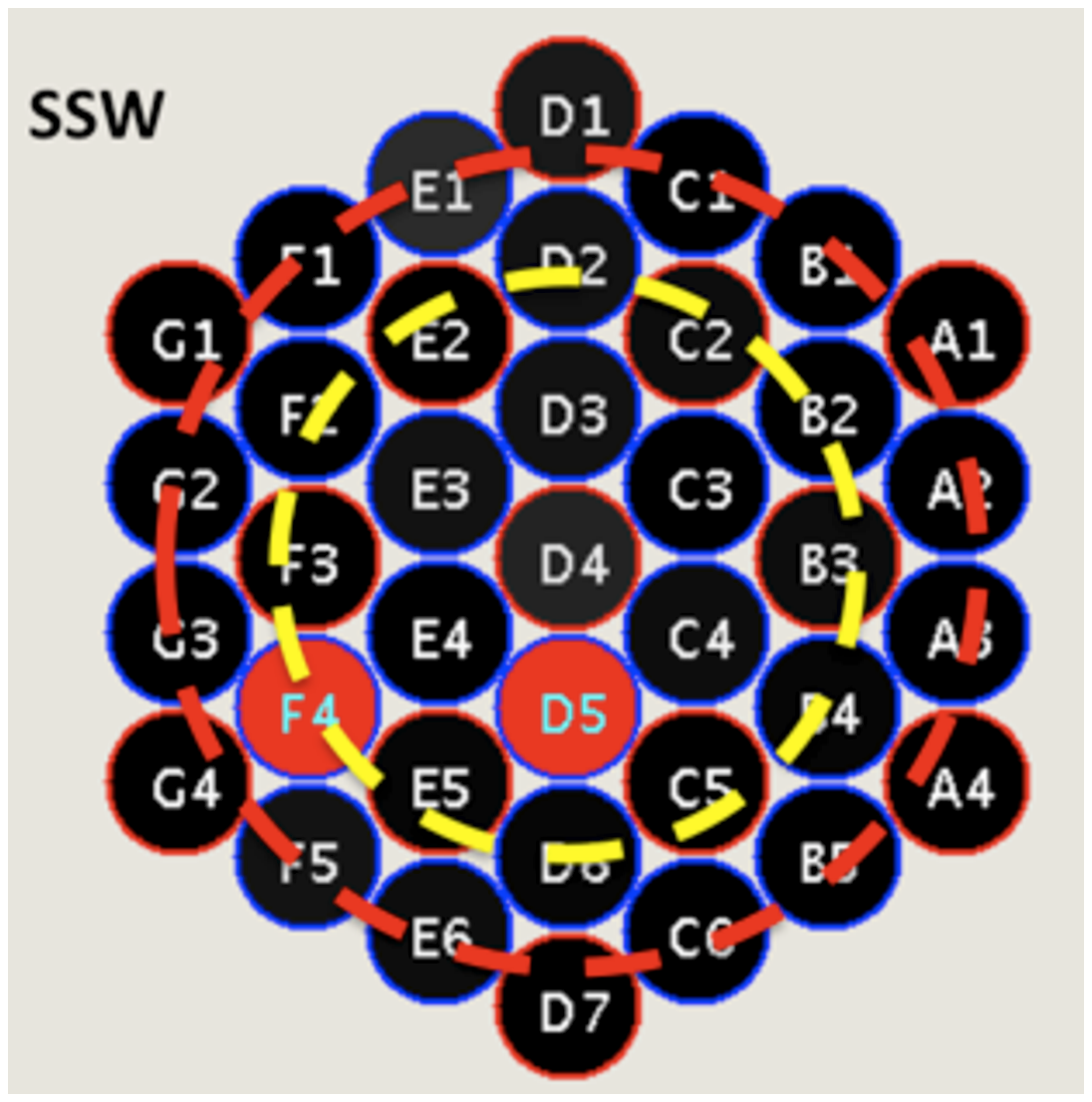}
\includegraphics[width=0.45\columnwidth,angle=0]{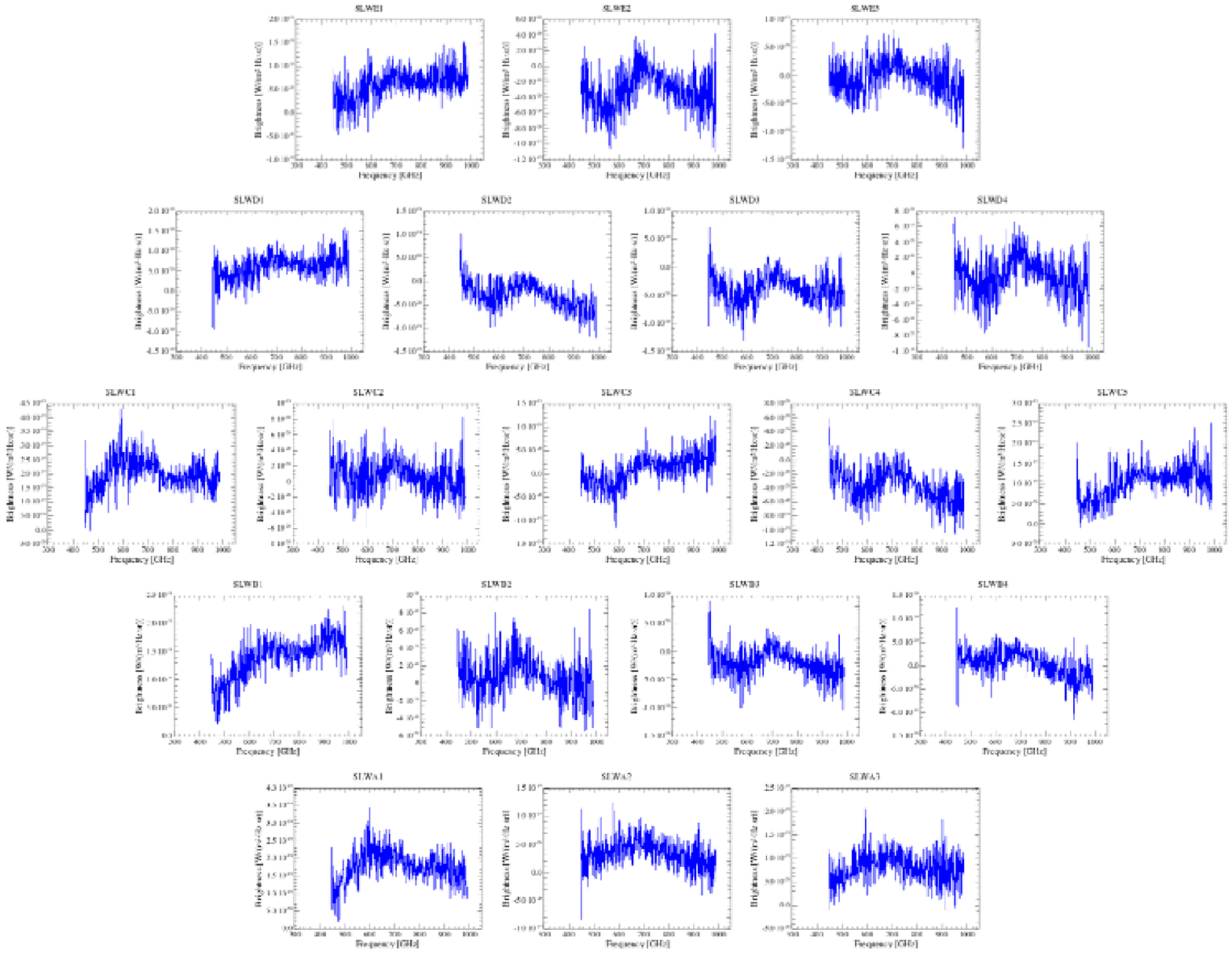}
\includegraphics[width=0.45\columnwidth,angle=0]{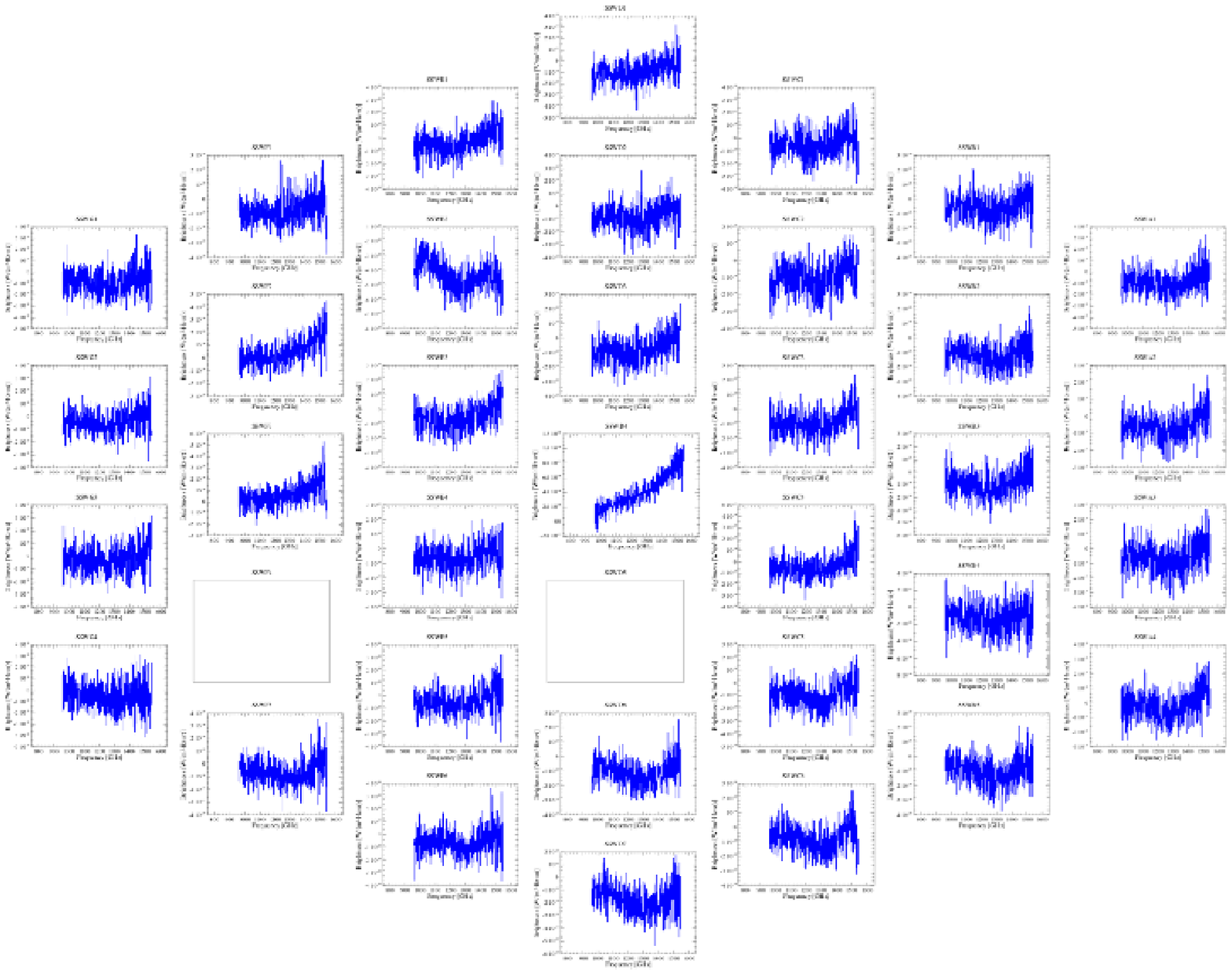}
\caption{{\it Top}: Array footprints for the SSW and SLW FTS arrays showing the central SSW D4 and SLW C3 detectors, the inner ring of detectors used for the off-axis background subtraction and the outer ring of vignetted detectors. {\it Bottom}: Detector readouts for the observation of IRAS 03158+4227 on OD804 (obsID=1342224764}
\label{fig:DetectorArrays}
\end{center}
\end{figure}

SPIRE spectroscopic observations were made in Point Source Spectroscopy mode ({\it SpireSpectroPoint}, SOF1,  \citealt{fulton10}) with high spectral resolution (HR, 0.048cm$^{-1}$), and sparse spatial sampling. Observations ranged from 45-100 repetitions (forward + reverse scans of the FTS, 6316 s -- 13752 s) depending on the predicted source continuum level. The SPIRE-FTS measures the Fourier transform of the spectrum of a source with two bolometer detector arrays (see Figure \ref{fig:DetectorArrays}); the Spectrometer Short Wavelength (SSW) and  Spectrometer Long Wavelength (SLW) simultaneously covering the entire wavelength range from  of 194--313$\mu$m (SSW)  and 303--671$\mu$m (SLW) respectively. Although the spectrometer arrays have 37 detectors (SSW) and   19 detectors (SLW), only the central SSWD4 and SLWC3 detectors are used for the point source spectrum.
 In addition to the target spectroscopic observations, the corresponding instrument {\it Dark Sky} {\it obsid} intended for background subtraction, for each spectroscopic observation is also listed (Section \ref{sec:Reduction}). These dark skies were supplied by the SPIRE Instrument Control Centre (ICC) on an observational day basis and were not attached specifically to the observation programme itself. Observations that were included in our data processing from {\it Herschel} archival data but were not within the HERUS programme itself are denoted by asterisks. The spectroscopic observation of IRAS07598+6508 had to be repeated due to a pointing error in the initial observation (obsid = 1342231979) but the photometric data were still usable.

Far-infrared spectroscopy was made using the {\it Herschel} Photodetector Array Camera and Spectrometer (PACS, ~\citealt{poglitsch10}). Observations with PACS for the 43 ULIRGs is split between the HERUS programme \citep{spoon13,farrah13} and the SHINING programme \citep{fischer10,sturm11,haileydunsheath10a,gonzalezalfonso13,mashi15}. A similar study of the intermediate redshift ULIRG population (0.21 $<$ z $<$ 0.88) has been presented by \citet{rigo14,magdis14}. The entire sample has already been observed \citep{armus07,farrah07} with the Infrared Spectrograph (IRS, \citealt{houck04}) onboard Spitzer \citep{wern04}.

\begin{table*}\tiny
\caption{Summary of HERUS observations of local ULIRGs}
\begin{tabular}{@{}llllllll}
\hline
Name           & Redshift & $\log$(L$_{IR}$) & \multicolumn{2}{c}{Photometer}& \multicolumn{3}{c}{Spectrometer}\\
               &          & L$_{\odot}$      & OD & ObsID & OD & ObsID & Dark ObsID\\
\hline
IRAS00397-1312 & 0.262 	  & 12.97            & 949 & 1342234696 & 1111 & 1342246257 & 1342246261	\\
Mrk1014        & 0.163    & 12.61            & 976 & 1342237540 & 998 & 1342238707\tablenotemark{a} & 1342238702	\\
3C273          & 0.158    & 12.72                        & 948  & 1342234882 & - & - & -	\\
IRAS03521+0028 & 0.152    & 12.56                        & 1022 & 1342239850 & 997 & 1342238704 & 1342231982	\\
IRAS07598+6508 & 0.148    & 12.46                        & 862  & 1342229642 & 1255 & 1342253659 & 1342253653	\\
IRAS10378+1109 & 0.136    & 12.38                        & 948  & 1342234867 & 1131 & 1342247118 & 1342247109	\\
IRAS03158+4227 & 0.134    & 12.6                         & 825  & 1342226656 & 804 & 1342224764 & 1342224758	\\
IRAS16090-0139 & 0.134    & 12.54                        & 862  & 1342229565 & 997 & 1342238699 & 1342238702	\\
IRAS20100-4156 & 0.13     & 12.63                        & 880  & 1342230817 & 1079 & 1342245106 & 1342245125	\\
IRAS23253-5415 & 0.13     & 12.4                         & 949  & 1342234737 & 1112 & 1342246277 & 1342246261	\\
IRAS00188-0856 & 0.128    & 12.47                        & 949  & 1342234693 & 1111 & 1342246259 & 1342246261	\\
IRAS12071-0444 & 0.128    & 12.4                         & 948  & 1342234858 & 1161 & 1342248239 & 1342248235	\\
IRAS13451+1232 & 0.122    & 12.32                        & 948  & 1342234792\tablenotemark{a} & 972 & 1342237024\tablenotemark{a} & 1342236999	\\
IRAS01003-2238 & 0.118    & 12.3                         & 949  & 1342234707 & 1111 & 1342246256 & 1342246261	\\
IRAS11095-0238 & 0.107    & 12.28                        & 948  & 1342234863 & 1151 & 1342247760 & 1342247753	\\
IRAS20087-0308 & 0.106    & 12.41                        & 880  & 1342230838 & 885 & 1342231049 & 1342231052	\\
IRAS23230-6926 & 0.106    & 12.32                        & 880  & 1342230806 & 1112 & 1342246276 & 1342246261	\\
IRAS08311-2459 & 0.1      & 12.46                        & 880  & 1342230796 & 879 & 1342230421 & 1342230416	\\
IRAS15462-0450 & 0.099    & 12.24                        & 989  & 1342238307 & 1178 & 1342249045 & 1342249068	\\
IRAS06206-6315 & 0.092    & 12.23                        & 825  & 1342226638 & 885 & 1342231038 & 1342231052	\\
IRAS20414-1651 & 0.087    & 12.24                        & 892  & 1342231345 & 1054 & 1342243623 & 1342243620	\\
IRAS19297-0406 & 0.086    & 12.38                        & 880  & 1342230837 & 886 & 1342231078 & 1342231052	\\
IRAS14348-1447 & 0.083    & 12.33                        & 989  & 1342238301 & 1186 & 1342249457 & 1342249454	\\
IRAS06035-7102 & 0.079    & 12.2                         & 353  & 1342195728\tablenotemark{a} & 879 & 1342230420 & 1342230416	\\
IRAS22491-1808 & 0.078    & 12.18                        & 949  & 1342234671 & 1080 & 1342245082 & 1342245125	\\
IRAS14378-3651 & 0.067    & 12.14                        & 989  & 1342238295 & 824 & 1342227456 & 1342227459	\\
IRAS23365+3604 & 0.064    & 12.17                        & 948  & 1342234919 & 804 & 1342224768 & 1342224758	\\
IRAS19254-7245 & 0.062    & 12.06                        & 515  & 1342206210\tablenotemark{a} & 885 & 1342231039 & 1342231052	\\
IRAS09022-3615 & 0.06     & 12.24                        & 880  & 1342230799 & 886 & 1342231063\tablenotemark{a} & 1342231052	\\
IRAS08572+3915 & 0.058    & 12.1                         & 880  & 1342230749 & 907 & 1342231978 & 1342231982	\\
IRAS15250+3609 & 0.055    & 12.03                        & 948  & 1342234775 & 998 & 1342238711 & 1342238702	\\
Mrk463         & 0.05     & 11.77                        & 963  & 1342236151 & 1178 & 1342249047 & 1342249068	\\
IRAS23128-5919 & 0.045    & 12                           & 544  & 1342209299\tablenotemark{a} & 1079 & 1342245110\tablenotemark{a} & 1342245125	\\
IRAS05189-2524 & 0.043    & 12.12                        & 467  & 1342203632\tablenotemark{a} & 317 & 1342192833\tablenotemark{a} & 1342192838	\\
IRAS10565+2448 & 0.043    & 12                           & 948  & 1342234869 & 1130 & 1342247096 & 1342247109	\\
IRAS17208-0014 & 0.043    & 12.38                        & 467  & 1342203587\tablenotemark{a} & 317 & 1342192829\tablenotemark{a} & 1342192838	\\
IRAS20551-4250 & 0.043    & 12.01                        & 880  & 1342230815 & 1079 & 1342245107\tablenotemark{a} & 1342245125	\\
Mrk231         & 0.042    & 12.49                        & 209  & 1342201218\tablenotemark{a} & 209 & 1342187893\tablenotemark{a} & 1342187890	\\
UGC5101        & 0.039    & 11.95                        & 495  & 1342204962\tablenotemark{a} & 544 & 1342209278\tablenotemark{a} & 1342208391	\\
Mrk273         & 0.038    & 12.13                        & 438  & 1342201217\tablenotemark{a} & 557 & 1342209850\tablenotemark{a} & 1342209858	\\
IRAS13120-5453 & 0.031    & 12.22                        & 829  & 1342226970 & 602 & 1342212342\tablenotemark{a} & 1342212320	\\
NGC6240        & 0.024    & 11.8                         & 467  & 1342203586\tablenotemark{a} & 654 & 1342214831\tablenotemark{a} & 1342214832	\\
Arp220         & 0.018    & 12.14                        & 229  & 1342188687\tablenotemark{a} & 275 & 1342190674\tablenotemark{a} & 1342190675	\\
\hline
\end{tabular}
\tablecomments{The operational day (OD) and observation identification (obsID) are tabulated for  the observations made with the SPIRE photometer and spectrometer. The obsID for the associated spectrometer dark sky is also listed. Indicative far-infrared luminosities ( L$_{IR}$= L$_{8-1000\mu m}$) calculated following \citet{sanders96} are included for reference. Note that 3C273 has photometer data but no corresponding spectroscopic data was taken. \\ $^{a}$ {\itshape Herschel} archival data.}
\label{tab:observations}
\end{table*}

\section{Data Reduction}\label{sec:Reduction}

\subsection{Pipeline}\label{sec:Pipeline}
\begin{figure} 
\begin{center}
\includegraphics[width=0.75\columnwidth,angle=0]{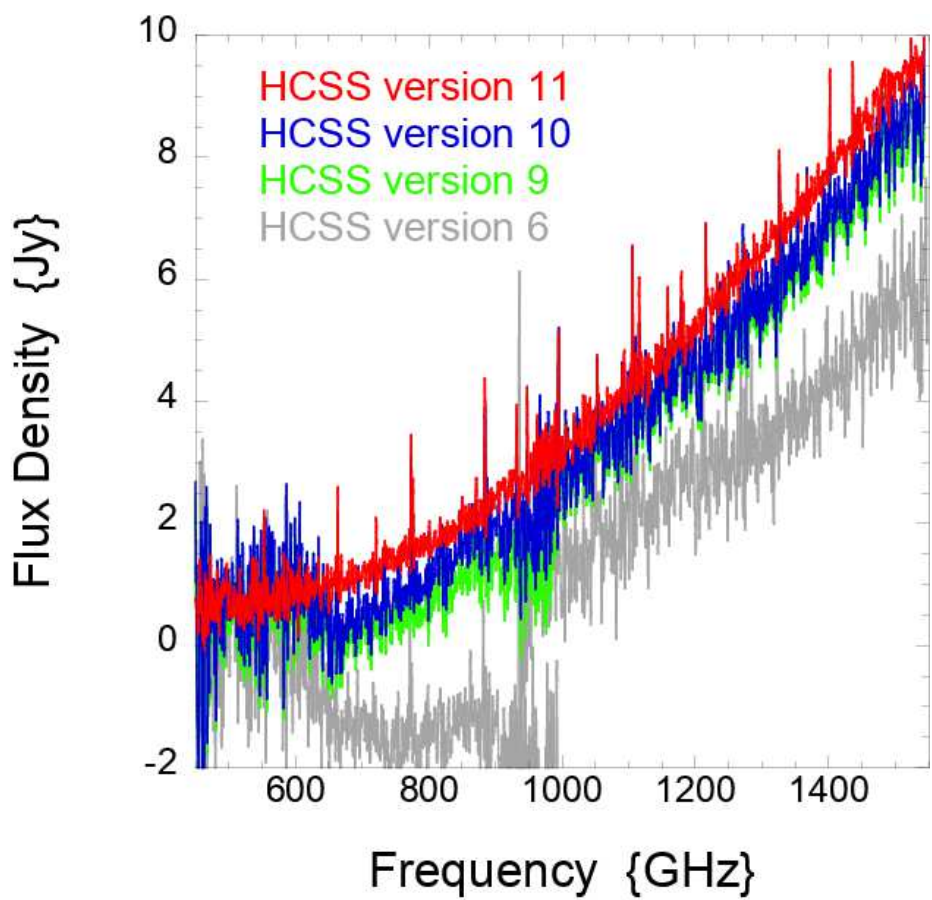}
\caption{Comparison of the improvement in pipeline processing within the HCSS HIPE environment, for Mrk 231, obsid=1342187893. The earliest pipeline processed spectrum shows an offset between the SSW and SLW bands due to sub-optimal subtraction of the telescope emission. The latest HIPE v11 data is processed with either new RSRFs + subktemp correction or new RSRFs + new telescope model correction. The HIPE v11 spectrometer pipeline also features significantly reduced noise compared to HIPE v10.}
\label{fig:pipelineSpectra}
\end{center}
\end{figure}

All spectrometer data was processed through the Spectrometer Single Pointing User Pipeline (\citealt{fulton10}, \citealt{fulton14}) within the Herschel Common Science System  {\it Herschel Interactive Processing Environment} (HIPE \citealt{ott10}). All the data were reprocessed with HIPE v11.2757 SPIRE Calibration Tree version 11.0. HIPE v11 generally improves the sensitivity levels predicted by HSpot \footnote{Herschel Observation Planning Tool} over the entire FTS wavelength range, by $\sim$0.05 Jy for a 1 hour observation (i.e. approximately 20$\%$ improvement). The error on the continuum shape also improves, with an average reduction in the offset of 0.08 Jy. This is due to the higher signal-to-noise in the flux calibration and is particularly important for the relatively faint lines in many of the HERUS sample (The improvements are most significant for observations taken after OD 998). Full details of the treatment of faint sources with the SPIRE FTS pipeline are explained in \citet{hopwood15}. In Figure ~\ref{fig:pipelineSpectra}, the evolution in data quality as a function of pipeline version is shown for the example of Mrk 231 (obsid=1342187893). By HIPE v10 the discontinuity between the SPIRE FTS SSW and SLW bands was improved, while the overall noise in the spectrum was further decreased between HIPE v10 and HIPE v11. Figure ~\ref{fig:pipelineSpectra} also shows the typical resulting spectrum from earlier versions (e.g. HIPE v6) of the spectrometer pipeline (e.g. Mrk 231: \citealt{vanderwerf10}, \citealt{gonzalezalfonso10}, Arp 220: \citealt{rangwala11}, NGC 6240: \citealt{meijerink13}). All the data, both HERUS and archival in Table ~\ref{tab:observations} have been re-reduced with the HIPE v11 pipeline, using the default values for all pipeline tasks. The pipeline produces {\it Level 1} products for all the array detectors in the form of extended emission calibrated spectra in W m$^{-2}$ Hz$^{-1}$ sr$^{-1}$ and the final  {\it Level 2} data products as point source calibrated unapodized spectra for the central detectors SLW C3 and SSW D4 measured in Jy as a function of frequency in GHz.

Although the current version of the SPIRE spectrometer pipeline is HIPE 14, there has not been a significant improvement in the sensitivity for faint point source spectroscopy since HIPE version 11. The most significant changes in HIPE 14 regard bright sources, extended mapping and low resolution spectroscopy \citep{hopwood15,fulton14}. It should be noted that much of the additional post pipeline processing described in Section 3.2 has since been incorporated into the standard HIPE 14 pipeline and therefore the results presented in the work would not significantly improve with re-processing using HIPE 14. 

\begin{figure} 
\begin{center}
\includegraphics[width=0.75\columnwidth,angle=0]{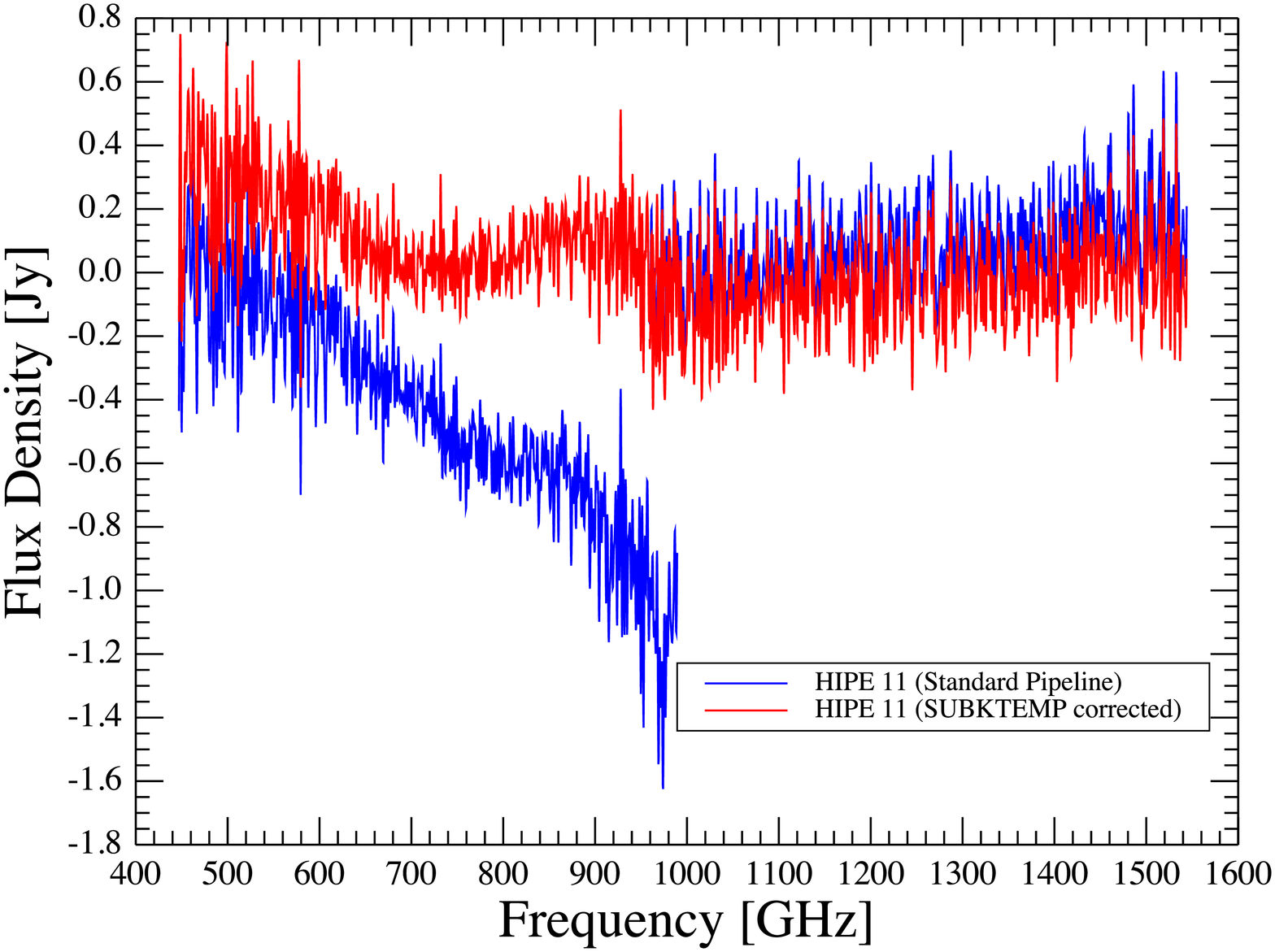}
\caption{Correction for the flux {\it drooping} effect exhibit by the SLW C3 central detector for FTS observations taken within 8 hours of a SPIRE cooler recycle with a cooler burp event (obsid = 1342246256). The discontinuity in the standard pipeline processed observation with HIPE v11 is compared with the corrected continuous spectrum obtained from an empirical correction utilising the correlation found between the flux density measured by the detector and the {\it SUBKTEMP} temperature.}
\label{fig:FluxDroop}
\end{center}
\end{figure}

An exception to the standard pipeline processing were FTS observations made within 8 hours of the beginning of an SPIRE FTS observation period, which were often found to suffer from a decrease or {\it drooping} in the SLW band flux. The SPIRE cooler was recycled at the beginning of every pair of SPIRE days and in some cases anomalous {\it Cooler Burps} (see Pearson et al., in preparation) caused an outlying low temperature in the instrument 0.3K cooler stage causing correspondingly low detector temperatures. There was found to be a correlation between this temperature change (measured by the instrument {\it  SUBKTEMP} sensor) and flux density. This effect was not corrected by the pipeline non-linearity correction and moreover cannot be corrected using a dark sky subtraction since the dark sky is more likely to have an average {\it  SUBKTEMP} and therefore no {\it droop}. For affected observations, the flux {\it droop} was empirically corrected within the pipeline using a linear correlation found between the SLW flux density and the {\it  SUBKTEMP}  temperature. This correction is shown in Figure ~\ref{fig:FluxDroop} for the example of IRAS 01003-2238 (obsid = 1342246256). Note that different detectors were found to have different sensitivities to this issue, e.g. SLWC3 is found to be very sensitive to  variations in {\it  SUBKTEMP}, but SSWD4 showed no clear correlation and therefore was not corrected. Observations that were corrected for the {\it  SUBKTEMP}  flux {\it drooping} were IRAS 00397-1312 (1342246256), IRAS 01003-2238 (1342246256), IRAS 06206-6315 (1342231038), IRAS 10565+2448 (1342247096), IRAS 16090-0139 (1342238699), IRAS 20100-4156 (1342245106), IRAS 20551-4250 (1342245107) and  IRAS 23128-5919 (1342245110).

Many of the corrections and techniques used in the data reduction in this work have also been incorporated into the standard processing for later versions of HIPE (13.0, 14.0 \citealt{hopwood15}) however, many such cases still need to be processed on a case by case basis.

\subsection{Post-Pipeline Processing}\label{sec:PostPipeline}
\begin{figure} 
\begin{center}
\includegraphics[width=0.75\columnwidth,angle=0]{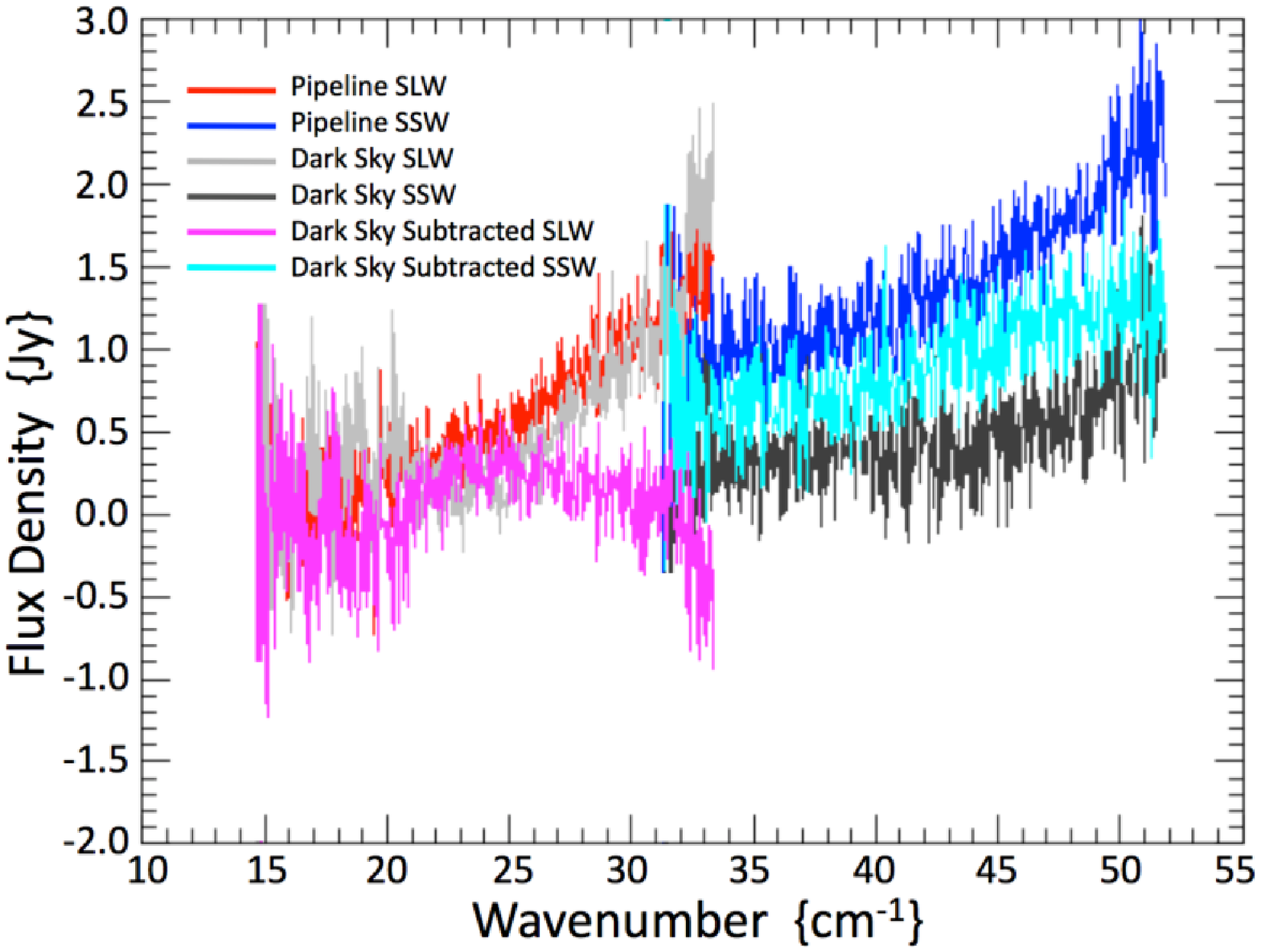}
\caption{Problems with post processing background removal using the OD803 dark sky observation associated with the FTS observation of IRAS 03158+4227 (obsid = 1342224764) taken on OD 804.  Pipeline SLW and Pipeline SSW are the spectra as naively processed by the standard pipeline exhibiting an offset bwteen the two spectral bands. The Dark Sky spectra themselves are labelled as {\it Dark Sky SLW} and {\it SSW} respectively. The post processed dark subtracted spectra are shown as {\it Dark Sky Subtracted SLW} and {\it  SSW} respectively. Note that the dark sky subtraction method has over compensated at the short wavelength end of the SLW band.}
\label{fig:pipelineDarkSpectra}
\end{center}
\end{figure}

For all but the brightest {\it Herschel} SPIRE FTS observations the total measured signal is always dominated by the emission from the $\sim$80K telescope. The contributed flux density corresponding to this emission is of the order of $\sim$200-800Jy. After pipeline processing residual telescope emission causes both a distortion in the overall spectral shape and a discontinuity between the spectra in the SSW and SLW bands due to the variation of beam size with frequency as shown in the {\it left} panel of Figure \ref{fig:pipelineDarkSpectra}. ~\citet{swinyard14} estimate that the associated offset for the central detectors is $\sim$0.4 Jy and $\sim$0.29Jy for SLWC3 and SSWD4 respectively meaning that the majority of the HERUS sample will be severely affected.

Therefore, in order to correct for discontinuities between the SSW and SLW spectra, post-processing for additional background subtraction is required. Several methods for background subtraction are possible;

\begin{enumerate}
\item To use the standard dark sky observation taken on or around the same OD.
\item To use a super-dark averaged from many dark sky observations.
\item To use the off-axis detectors on the spectrometer arrays to produce an effective local dark sky measurement.
\end{enumerate}

For every pair of SPIRE spectrometer days, a generic dark sky observation was taken by the SPIRE ICC with a duration corresponding to the longest FTS observation from any programme taken on that pair of ODs. For the HERUS sample the corresponding dark sky observations are listed in Table \ref{tab:observations}. In Figure~\ref{fig:pipelineDarkSpectra} the results for the pipeline processed data for the FTS observation of IRAS 03158+4227 (obsid = 1342224764) taken on OD 804 show an offset between the two spectrometer bands SSW, SLW. This offset is due to differences in the background (including telescope) subtraction. In some cases this offset is overcome by subtracting a spectrometer observation of dark sky of similar integration. In the case of Figure~\ref{fig:pipelineDarkSpectra}, the dark sky observation taken on OD803 (obsID=1342224758) was used. Unfortunately, the subtraction is unsatisfactory, as sometimes happens when the dark from a different day or different number of repetitions is used (in this case the observation was taken on OD804 whilst the dark was taken on OD803). Processing using a {\it Super Dark} averaged from many independent dark sky observations also exhibits similar issues.

\begin{figure} 
\begin{center}
\includegraphics[width=0.45\columnwidth,angle=0]{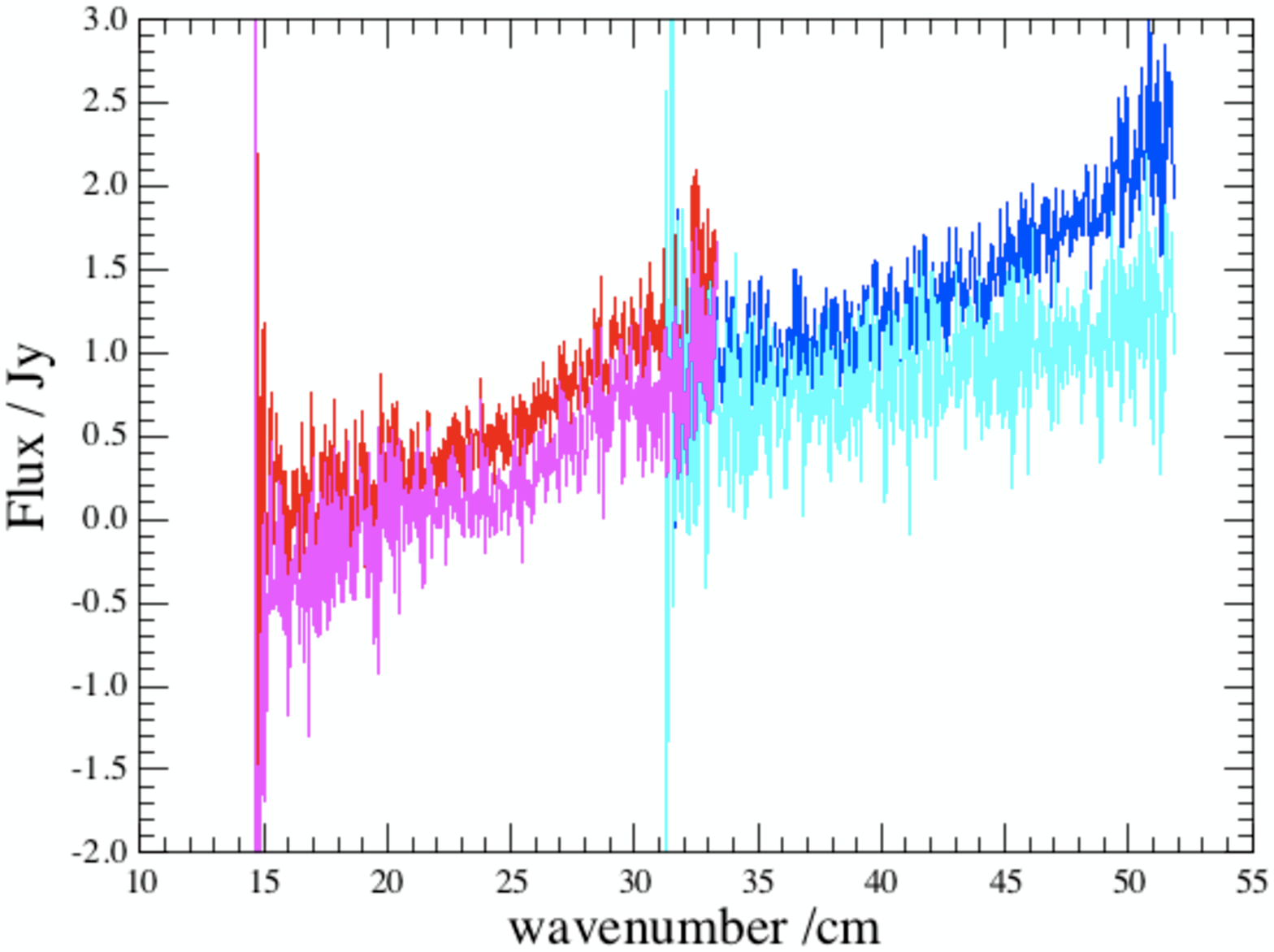}
\includegraphics[width=0.45\columnwidth,angle=0]{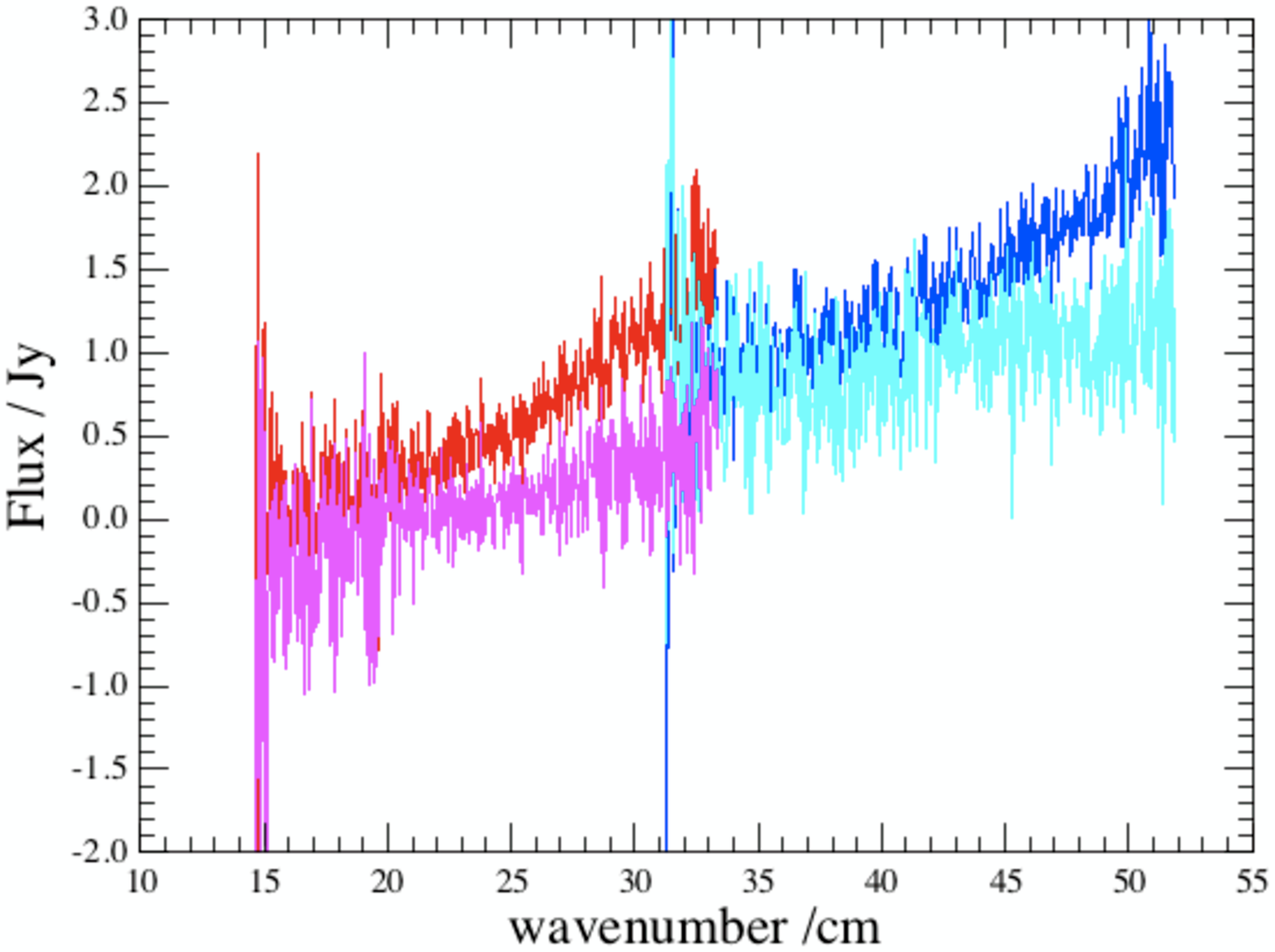}
\includegraphics[width=0.45\columnwidth,angle=0]{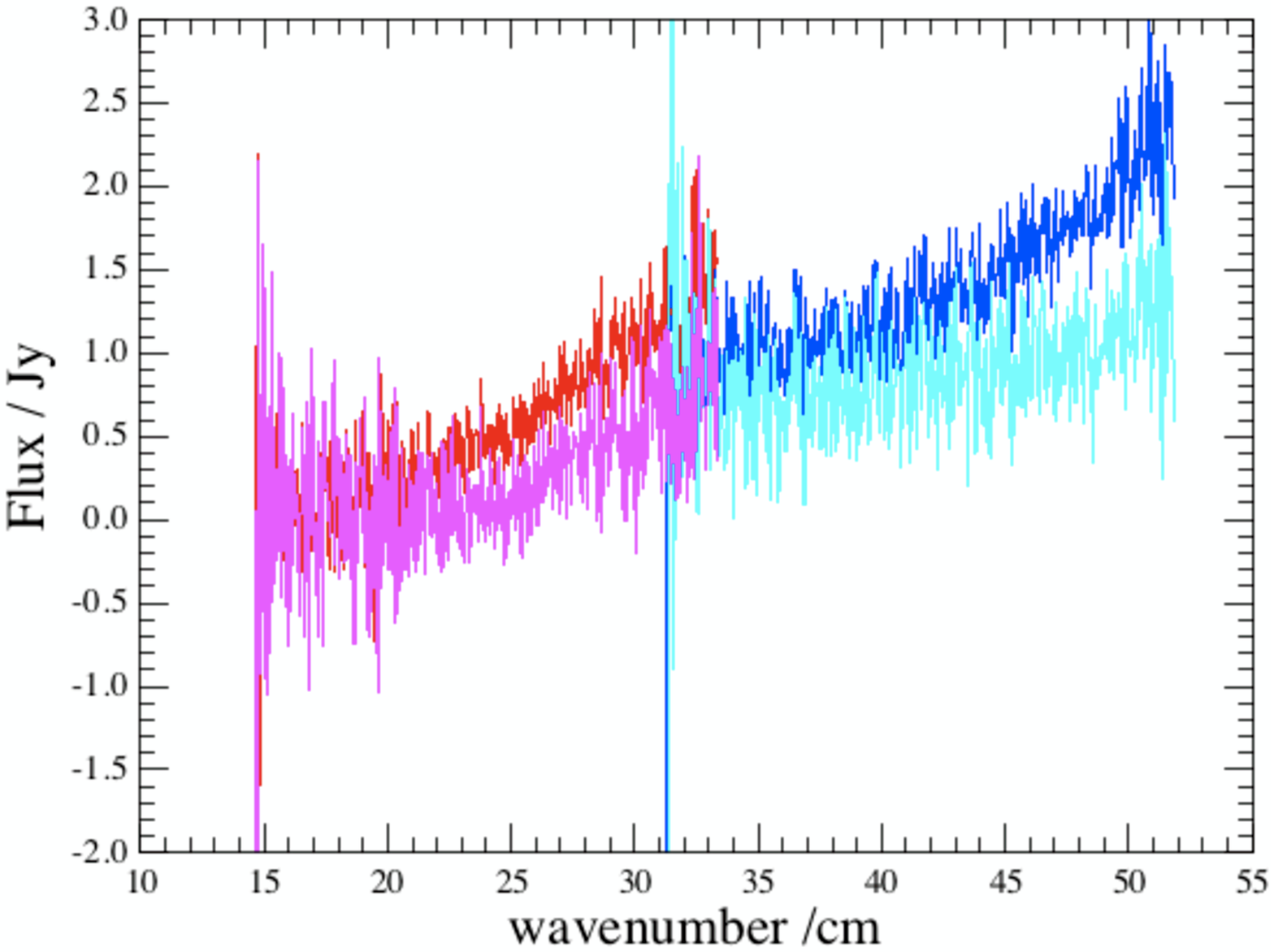}
\includegraphics[width=0.45\columnwidth,angle=0]{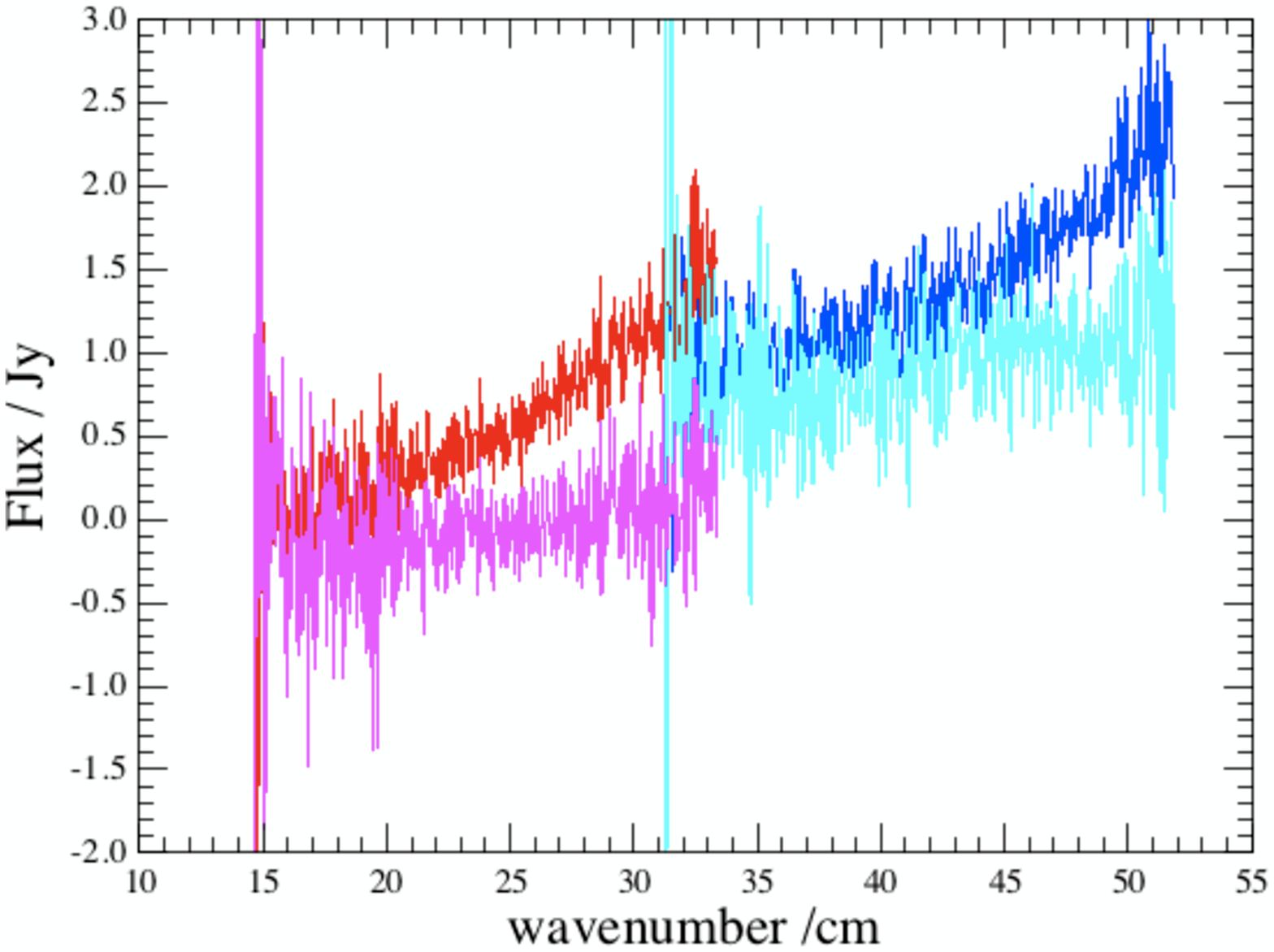}
\caption{Using the off-axis detectors for the dark subtraction for the observation. Upper lines ({\it red, blue}) show original pipeline spectrum and lower lines ({\it magenta, cyan}) show the spectra after correction by dark subtraction using a specific pair of SSW, SLW off-axis co-aligned detectors that view the same sky as each other detectors. From left to right, using SSW C5 \& SLW C4 (co-aligned detectors), SSW C5 \& SLWB3 (co-aligned detectors), SSW D1 \& SLW E2 (co-aligned detectors), SSW D7 \& SLW A2 (co-aligned detectors).}
\label{fig:SingleOffAxisDetectors}
\end{center}
\end{figure}

An alternative to using the dedicated dark sky or super dark observations is to use the off-axis (non-central, see Figure ~\ref{fig:DetectorArrays}) detectors to produce an effective {\it local dark} observation. The advantage with this option is that the dark observation is effectively taken at the same time as the observation. The disadvantage is that the dark is taken with a different detector to that of the target central detector. A number of detectors on the SSW array are co-aligned (see the same area of sky) so a corresponding detector on the SLW array and these co-aligned detectors were used to evaluate the quality of off-axis detector background subtraction. Figure~\ref{fig:SingleOffAxisDetectors} shows the results of using different pairs of co-aligned off-axis pairs for the dark subtraction. Unfortunately it was found that the subtraction varies from off-axis detector pair to off-axis detector pair.


In order to overcome the differences from detector to detector, an average background can be computed using the outer rings of non-vignetted detectors on the SSW and SLW arrays, as shown in Figure ~\ref{fig:DetectorArrays} for the example observation of IRAS03158+4227 on OD804 (obsID=1342224764). Note that Figure ~\ref{fig:DetectorArrays} also shows the footprints of each array with the detector read-outs for processed spectra with the target on the central SSW D4 and SLW C3 detectors. It can be seen that there can still be non-negligible outliers within the off-axis detectors. These differences can be due to individual detector performance during an observation or due to emission from some source or astrophysical background that serendipitously falls on that particular off-axis detector.

\begin{figure} 
\begin{center}
\includegraphics[width=0.32\columnwidth,angle=0]{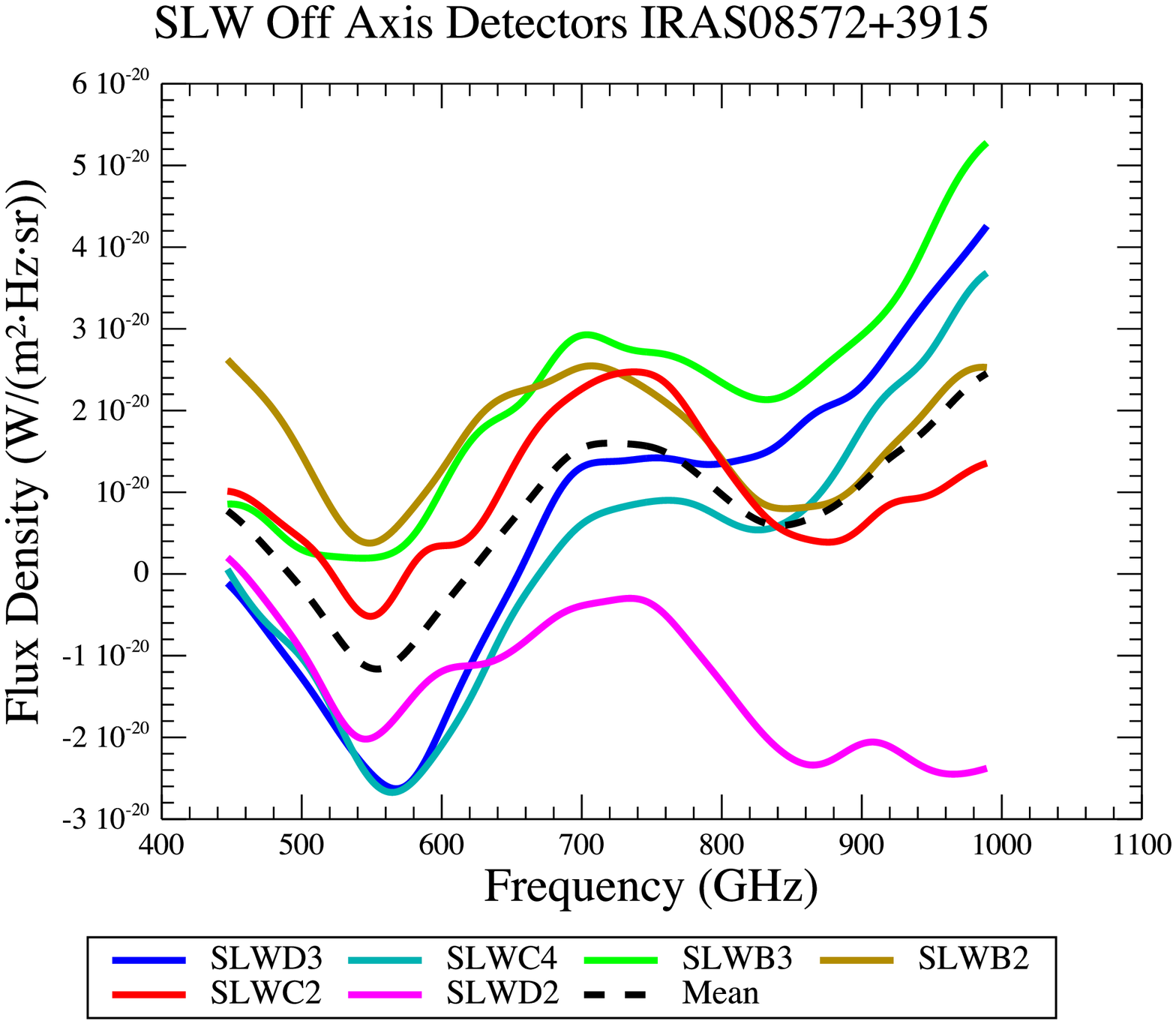}
\includegraphics[width=0.32\columnwidth,angle=0]{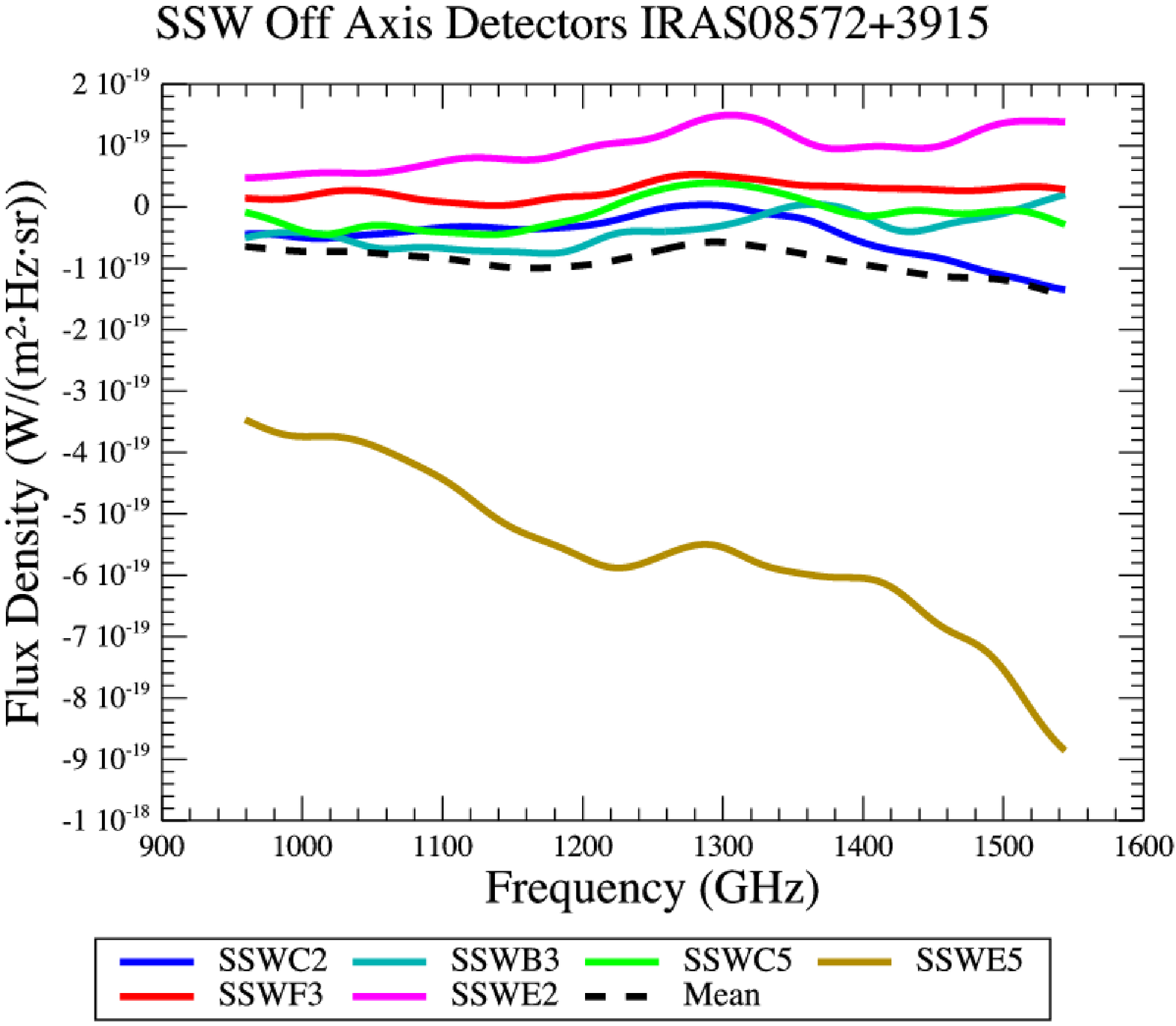}
\includegraphics[width=0.32\columnwidth,angle=0]{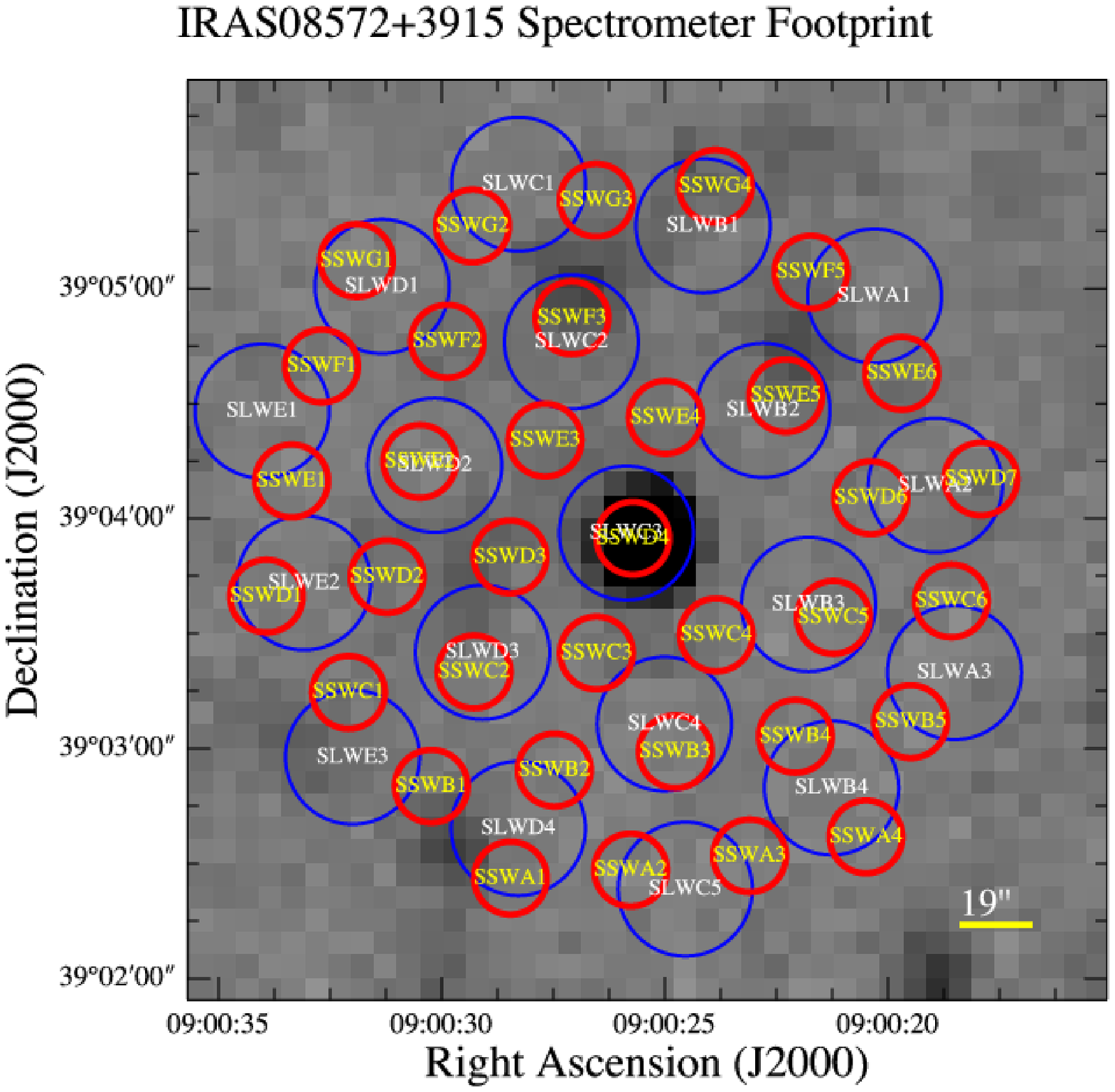}
\includegraphics[width=0.32\columnwidth,angle=0]{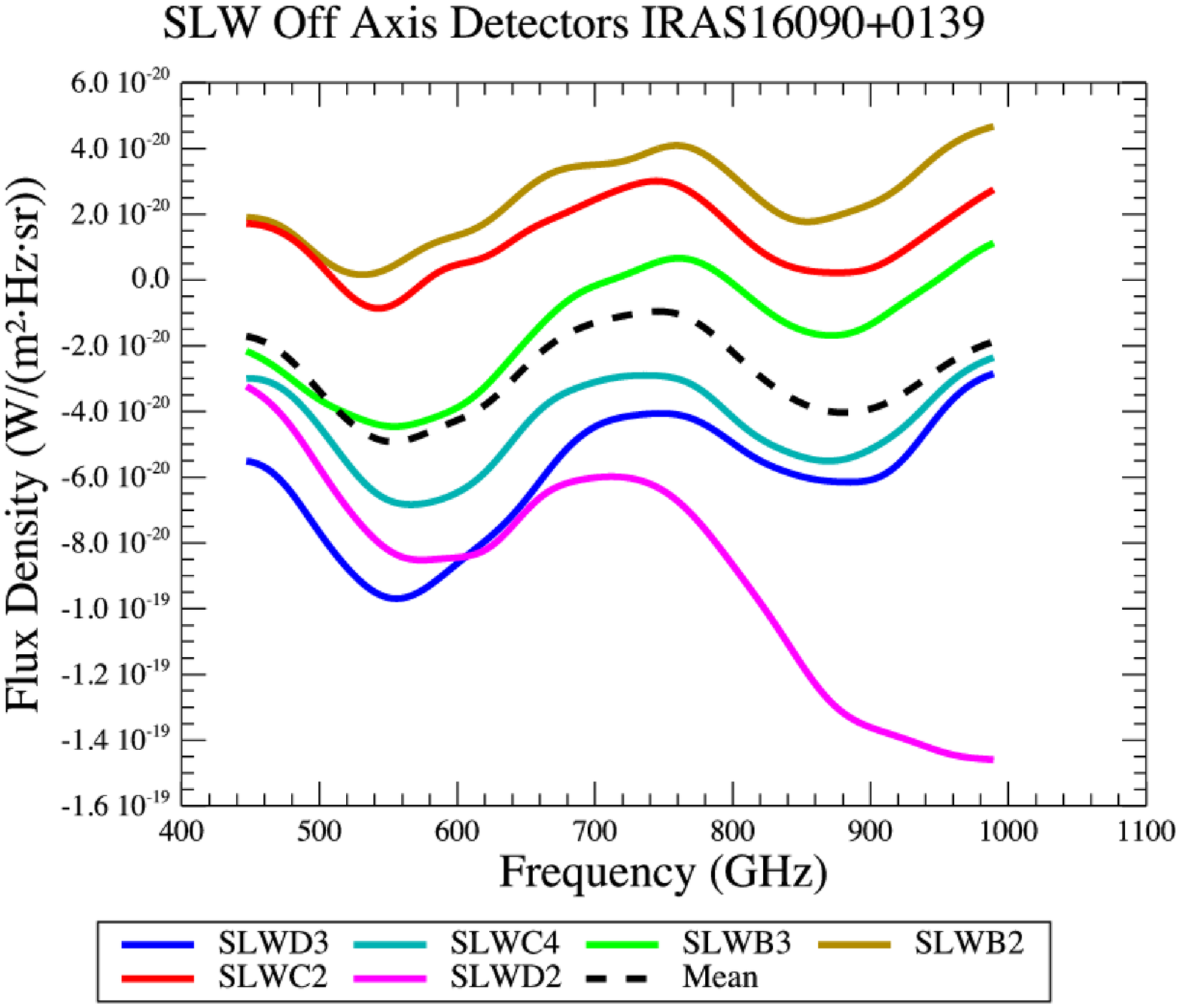}
\includegraphics[width=0.32\columnwidth,angle=0]{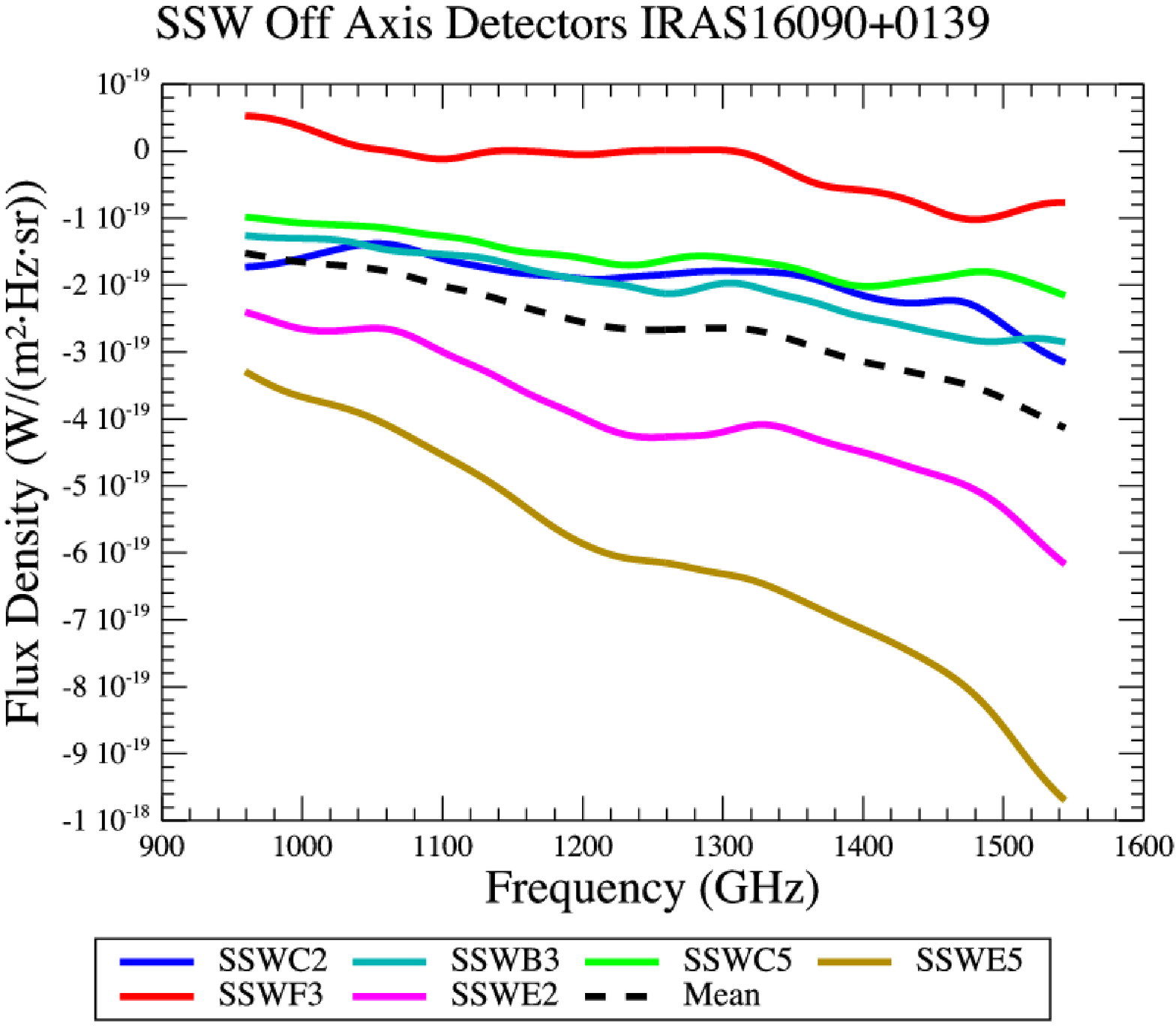}
\includegraphics[width=0.32\columnwidth,angle=0]{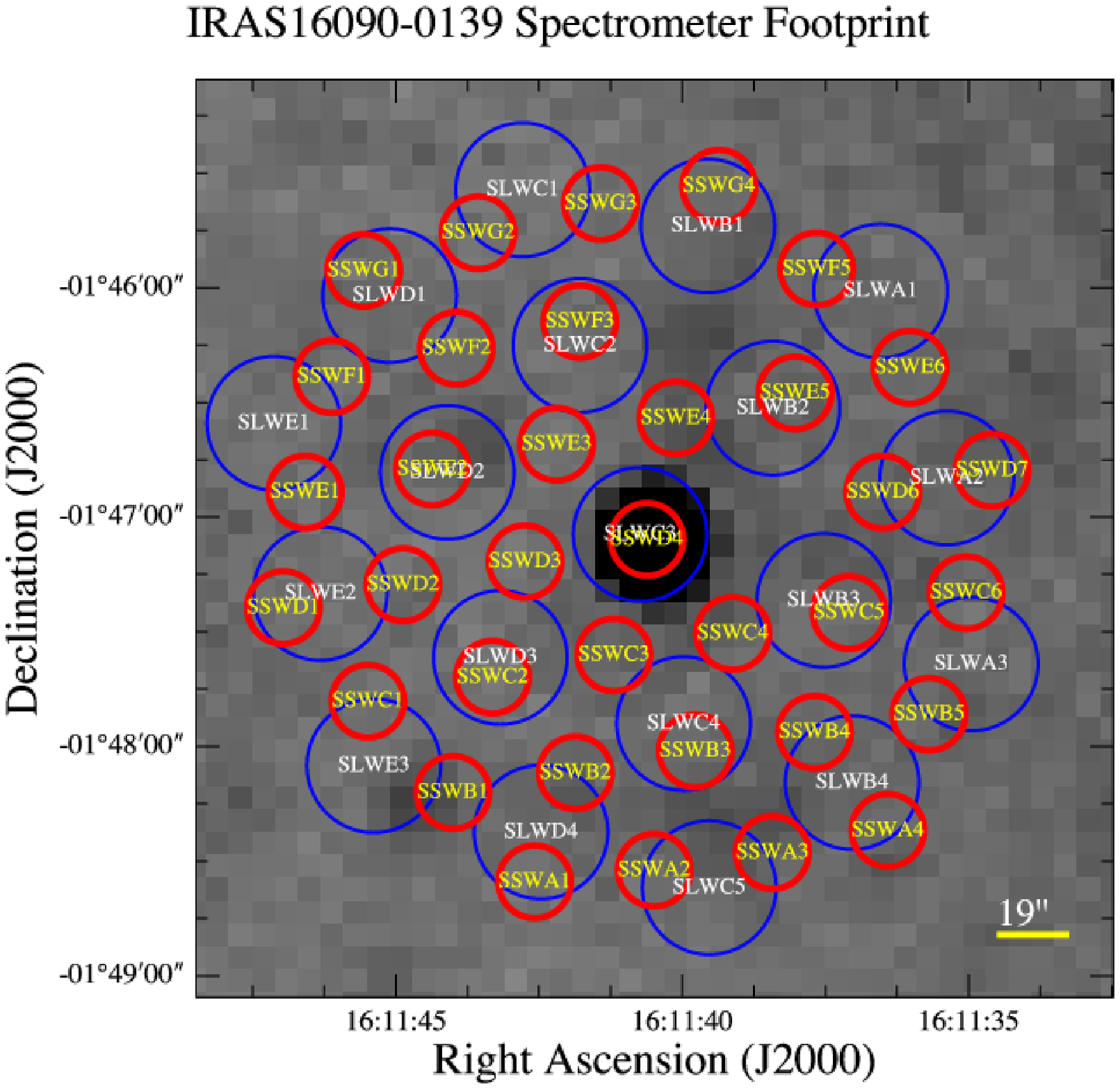}
\caption{Example of the selection of median smoothed off-axis detector spectra for background subtraction for the cases of IRAS 08572+3915 and IRAS 16090+0139. Using a combination of the off-axis detector spectra and the inspection of the spectrometer array footprint overlaid on the photometry maps, outliers are rejected and the remaining off-axis detectors used to form an averaged off-axis super-dark. For IRAS 08572+3915 the SLWD3, SLWB2 \& SSWE5 detectors are excluded. For IRAS 16090+0139 the SLWD2, SSWE5 and SSWE2 detectors are excluded. The resulting spectra are shown in Figure ~\ref{fig:offAxisCorrected}.}
\label{fig:outlierRejection}
\end{center}
\end{figure}

Outlier rejection of anomalous off-axis detectors was carried out by inspecting the smoothed spectra from each off-axis detector by eye to reject any outliers as shown in Figure~\ref{fig:outlierRejection}. In addition, the spectrometer array footprint was overlaid on the SPIRE photometer maps to identify cases where the off-axis detectors were adversely affected by any background emission or source serendipitously lying within the detector beam on the sky as shown in the {\it right}-panels of Figure~\ref{fig:outlierRejection}.

The processing steps to obtain the final Level 2 point source calibrated spectra using the off-axis detectors for the background subtraction are;
\begin{enumerate}
\item Begin with Level 1 product spectra (W m$^{-2}$ Hz$^{-1}$ sr$^{-1}$)
\item Average all (forward \& reverse) FTS scans for each individual detector
\item Inspect the smoothed spectra from the off-axis detectors by eye to reject any outliers as shown in Figure~\ref{fig:outlierRejection}
\item Average all selected off-axis detector spectra to form an off-axis super-dark
\item Subtract off-axis super-dark from central detector spectra for both the SSW and SLW arrays
\item Filter Channels (select the central detectors only)
\item Carry out the point source calibration on the 2 central detector channels
\item Final Level 2 point source calibrated spectra (Jy)
\end{enumerate}

\begin{figure} 
\begin{center}
\includegraphics[width=0.32\columnwidth,angle=0]{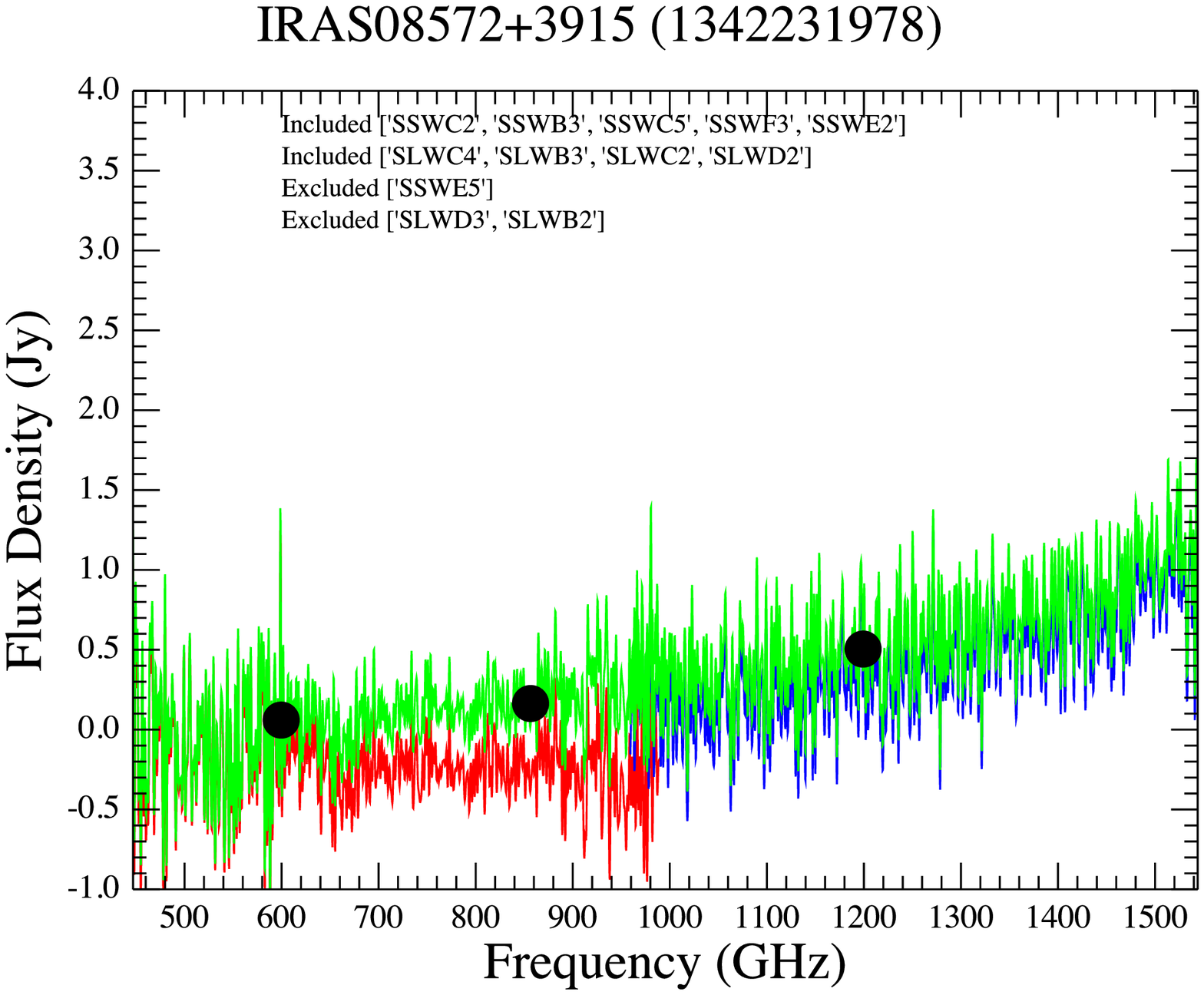}
\includegraphics[width=0.32\columnwidth,angle=0]{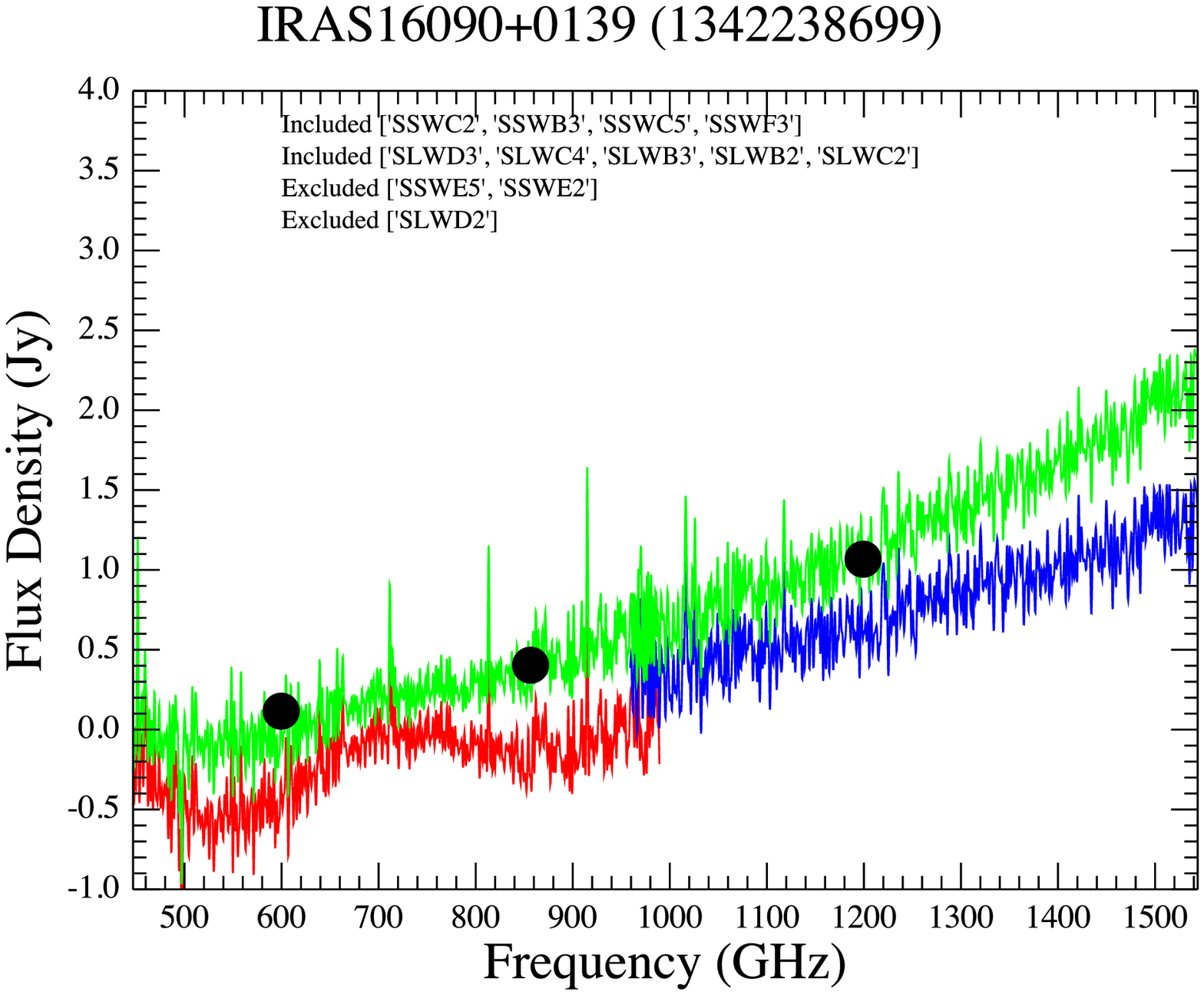}
\includegraphics[width=0.32\columnwidth,angle=0]{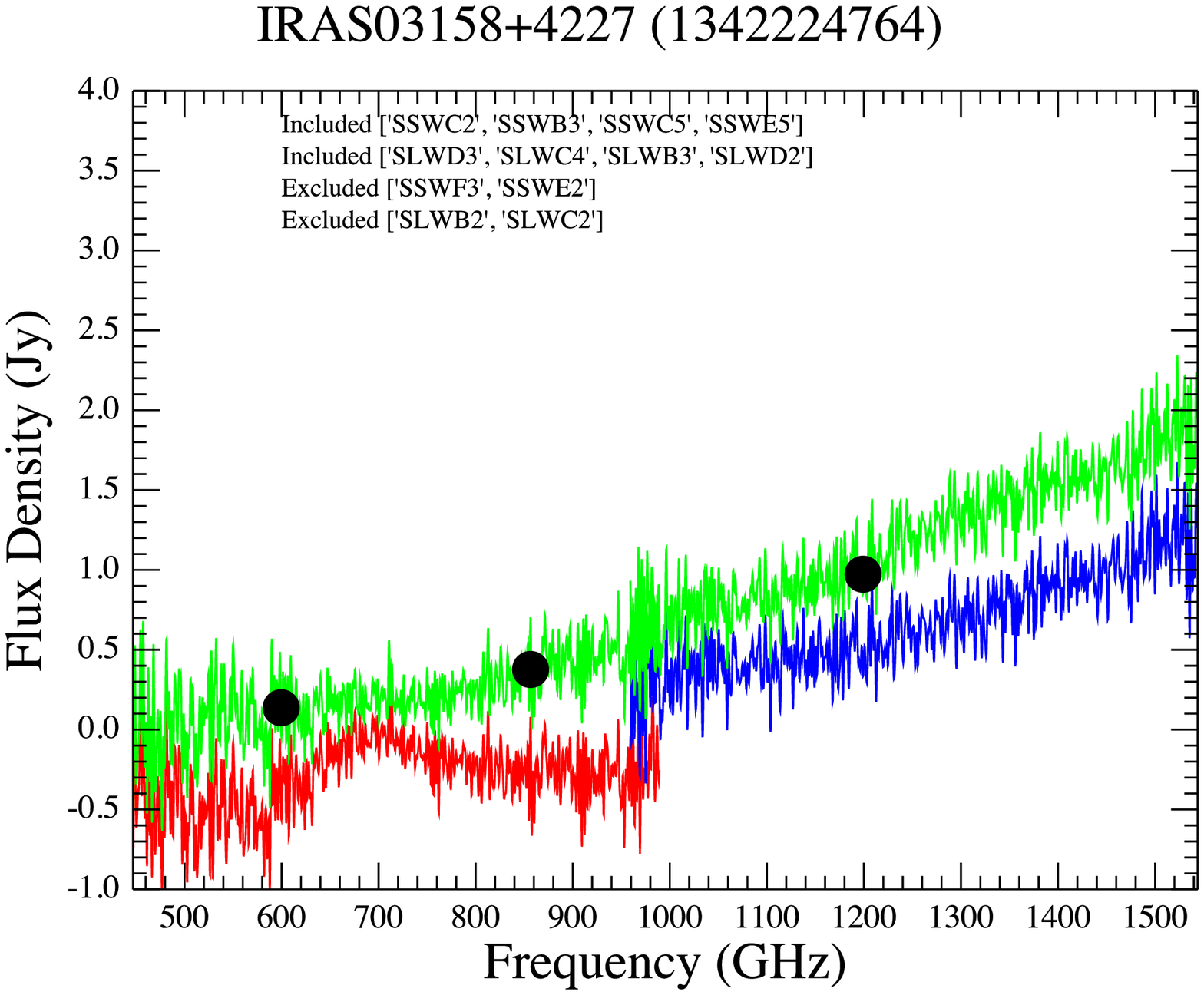}
\caption{Final spectra (continuous green line) after all post processing steps described in the text compared with the original pipeline processed spectra for 3 HERUS galaxies, IRAS 08572+3915, IRAS 16090+0139 \& IRAS 03158+4227 (obsid = 1342231978, 1342238699 \&  1342224764 respectively). After the post processing steps we find the SSW and SLW band spectra are well aligned, and that the photometry is in good agreement with the spectra. The SSW and SLW off-axis detectors included and excluded for the dark subtraction are listed. For comparison, the detector spectra before post processing are shown for the SLW (red) and SSW (blue) arrays respectively.}
\label{fig:offAxisCorrected}
\end{center}
\end{figure}

Applying this method of off-axis detectors for the dark subtraction results in a much better agreement for the SSW and SLW spectra (e.g. the 3 representative observations shown in Figure~\ref{fig:offAxisCorrected},  (obsid = 1342231978, 1342238699 \& 1342224764). In almost all cases  the corrected spectral continuum level agrees well with the measured SPIRE photometry (see Section ~\ref{sec:Photometer}) for SSW (250$\mu$m band) and SLW (350$\mu$m, 500$\mu$m bands). In order to remove any residual background, the baseline is subtracted by fitting a 3 order polynomial to the continuum and then fit to the SPIRE photometry points using a polynomial fit for the photometer to normalize the spectrum baseline. 

The standard spectrometer pipeline assumes that the source is a point source on the central detector.  However, if a source is extended with respect to the beam there may be a discontinuity in the overlapping spectral region between the short and long wavelength bands due to the sudden change in the effective beam size at those frequencies, given that the SLW beam diameter is approximately twice that of SSW beam diameter. The discontinuities found for extended sources can be corrected using the dedicated Semi-Extended Correction Tool (SECT) within HIPE. The SECT tool allows the fitting of a Gaussian model for a source and corrects the spectra continuum and flux accordingly.
 
We have identified possible extended sources within our sample by comparing the photometer maps with either the Photometer beam or the FTS footprint as shown in Figure ~\ref{fig:outlierRejection}. A source is flagged as being possibly extended if it falls within the first external ring of FTS detectors. Sources that fall within the category are IRAS 10565+2448, IRAS 13120-5453, IRAS 17208-0014, NGC6240 and UGC 5101. To correct them, we have followed the procedure detailed in  \citet{wu13} using the SECT tool within HIPE but find only small differences, of 1 to $<$5$\%$.

\subsection{Final Spectra}\label{sec:spectra}
\begin{figure} 
\begin{center}
\includegraphics[width=0.85\columnwidth,angle=0]{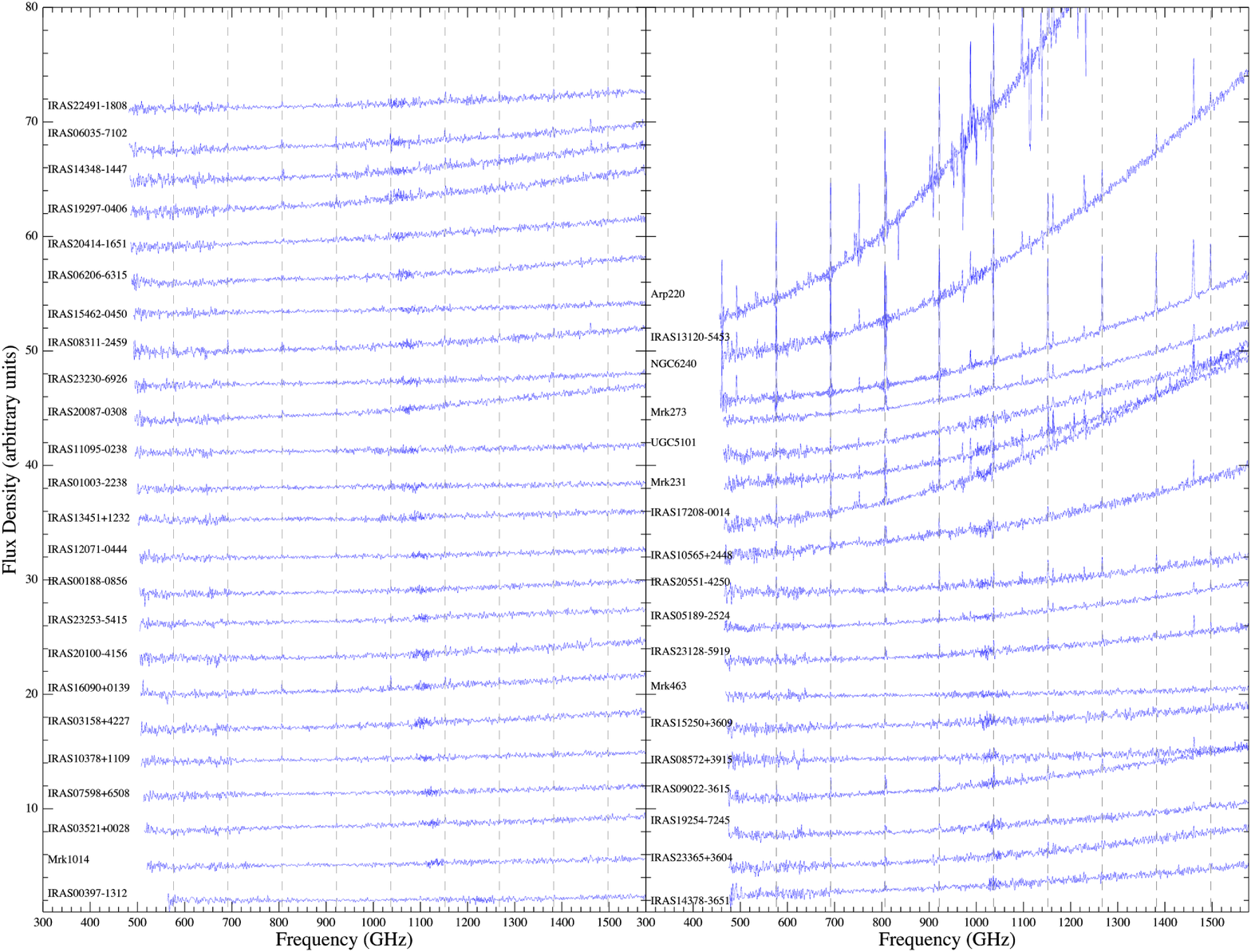}
\caption{Final processed spectra for all HERUS targets in the rest frame. Targets are ordered from bottom to top and left to right in decreasing redshift. Flux density is in arbitrary offsets for clarity. Also overlaid as dotted lines are the lines on the CO ladder from the CO(5-4), CO(6-5), CO(7-6), CO(8-7), CO(9-8), CO(10-9), CO(11-10), CO(12-11), CO(13-12) transitions at 576, 691, 806, 921, 1036, 1151, 1267, 1381, 1496 GHz (519, 432, 371, 324, 288, 260, 236, 216, 200$\mu$m).}
\label{fig:allSpectra}
\end{center}
\end{figure}

After the pipeline processing (including {\it SUBKTEMP} correction where necessary) and the post processing including the background subtraction using the off-axis detectors, fitting to the SPIRE photometry and correction for any extension, the final point source calibrated spectra are produced in Jy as a function of frequency (GHz). In Figure ~\ref{fig:allSpectra} a summary of the final spectra for all HERUS targets are shown. The spectra are in the galaxy rest frame ordered by increasing redshift. Many lines are visible in the spectra: we overlay vertical lines for the $^{12}$CO ladder transitions between CO(5-4) to CO(13-12) at 576-1496 GHz (519-200$\mu$m).

\subsection{Photometry of HERUS sources}\label{sec:Photometer}
All SPIRE photometry observations (listed in Table ~\ref{tab:observations}) were  processed through the standard Small Map User Pipeline with HIPE 11.2825, using SPIRE Calibration Tree 11.0 with default values for all pipeline tasks. Target positions in the map were found by the HIPE SUSSEXtractor task \citep{savage07}, assuming a Full Width Half Maximum (FWHM) of 18.15\arcsec, 25.2\arcsec, 36.9\arcsec\ for the PSW, PMW, PLW bands respectively. These positions were then input the SPIRE Timeline Fitter task within HIPE \citep{pearson14} that fits a Gaussian function to the baseline-subtracted SPIRE timelines. The background is measured within an annulus between of 300 and 350 arcsec and then an elliptical Gaussian function is fit to both the central 22\arcsec, 32\arcsec, 40\arcsec (for the PSW, PMW, PLW bands respectively) and the background annulus. The results for the photometry of all the HERUS galaxies is shown in Table ~\ref{tab:photometry}. Full details will be given in Clements et al. (in preparation).

\begin{table*}\tiny
\caption{Photometry of the HERUS galaxies in the SPIRE bands PSW (250$\mu$m), PMW (350$\mu$m), PLW (500$\mu$m).}
\centering
\begin{tabular}{@{}llllllll}
\hline
Target         &  obsid     &  \multicolumn{2}{c}{PSW} &  \multicolumn{2}{c}{PMW}  &  \multicolumn{2}{c}{PLW} \\
               &            &  Flux Density /Jy   & Error /Jy & Flux Density /Jy & Error /Jy & Flux Density /Jy & Error /Jy	\\
\hline
IRAS00397-1312 & 1342234696 & 0.389  & 0.004 & 0.130  & 0.004 & 0.040 & 0.005	\\
Mrk1014        & 1342237540 & 0.460  & 0.004 & 0.175  & 0.004 & 0.063 & 0.005	\\
3C273          & 1342234882 & 0.437  & 0.004 & 0.633  & 0.004 & 0.994 & 0.005	\\
IRAS03521+0028 & 1342239850 & 0.684  & 0.004 & 0.270  & 0.004 & 0.094 & 0.004	\\
IRAS07598+6508 & 1342229642 & 0.500  & 0.004 & 0.197  & 0.004 & 0.058 & 0.005	\\
IRAS10378+1109 & 1342234867 & 0.480  & 0.004 & 0.183  & 0.004 & 0.050 & 0.005	\\
IRAS03158+4227 & 1342226656 & 0.973  & 0.004 & 0.377  & 0.004 & 0.137 & 0.005	\\
IRAS16090-0139 & 1342229565 & 1.067  & 0.004 & 0.404  & 0.004 & 0.116 & 0.005	\\
IRAS20100-4156 & 1342230817 & 1.001  & 0.004 & 0.349  & 0.004 & 0.102 & 0.005	\\
IRAS23253-5415 & 1342234737 & 1.044  & 0.005 & 0.437  & 0.004 & 0.165 & 0.005	\\
IRAS00188-0856 & 1342234693 & 0.877  & 0.004 & 0.345  & 0.004 & 0.111 & 0.005	\\
IRAS12071-0444 & 1342234858 & 0.471  & 0.004 & 0.163  & 0.004 & 0.044 & 0.005	\\
IRAS13451+1232 & 1342234792 & 0.503  & 0.005 & 0.256  & 0.004 & 0.197 & 0.006	\\
IRAS01003-2238 & 1342234707 & 0.222  & 0.004 & 0.070  & 0.004 & 0.026 & 0.006	\\
IRAS23230-6926 & 1342230806 & 0.617  & 0.004 & 0.204  & 0.004 & 0.064 & 0.005	\\
IRAS11095-0238 & 1342234863 & 0.380  & 0.004 & 0.119  & 0.004 & 0.036 & 0.005	\\
IRAS20087-0308 & 1342230838 & 1.804  & 0.006 & 0.687  & 0.004 & 0.210 & 0.005	\\
IRAS15462-0450 & 1342238307 & 0.492  & 0.004 & 0.162  & 0.004 & 0.050 & 0.008	\\
IRAS08311-2459 & 1342230796 & 1.246  & 0.005 & 0.464  & 0.004 & 0.148 & 0.005	\\
IRAS06206-6315 & 1342226638 & 1.248  & 0.005 & 0.477  & 0.004 & 0.158 & 0.005	\\
IRAS20414-1651 & 1342231345 & 1.315  & 0.005 & 0.519  & 0.004 & 0.168 & 0.005	\\
IRAS19297-0406 & 1342230837 & 2.039  & 0.006 & 0.752  & 0.004 & 0.244 & 0.005	\\
IRAS14348-1447 & 1342238301 & 1.842  & 0.006 & 0.666  & 0.005 & 0.197 & 0.006	\\
IRAS06035-7102 & 1342195728 & 1.226  & 0.022 & 0.397  & 0.001 & 0.130 & 0.008	\\
IRAS22491-1808 & 1342234671 & 0.862  & 0.004 & 0.305  & 0.004 & 0.097 & 0.005	\\
IRAS14378-3651 & 1342238295 & 1.330  & 0.005 & 0.478  & 0.005 & 0.135 & 0.006	\\
IRAS23365+3604 & 1342234919 & 1.849  & 0.006 & 0.669  & 0.004 & 0.210 & 0.005	\\
IRAS19254-7245 & 1342206210 & 1.545  & 0.005 & 0.587  & 0.004 & 0.185 & 0.005	\\
IRAS09022-3615 & 1342230799 & 2.449  & 0.007 & 0.823  & 0.004 & 0.252 & 0.005	\\
IRAS08572+3915 & 1342230749 & 0.504  & 0.004 & 0.164  & 0.004 & 0.060 & 0.004	\\
IRAS15250+3609 & 1342234775 & 0.966  & 0.004 & 0.368  & 0.004 & 0.136 & 0.005	\\
Mrk463         & 1342236151 & 0.344  & 0.004 & 0.134  & 0.004 & 0.052 & 0.005	\\
IRAS23128-5919 & 1342209299 & 1.565  & 0.008 & 0.556  & 0.006 & 0.176 & 0.007	\\
IRAS10565+2448 & 1342234869 & 3.619  & 0.011 & 1.319  & 0.004 & 0.407 & 0.005	\\
IRAS20551-4250 & 1342230815 & 1.629  & 0.005 & 0.556  & 0.004 & 0.170 & 0.005	\\
IRAS05189-2524 & 1342203632 & 1.963  & 0.011 & 0.717  & 0.007 & 0.211 & 0.009	\\
IRAS17208-0014 & 1342203587 & 7.918  & 0.037 & 2.953  & 0.010 & 0.954 & 0.009	\\
Mrk231         & 1342201218 & 5.618  & 0.019 & 2.011  & 0.008 & 0.615 & 0.008	\\
UGC5101        & 1342204962 & 6.071  & 0.039 & 2.327  & 0.018 & 0.746 & 0.009	\\
Mrk273         & 1342201217 & 4.190  & 0.011 & 1.493  & 0.006 & 0.471 & 0.006	\\
IRAS13120-5453 & 1342226970 & 12.097 & 0.036 & 4.441  & 0.010 & 1.355 & 0.006	\\
NGC6240        & 1342203586 & 5.166  & 0.029 & 2.031  & 0.009 & 0.744 & 0.008	\\
Arp220         & 1342188687 & 30.414 & 0.132 & 12.064 & 0.036 & 4.145 & 0.015	\\
\hline
\end{tabular}
\label{tab:photometry}
\end{table*}

\section{Line Fitting}\label{sec:LineFitting}

\subsection{Fitting Lines to the data}\label{sec:LineFitData}
Line fitting to our final spectra is carried out using a dedicated line fitting algorithm  for the SPIRE FTS within the HIPE framework. The task does not search for lines but rather fits spectral features at expected line positions. The task performs a global fit to a specified list of lines in the Level 2 point source calibrated spectrum using the HIPE Spectrum Fitter (dedicated spectrum fitting Java task within HIPE).

Since the Spectrum Fitter task makes a global fit for all lines, the results were found to be sensitive to the number of input lines to the fit. Therefore it was found that the best way to run this fitter followed an iterative approach where a long list of lines for the initial fit was used. The resulting fits were then inspected visually. Lines that were clearly not real were discarded from the input file. The Spectrum Fitter was then run again on the reduced set of lines to produce a final list of measurements. This method does not find lines that are not in the original input table, but no such lines were apparent from visual inspection. The errors on the line fluxes are provided by the HIPE line fitter task as the standard deviation on the fit (assuming a sinc function using the Levenberg-Marquardt algorithm). Note that there is an additional systematic error component of 2.6$\%$ due to the line asymmetry caused by a residual phase shift in the interferogram \citep{hopwood15}.

\begin{figure} 
\begin{center}
\includegraphics[width=0.65\columnwidth,angle=0]{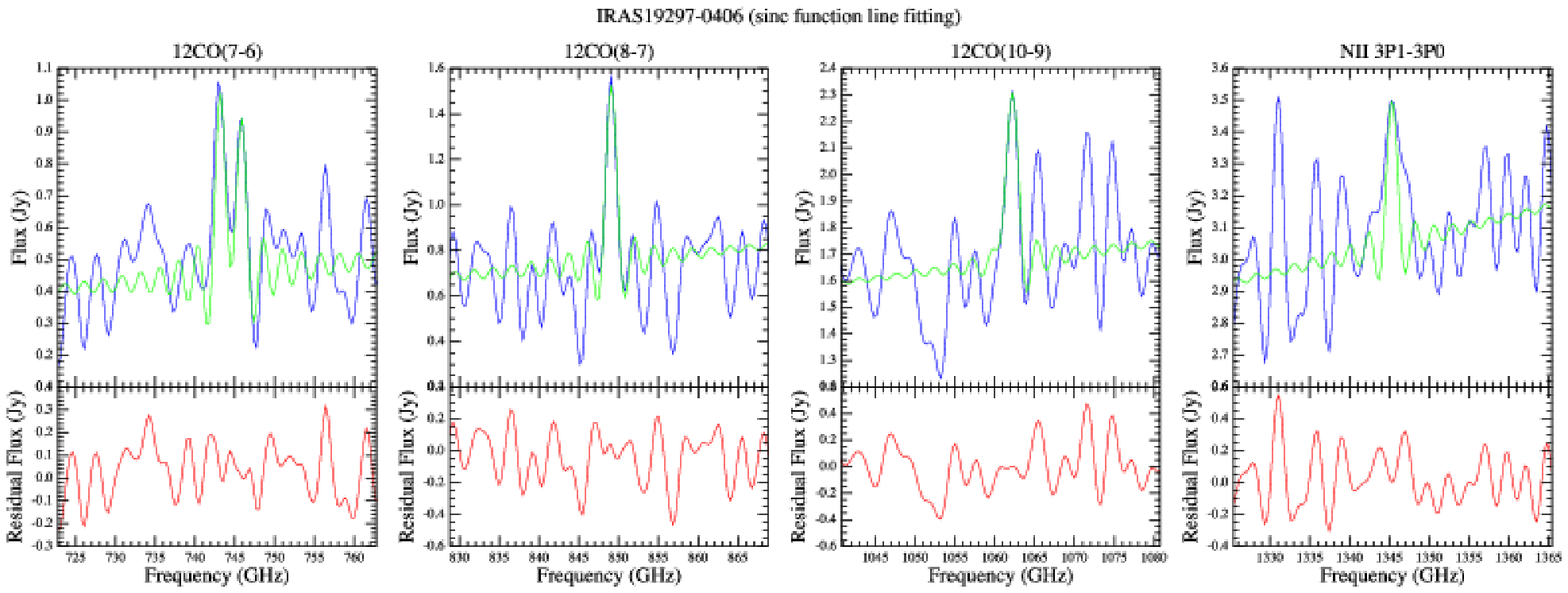}
\includegraphics[width=0.65\columnwidth,angle=0]{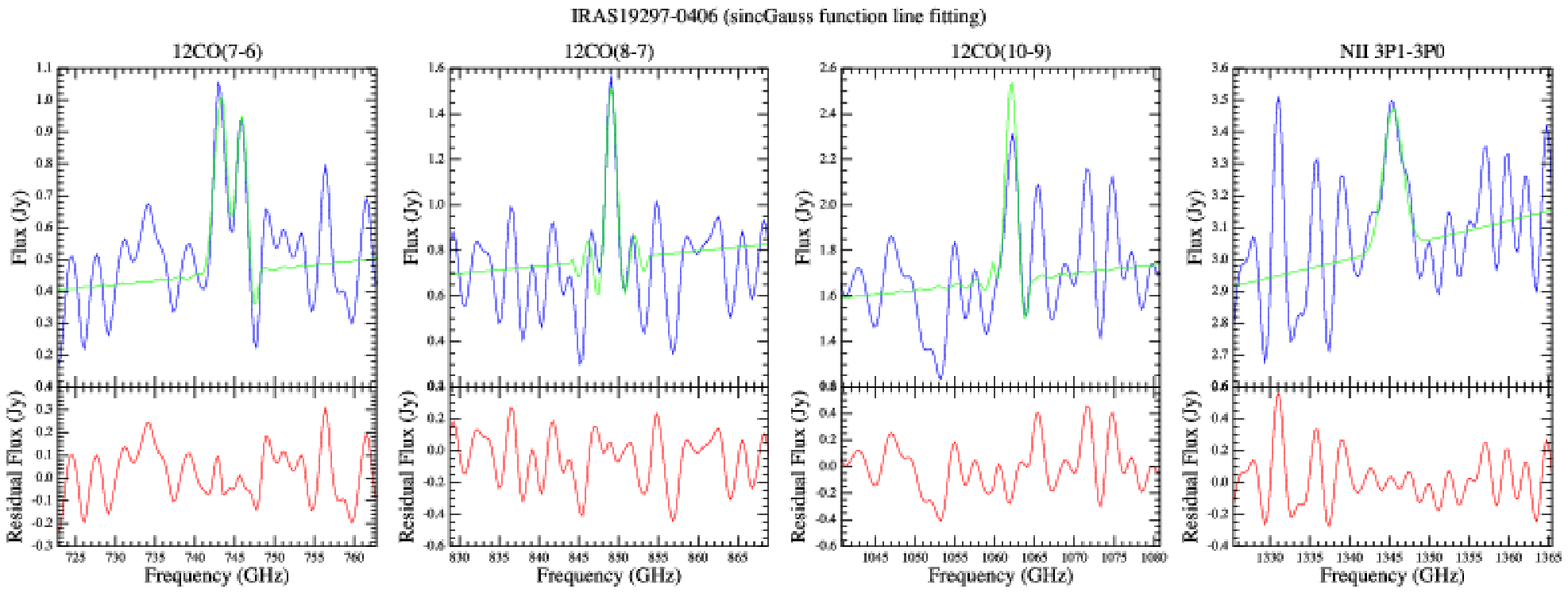}
\includegraphics[width=0.65\columnwidth,angle=0]{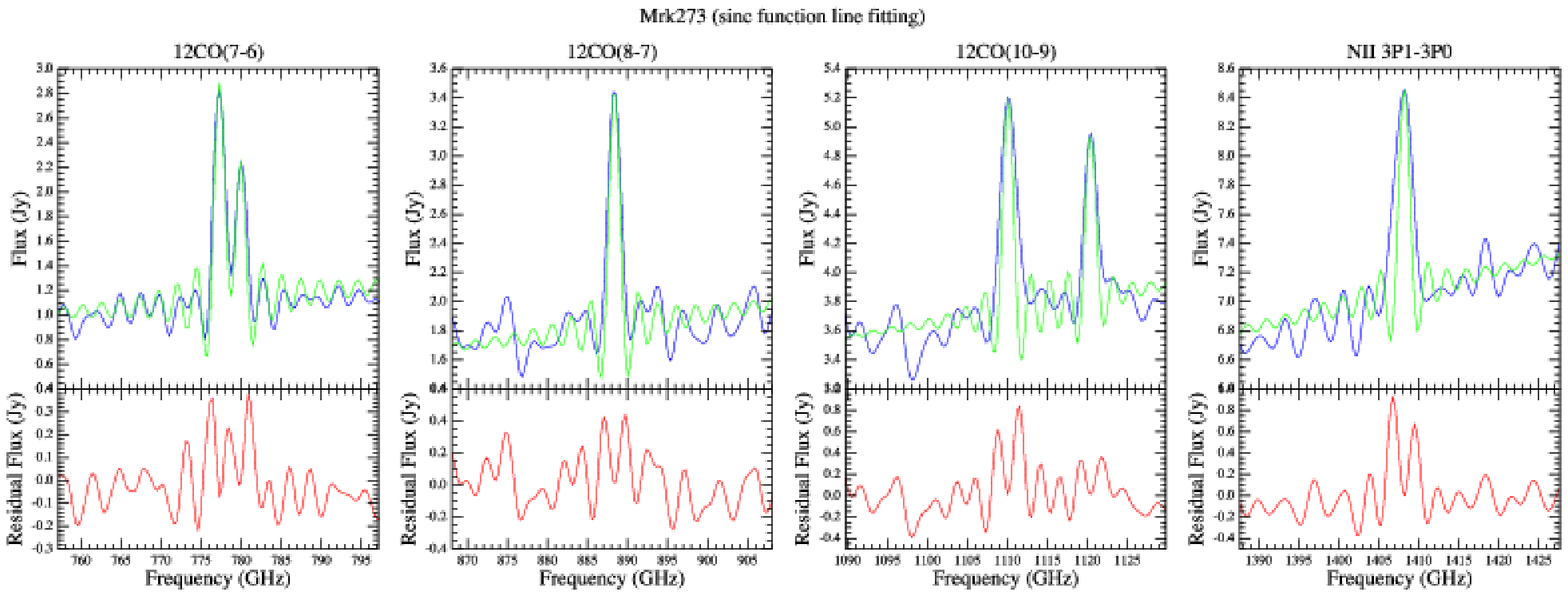}
\includegraphics[width=0.65\columnwidth,angle=0]{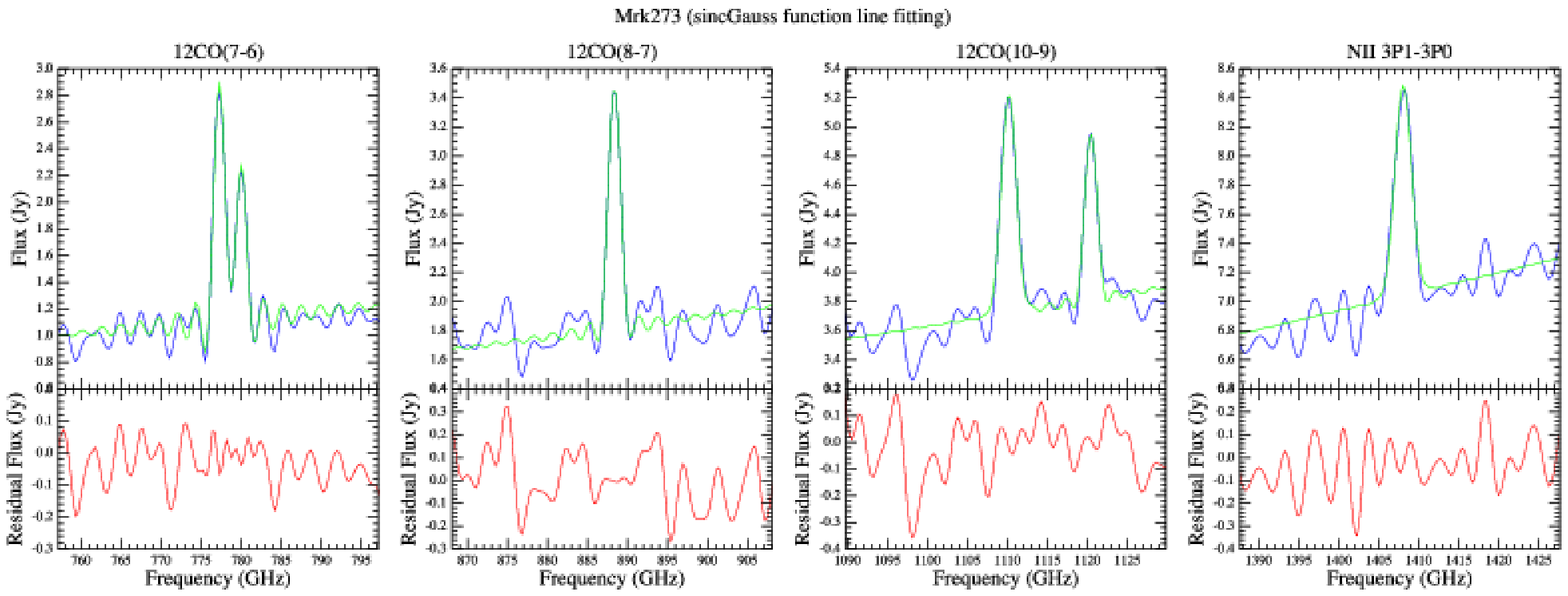}
\caption{Line fitting results for IRAS19297-0406 ({\it top rows}) and Mrk 273 ({\it bottom rows}). {\it Blue} lines are the final spectra, {\it green} lines the fit and {\it red} lines, the residual after model fit subtraction.The {\it top}-panel shows an example where the sinc-fit to the CO lines has been made. The {\it lower}-panels for each object shows an example where the sinc-Gauss-fit to the CO lines has been made. For IRAS 19297-0406, the sinc-fit is better than the sinc-Gauss-fit. For Mrk 273, the sinc-Gauss-fit produces a better fit. Note that in both cases the sinc-Gauss profile provides a better fit to [NII] 205$\, \mu$m.}
\label{fig:SincSincGauss}
\end{center}
\end{figure}

The FTS instrument line shape can be approximated by a sinc function \citep{swinyard14,hopwood15} however for partially resolved lines, a sinc-Gaussian may be more appropriate. Most line widths for our sources were well fitted by the sinc function and have widths less than $\sim$300km$^{-1}$. In principle the resolution at the highest frequencies is comparable with some of the ground based measurements but we see very few cases where an alternative sinc-Gauss improves the CO data. In Figure ~\ref{fig:SincSincGauss} the effectiveness of the sinc profile fit for the HERUS ULIRGs is shown for two examples, IRAS 19297-0406 and Mrk273. For IRAS 19297-0406, the profiles for the CO(7-6), CO(8-7) \& CO(10-9) lines are all well fitted by the sinc function with the sinc-Gauss  overestimating the the peak flux for the CO(10-9) line. For Mrk273, the CO(10-9) line and possibly the CO(8-7) line appear partially resolved and the sinc-Gauss profile provides a better fit. Note however that this line may suffer contamination from an H2O line (H2O 3$\_$12-2$\_$21 on the blue side of the COline), which could broaden the 2CO(10-9) line (\citealt{gonzalezalfonso10,gonzalezalfonso14}).
In both cases a broad Nitrogen II line at 205$\, \mu$m (1461 GHz) is present and is better fit by the sinc-Gauss profile.

We obtain a measure of the signal to noise of the line detection by using the residual of the continuum fit to the final spectrum (calculated by the line fitting task, see lower panels in each plot of Figure ~\ref{fig:SincSincGauss}). By selecting an appropriate frequency range around the line of interest (50 spectral pixels around the line), the  standard deviation of the residual in that range is calculated. The signal to noise is then estimated by dividing the fitted line height (i.e. peak flux) by the standard deviation of the residual.

\subsection{Jack Knife Tests}\label{sec:JackKnife}
Many of the HERUS sample are considered faint sources for the SPIRE FTS (250$\, \mu$m flux density $<$500mJy) and the corresponding line detections are often $<$5$\sigma$. In order to differentiate a true line detection from spurious detections manifested by low frequency noise in the FTS spectrum a Jack Knife test was performed to confirm statistically robust detections of the lines.

\begin{figure} 
\begin{center}
\includegraphics[width=0.75\columnwidth,angle=0]{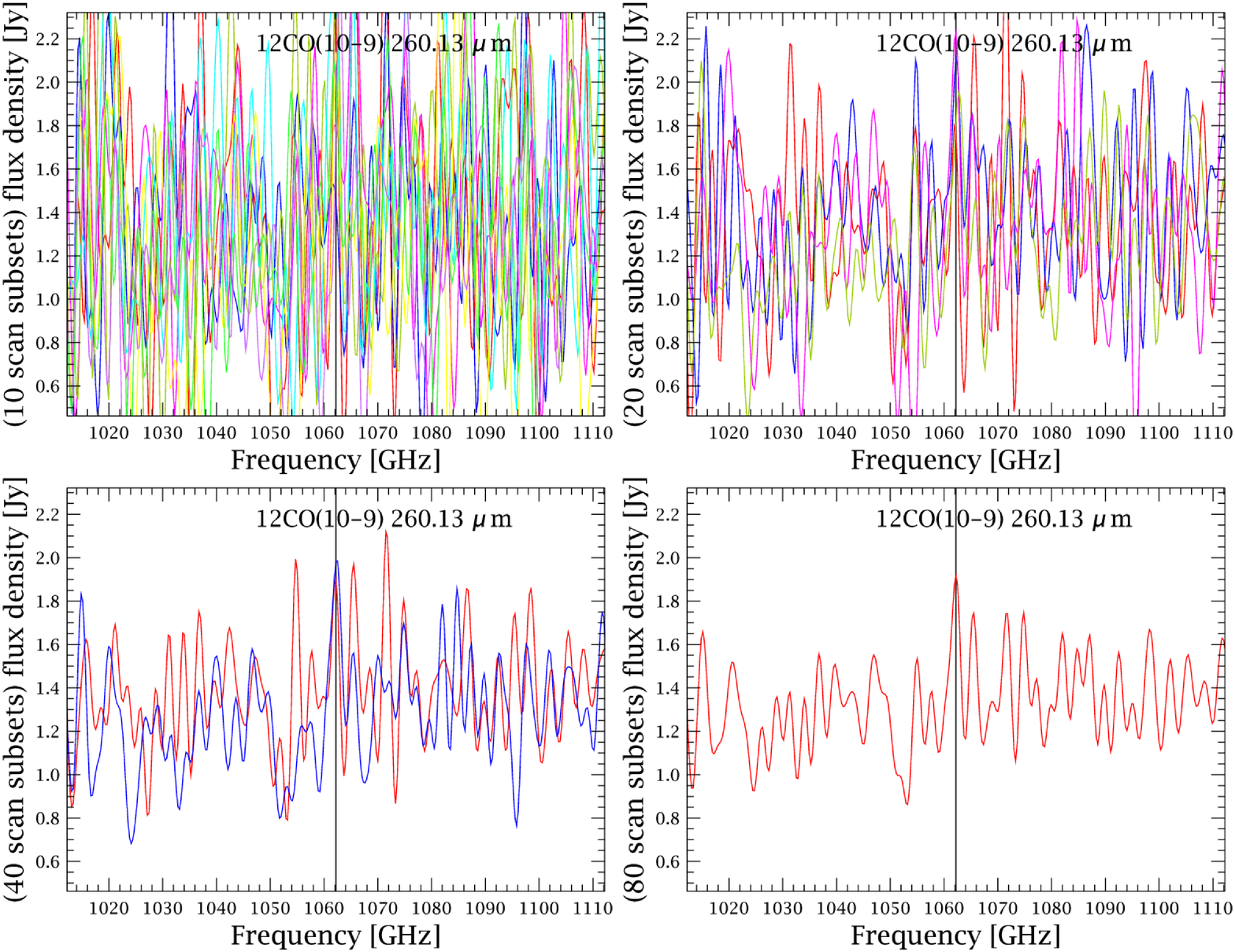}
\caption{Example of a jack knife test for the CO(10-9) line, for IRAS 19297-0406. The total number of FTS scans is 80 and subsets of 60, 40, 30, 20 and 10 averaged scans are created. Except for the lowest subset of 10 scans, the CO(10-9) is seen in all scans in all subsets (only the subsets 10, 20, 40, 80 are shown in the figure), implying a robust detection of the line. The jack knife test can be used to distinguish between real detections such as the redshifted CO(10-9) line around 1060GHz and spurious artefacts such as the line around 1055 GHz.}
\label{fig:jackknife}
\end{center}
\end{figure}

The Jack Knife tests can be performed by either treating the forward and reverse scans of the FTS separately or by dividing the total number of scans into smaller subsets. For our data we perform Jack knife tests by first dividing the total number of scans into two subsets and then consecutively halving the subsets to create further subsets (i.e. for an observation with a total of 200 scans, it is first divided into two subsets of 100 scans, then 4 subsets of 50, down to subsets of 10 scans). Visually inspecting plots of the scan subsets allows confirmation of the existence of a line. The smallest subsets should be indistinguishable from noise unless there is some systematic noise manifesting as a line detection. Increasing the number of scan lines per subset should highlight any real line whilst the surrounding samples around the line are still consistent with noise. Using this method a statistically robust selection of lines can be realised. The Jack Knife test for the CO(10-9) line in IRAS 19297-0406 is shown in Figure \ref{fig:jackknife}, which shows results for scan subsets of 80, 40, 20, 10 FTS scans. The CO(10-9) is seen in all scans in all subsets, except the lowest 10 scan subset plot, giving confidence in the detection of the line.

\section{Results}\label{sec:results}

\subsection{Fitted Spectra}\label{sec:fittedspectra}
\begin{figure} 
\begin{center}
\includegraphics[width=0.40\columnwidth,angle=0]{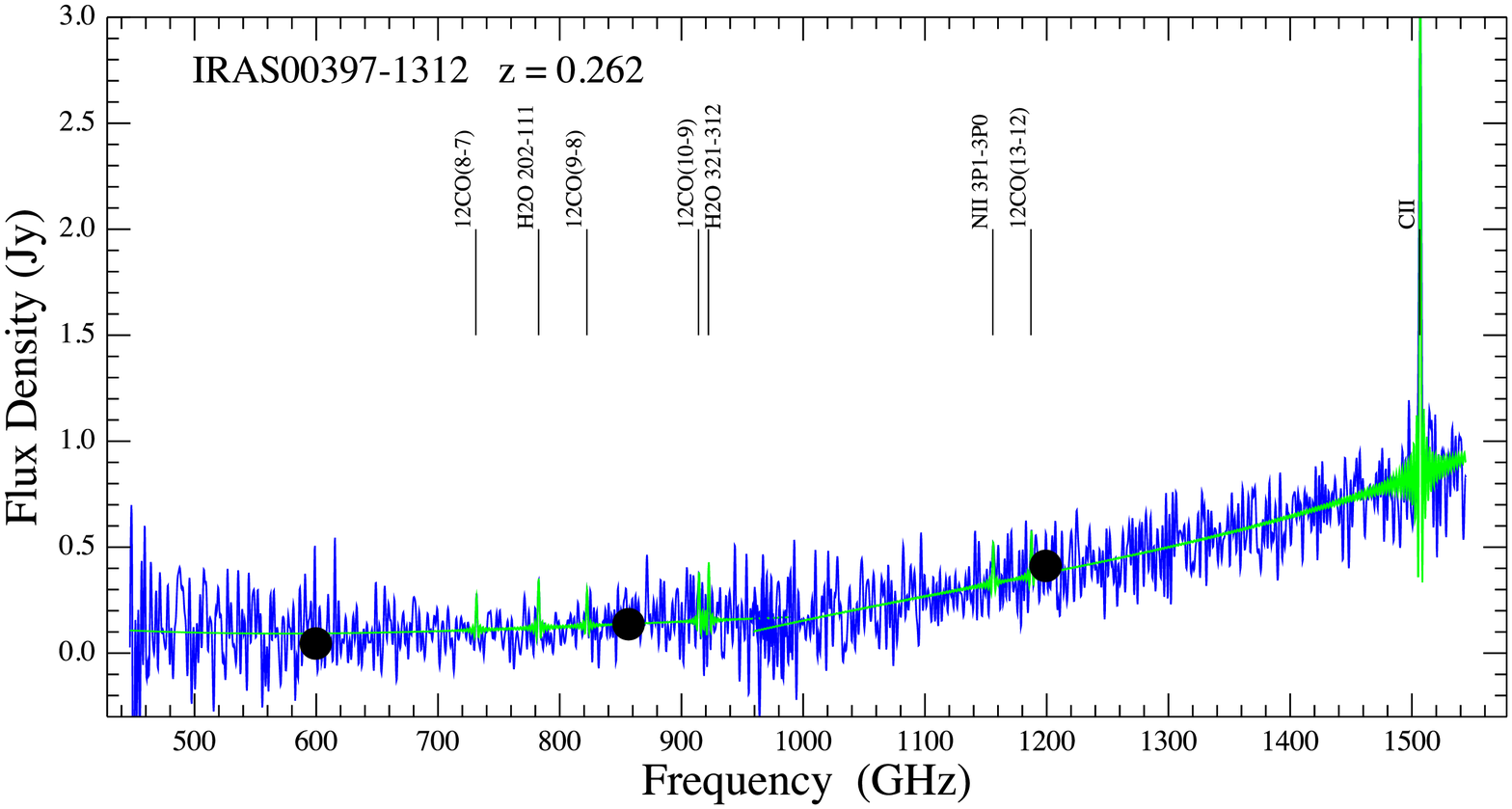}
\includegraphics[width=0.40\columnwidth,angle=0]{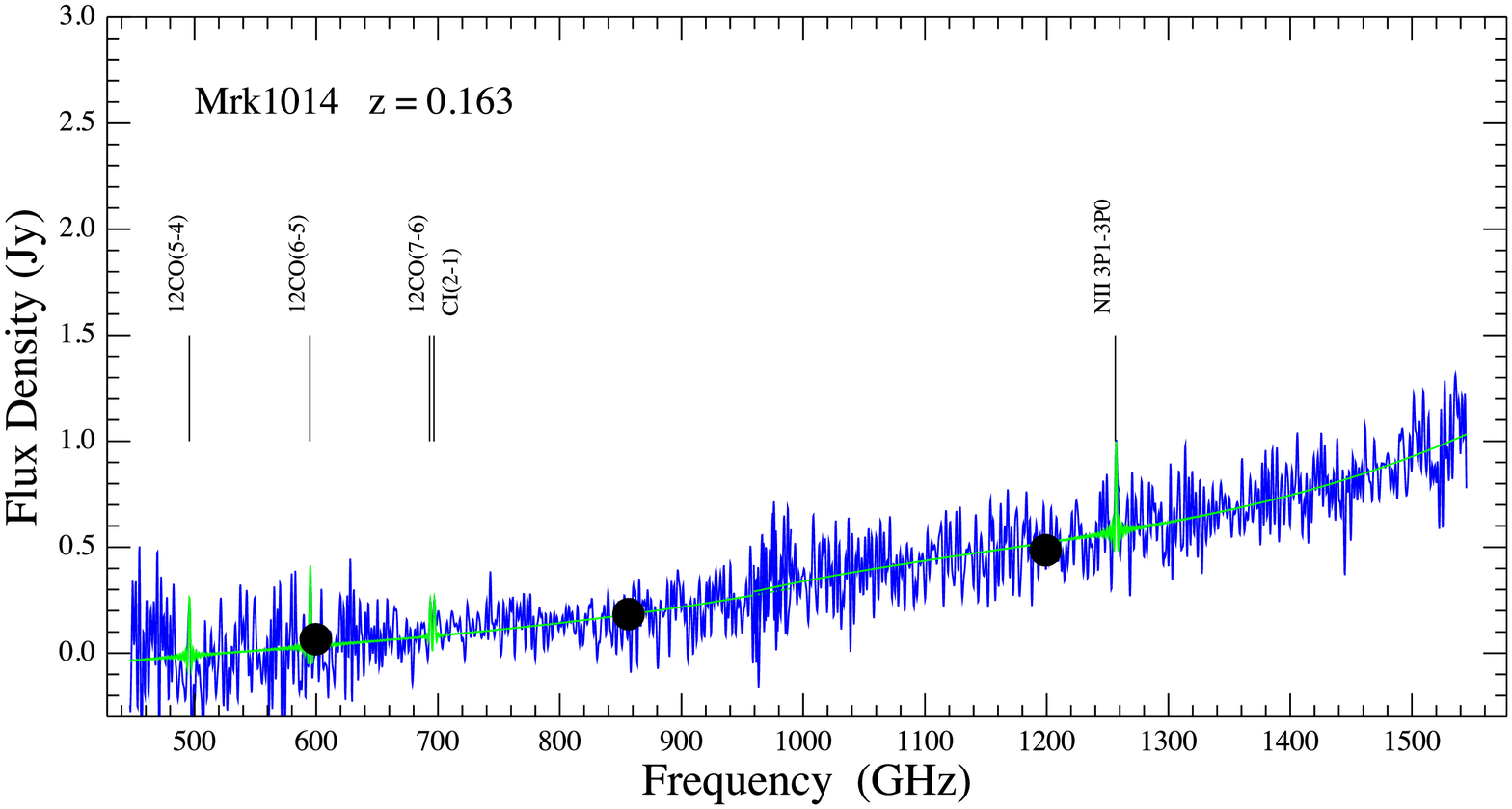}
\includegraphics[width=0.40\columnwidth,angle=0]{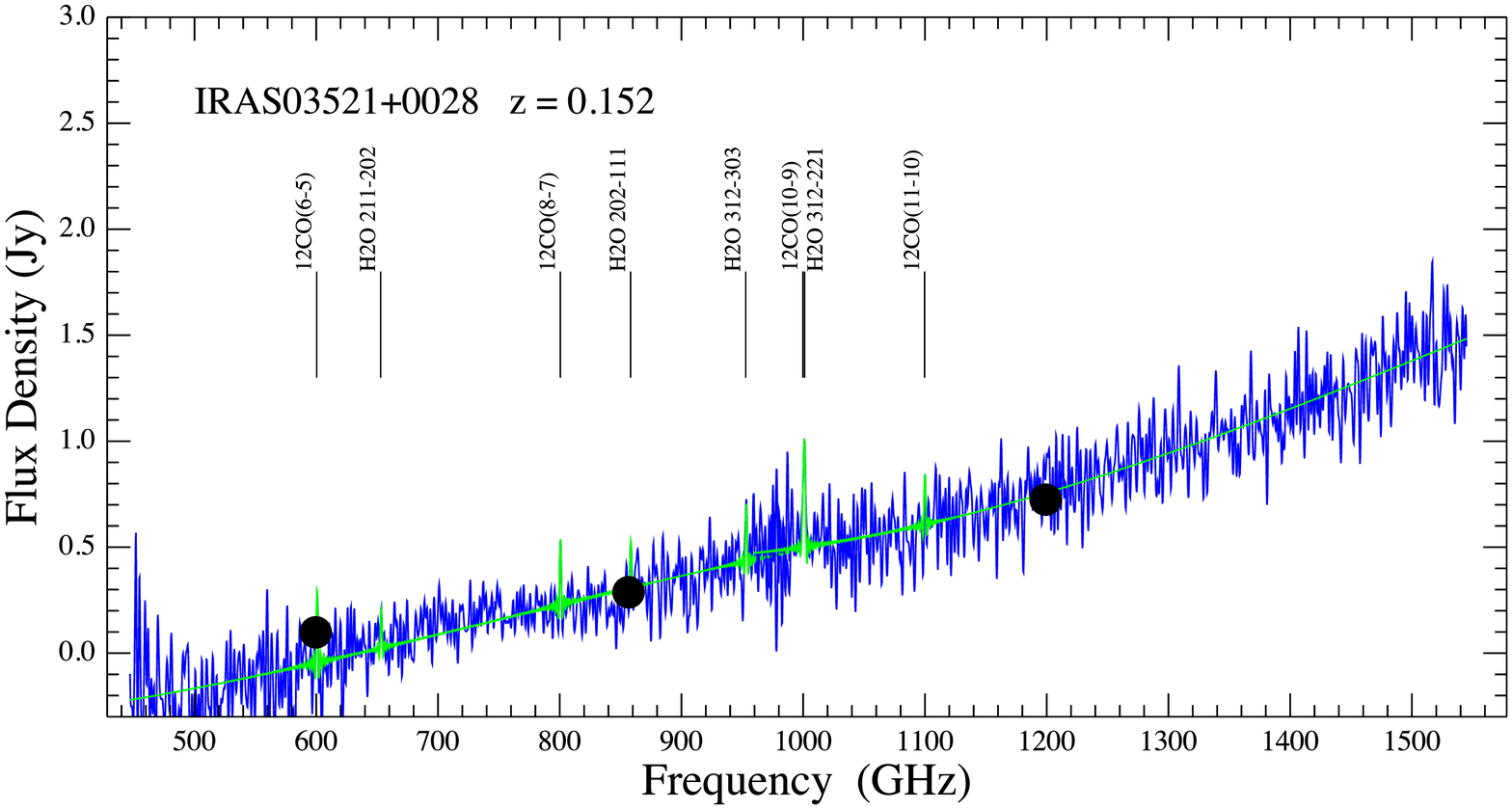}
\includegraphics[width=0.40\columnwidth,angle=0]{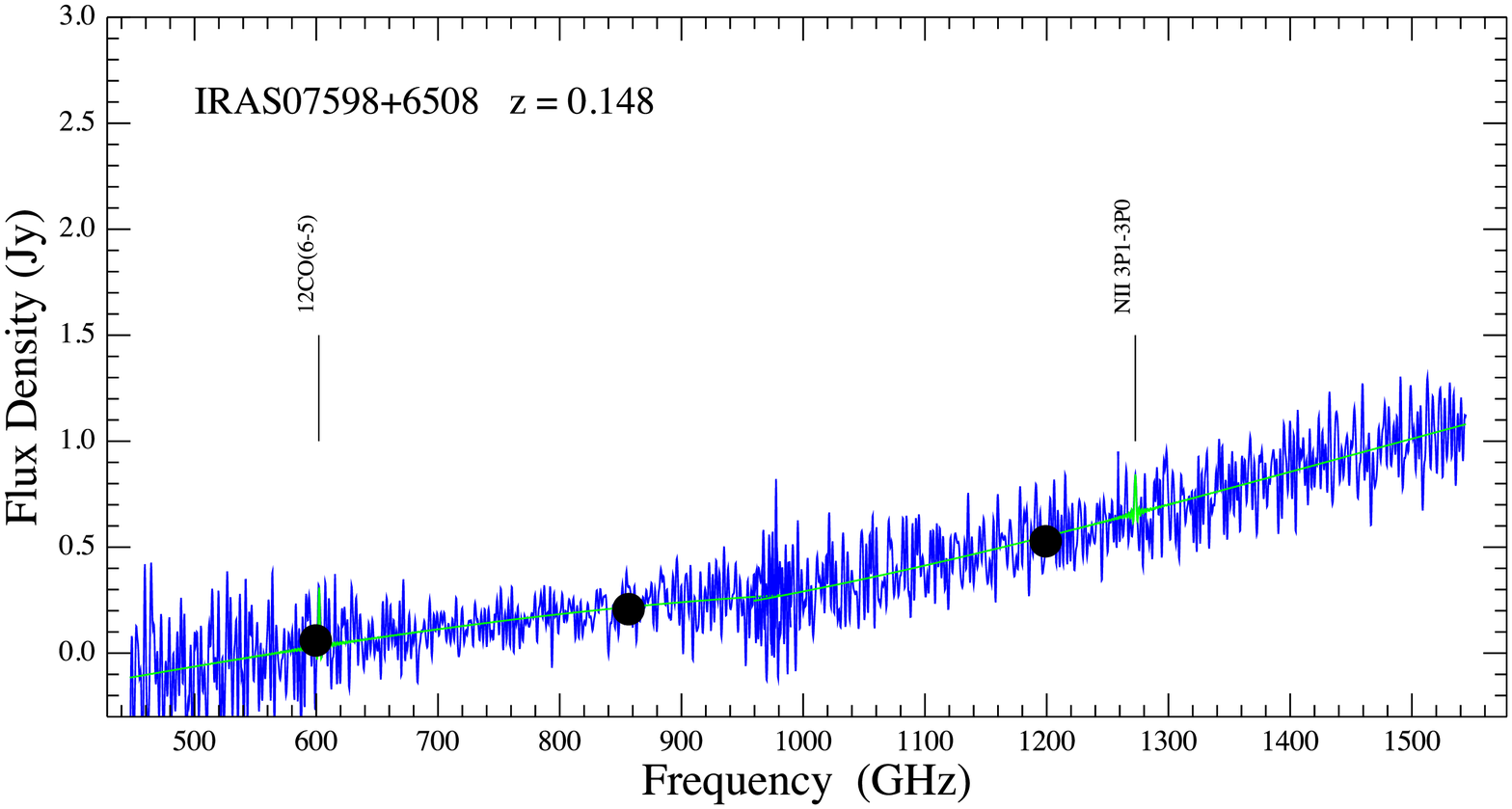}
\includegraphics[width=0.40\columnwidth,angle=0]{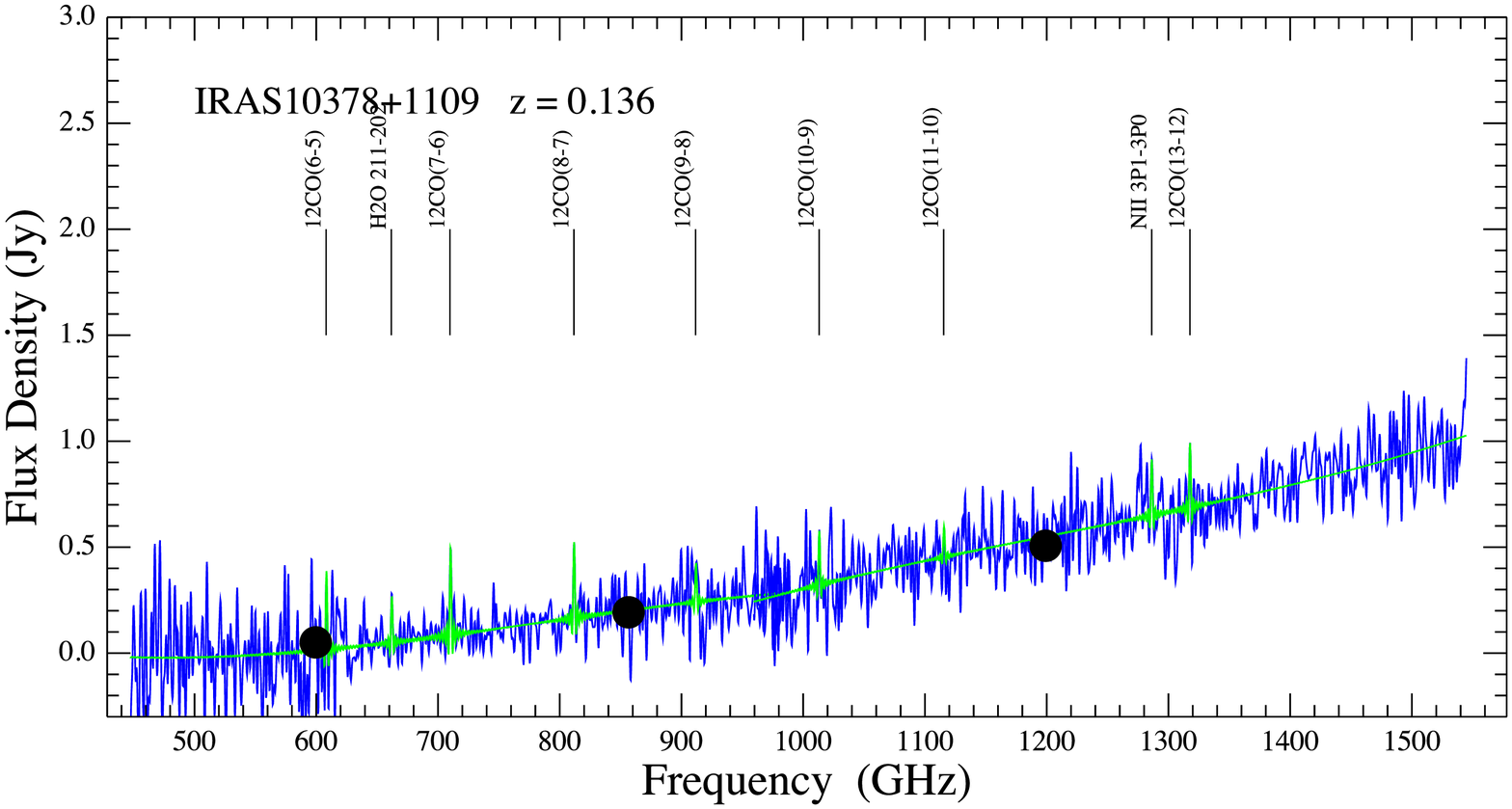}
\includegraphics[width=0.40\columnwidth,angle=0]{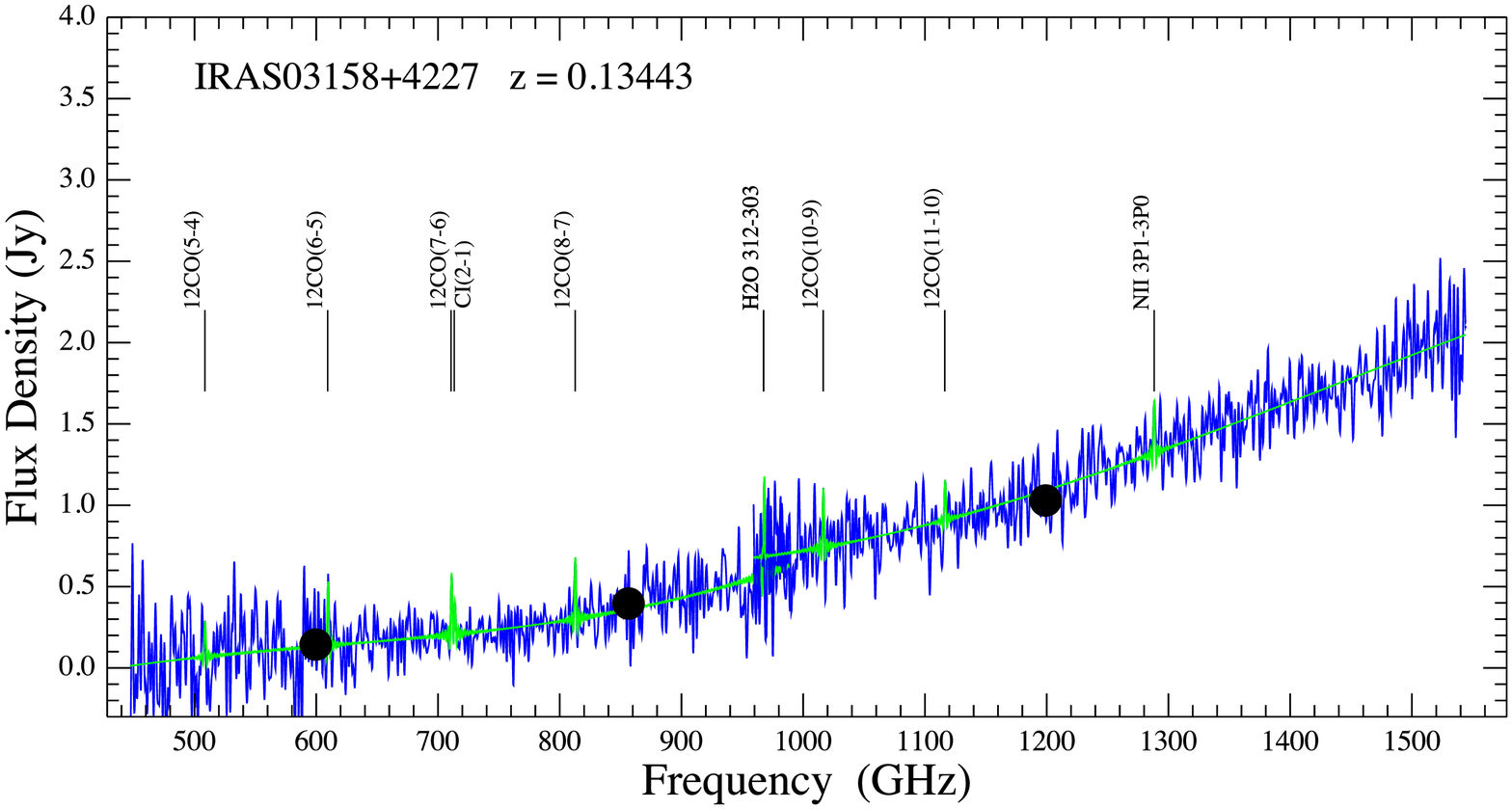}
\includegraphics[width=0.40\columnwidth,angle=0]{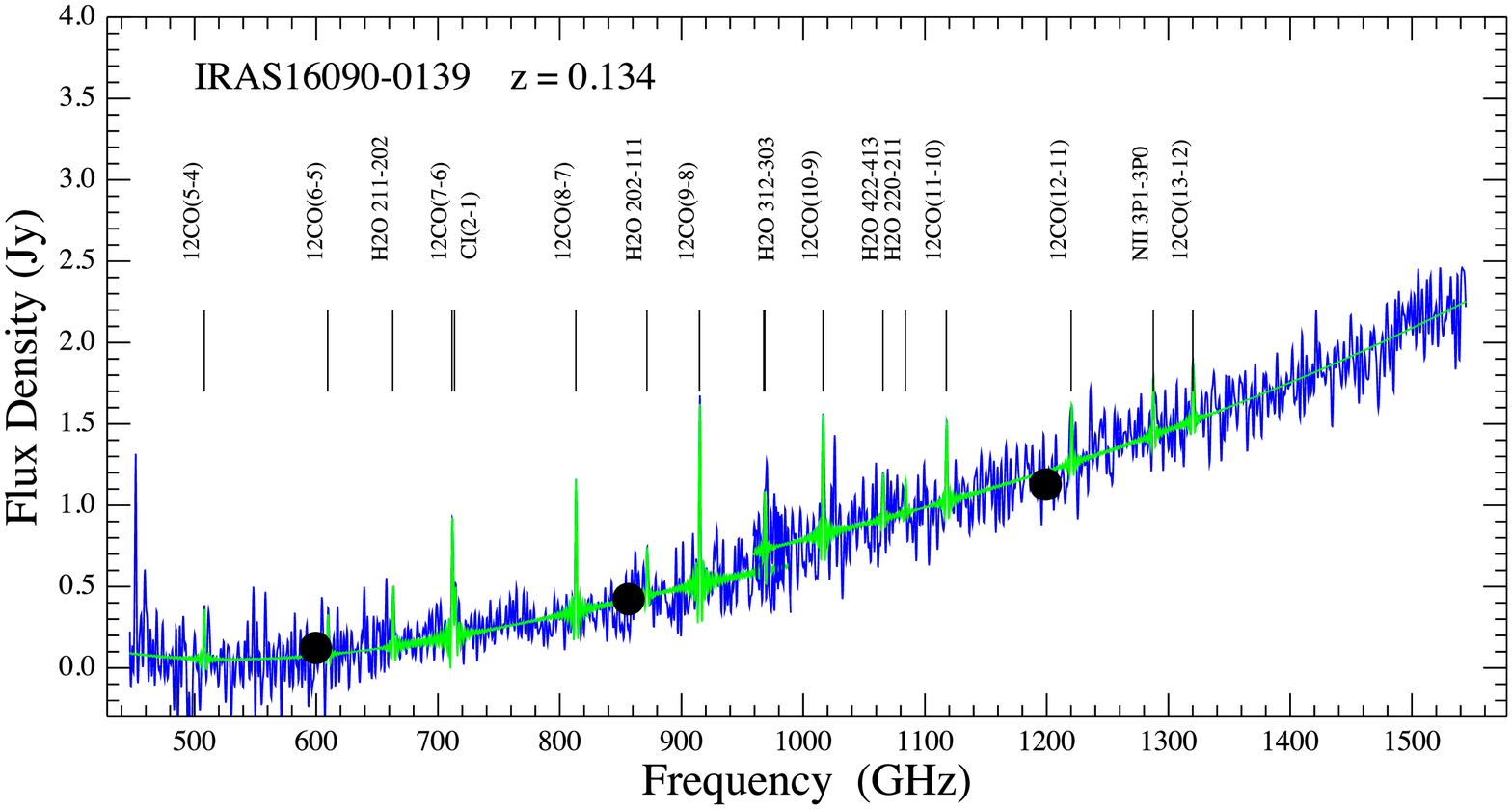}
\includegraphics[width=0.40\columnwidth,angle=0]{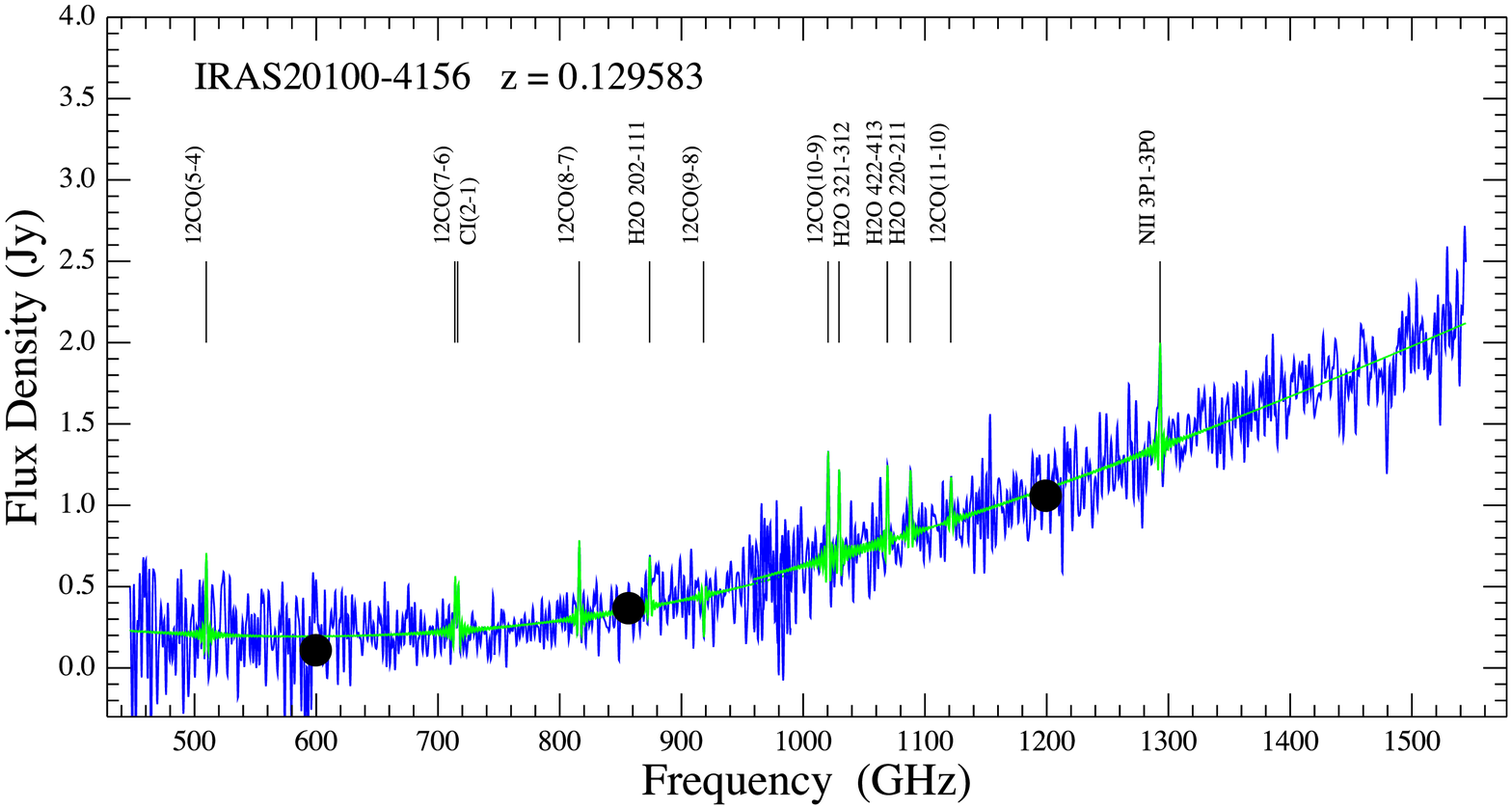}
\includegraphics[width=0.40\columnwidth,angle=0]{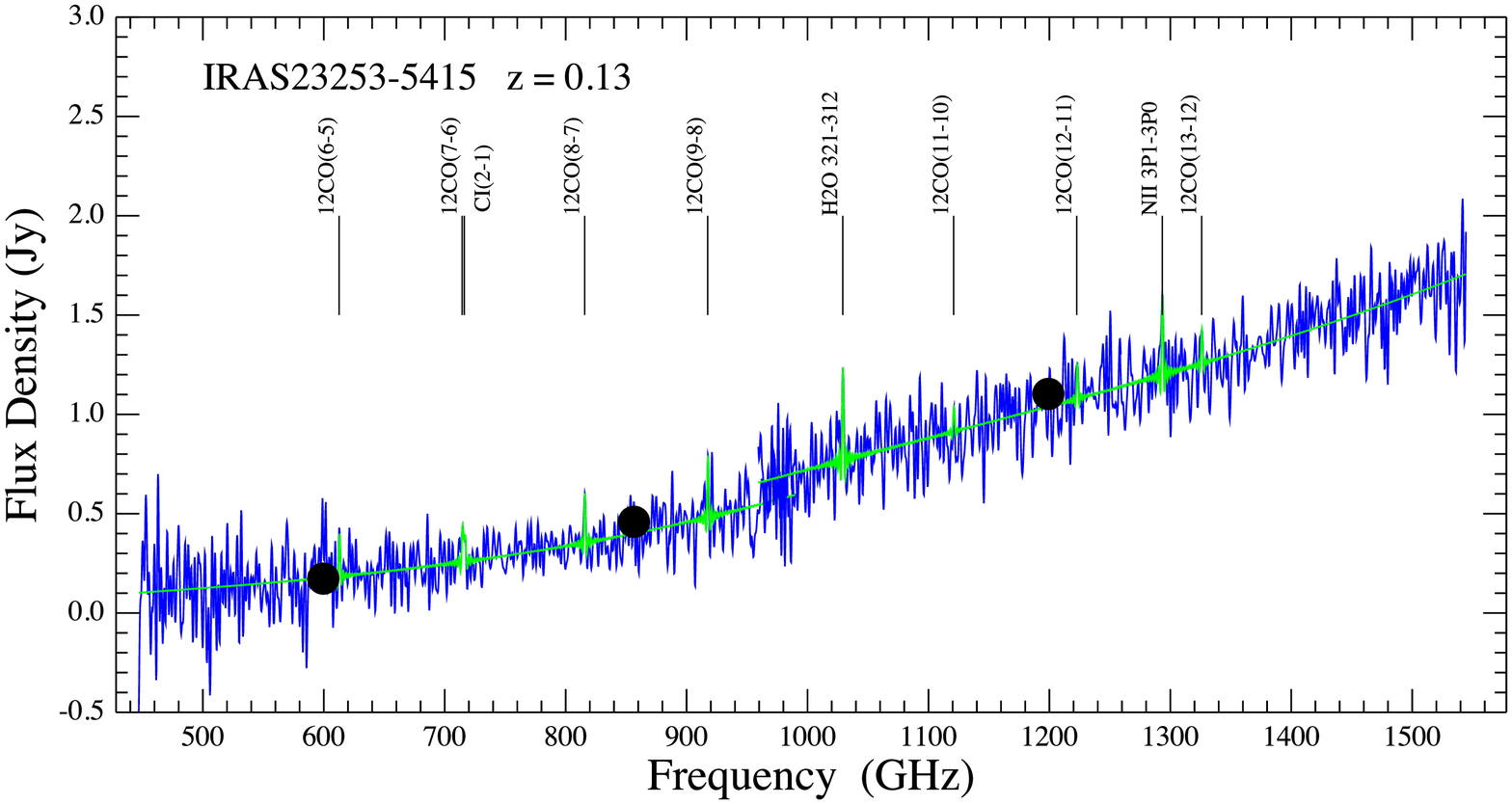}
\includegraphics[width=0.40\columnwidth,angle=0]{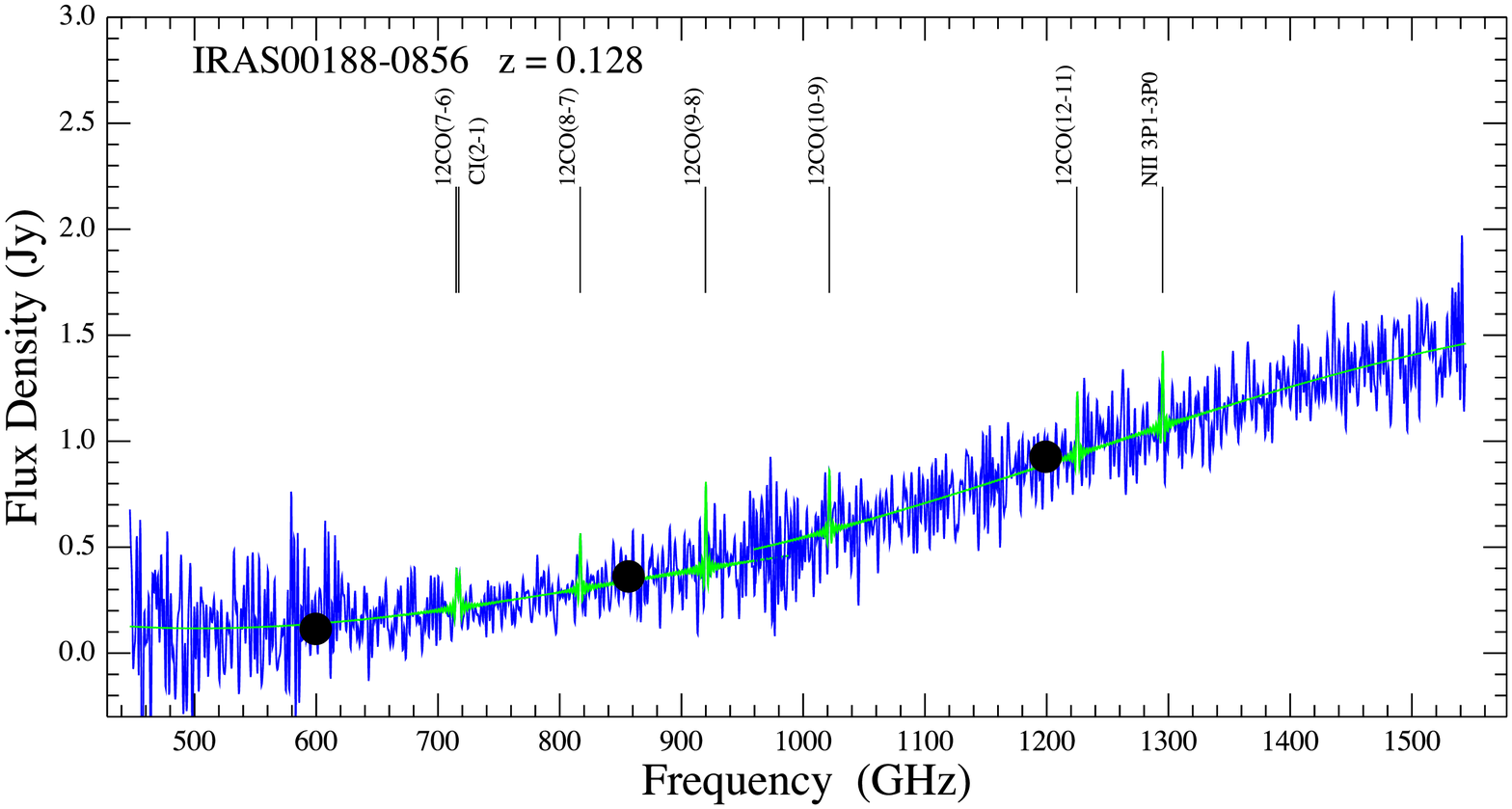}
\caption{Line fitting results for IRAS 00397-1312, Mrk1014, IRAS 03521+0028, IRAS 07598+6508, IRAS 10378+1109, IRAS 03158+4227, IRAS 16090-0139, IRAS 20100-4156, IRAS 23253-5415, IRAS 00188-0856. The processed spectra (blue) are shown with the model fits overlaid (green). Photometry points are also shown (black dots). Fitted line species are indicated by the vertical bars.}
\label{fig:fittedlines1}
\end{center}
\end{figure}

\begin{figure} 
\begin{center}
\includegraphics[width=0.40\columnwidth,angle=0]{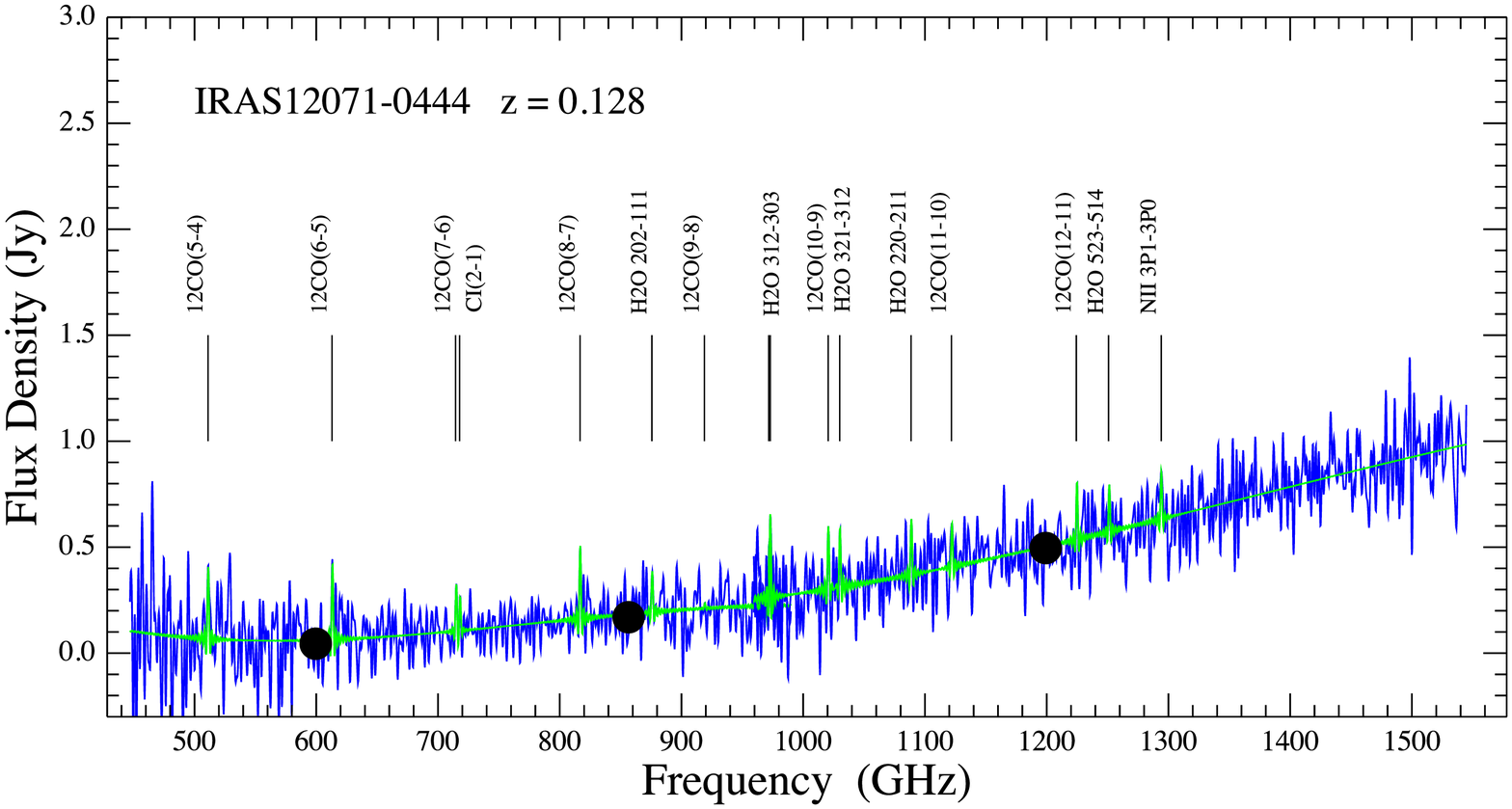}
\includegraphics[width=0.40\columnwidth,angle=0]{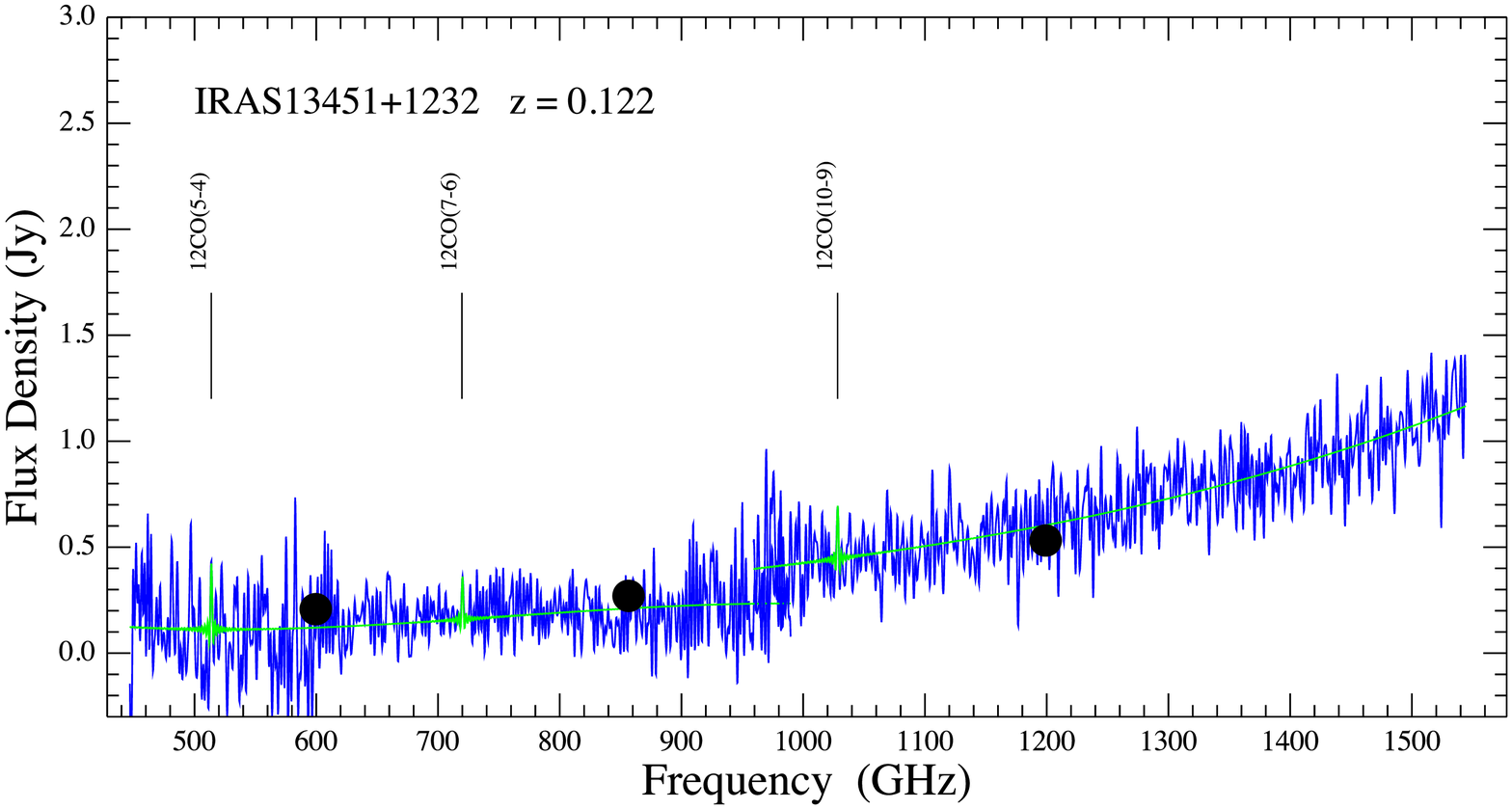}
\includegraphics[width=0.40\columnwidth,angle=0]{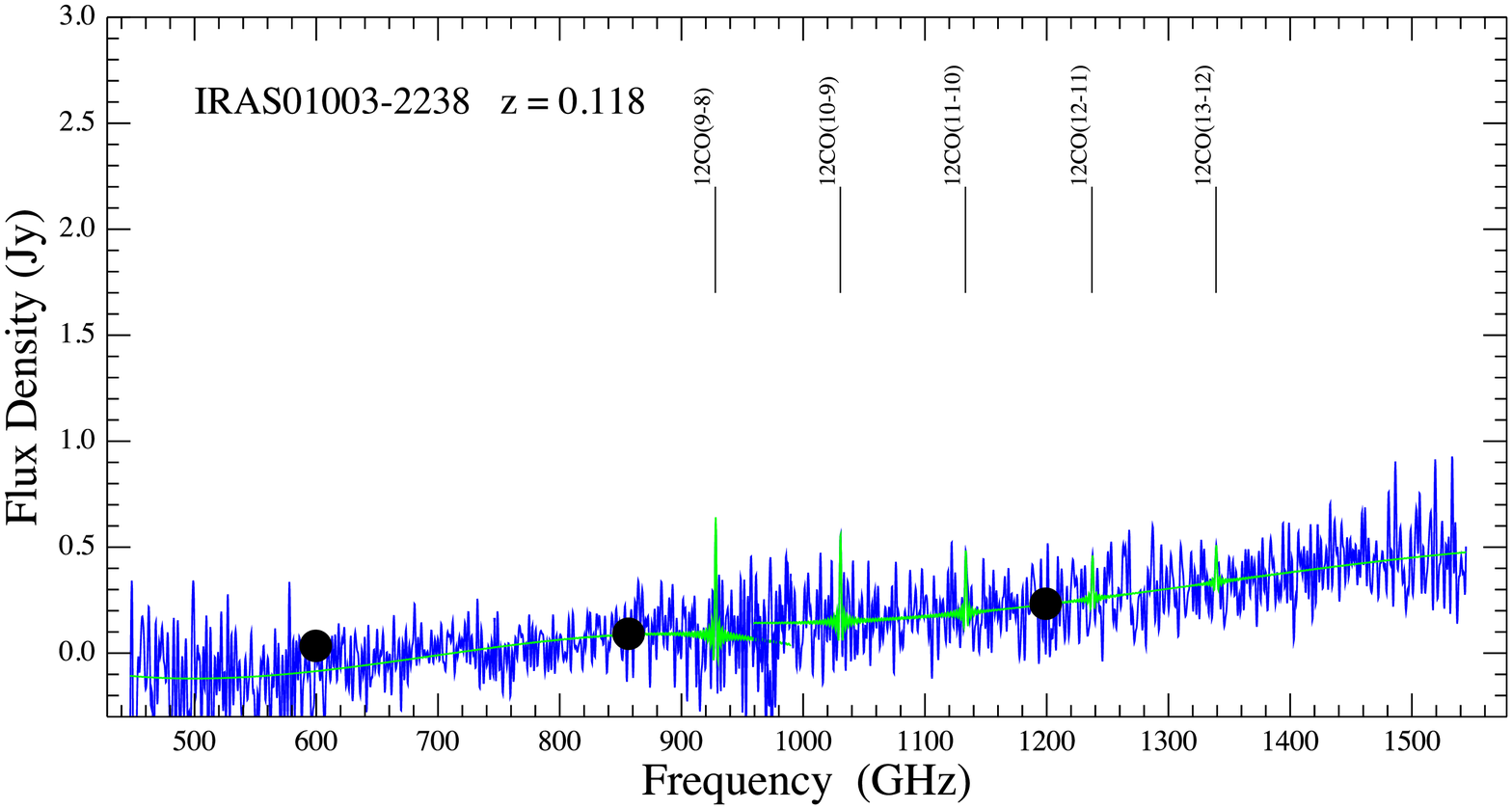}
\includegraphics[width=0.40\columnwidth,angle=0]{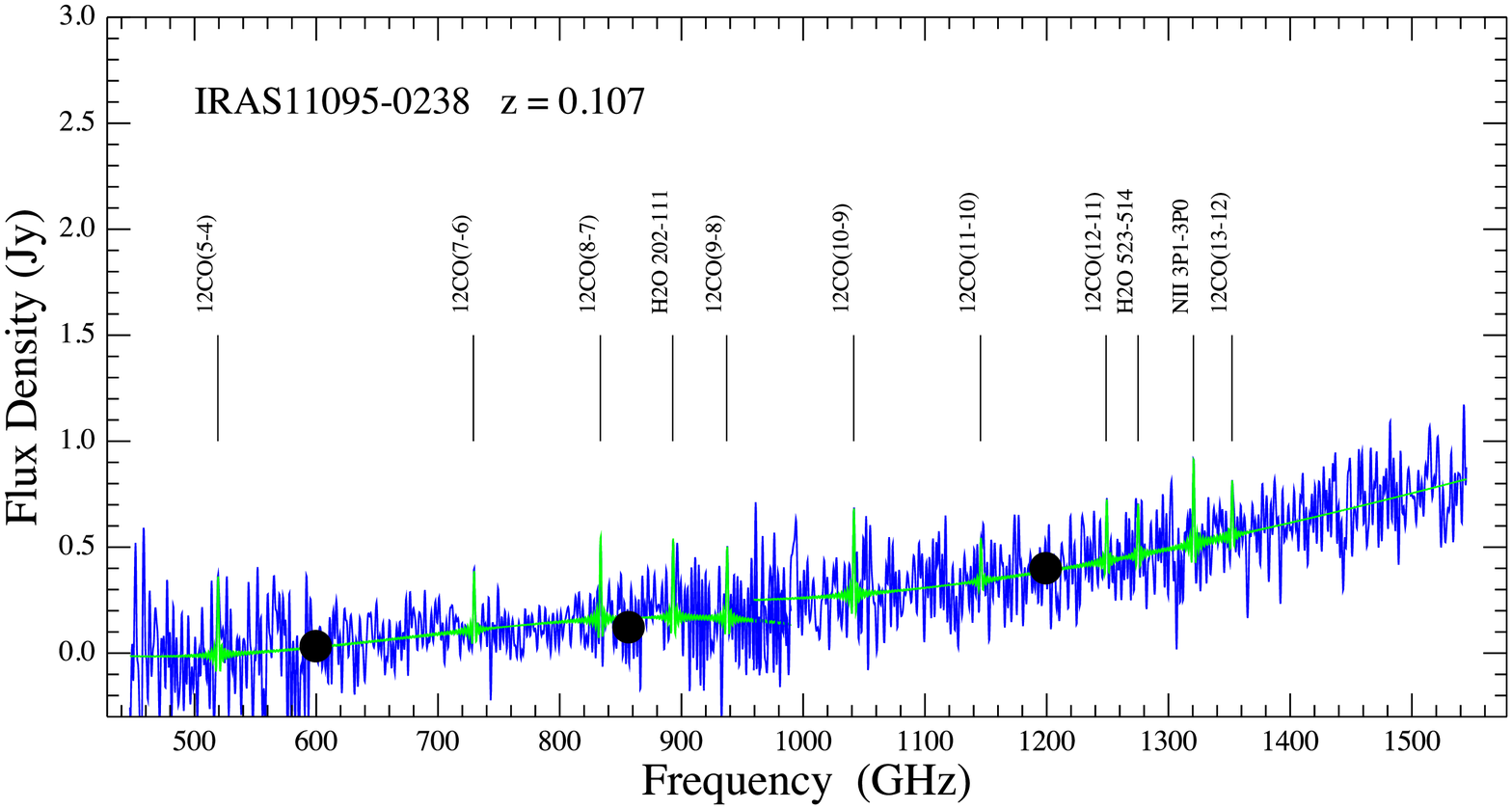}
\includegraphics[width=0.40\columnwidth,angle=0]{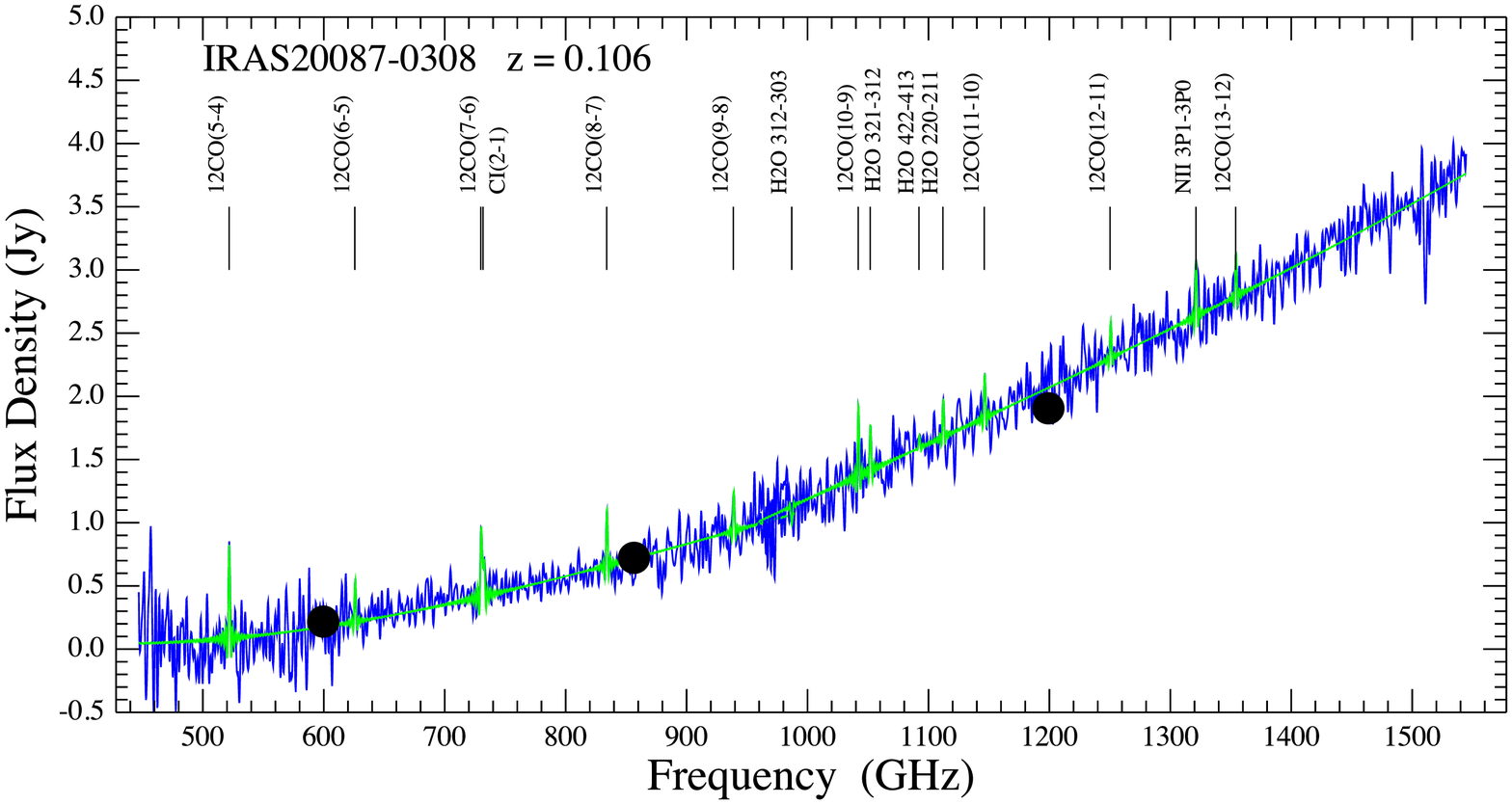}
\includegraphics[width=0.40\columnwidth,angle=0]{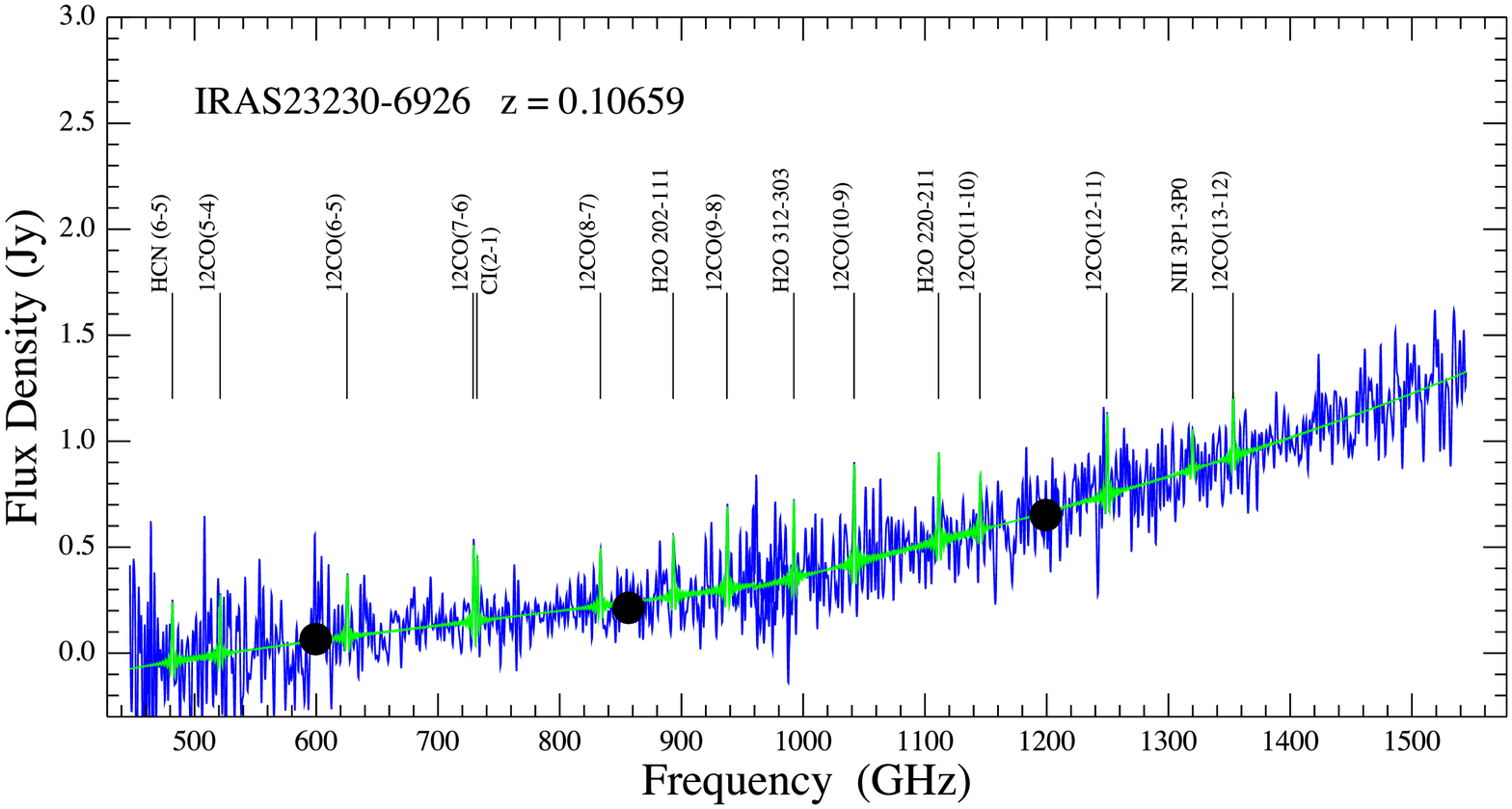}
\includegraphics[width=0.40\columnwidth,angle=0]{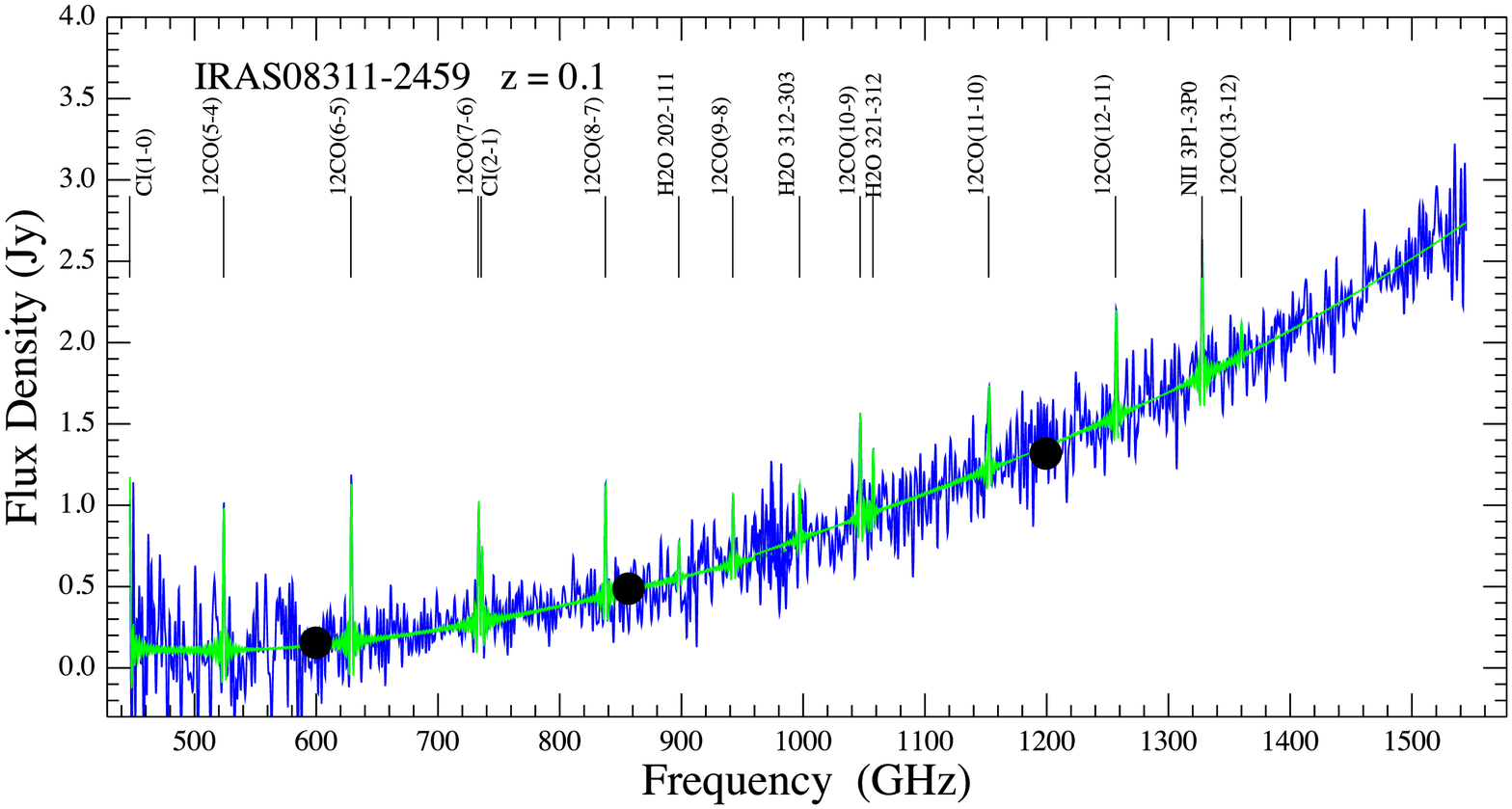}
\includegraphics[width=0.40\columnwidth,angle=0]{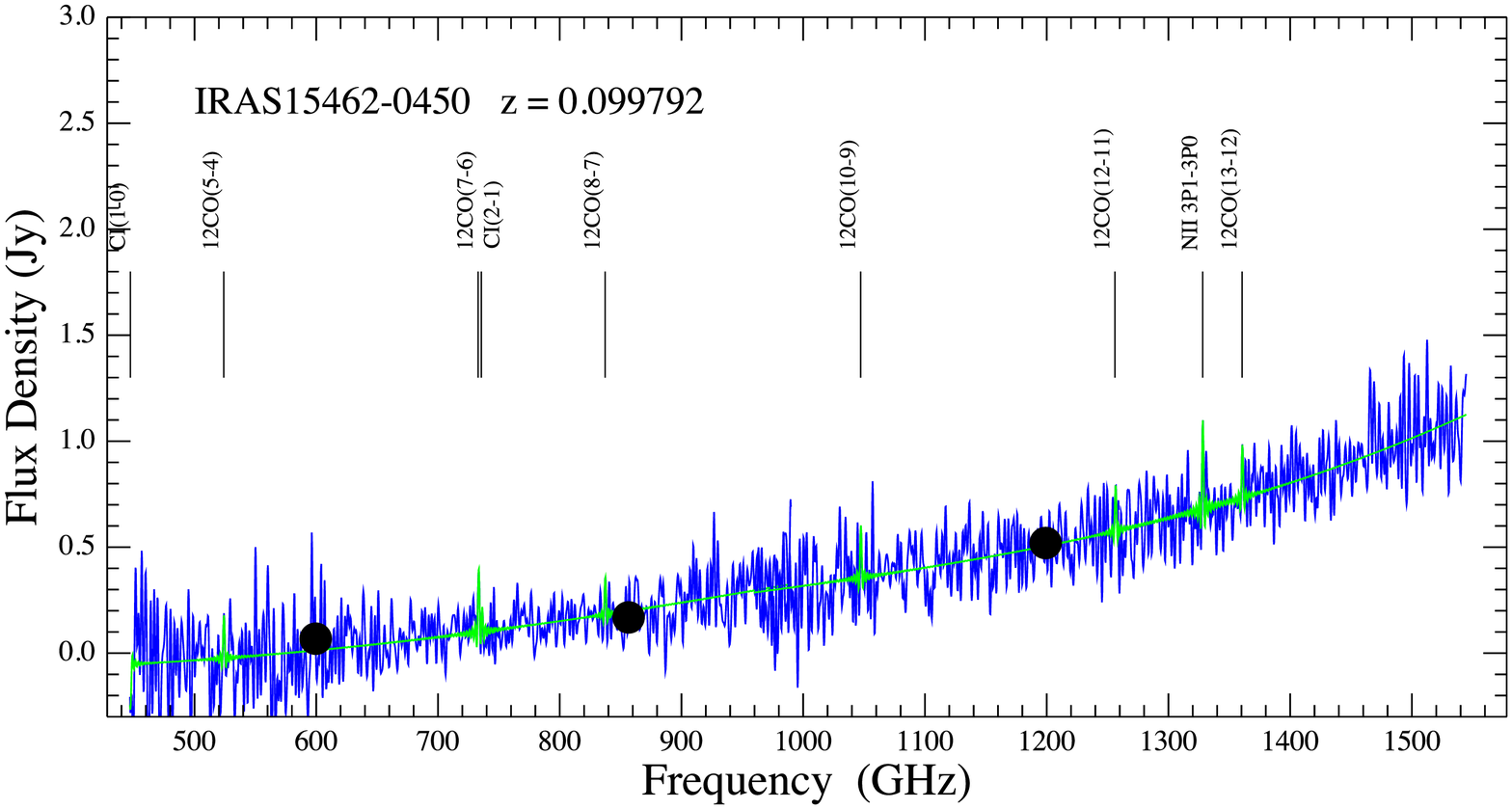}
\includegraphics[width=0.40\columnwidth,angle=0]{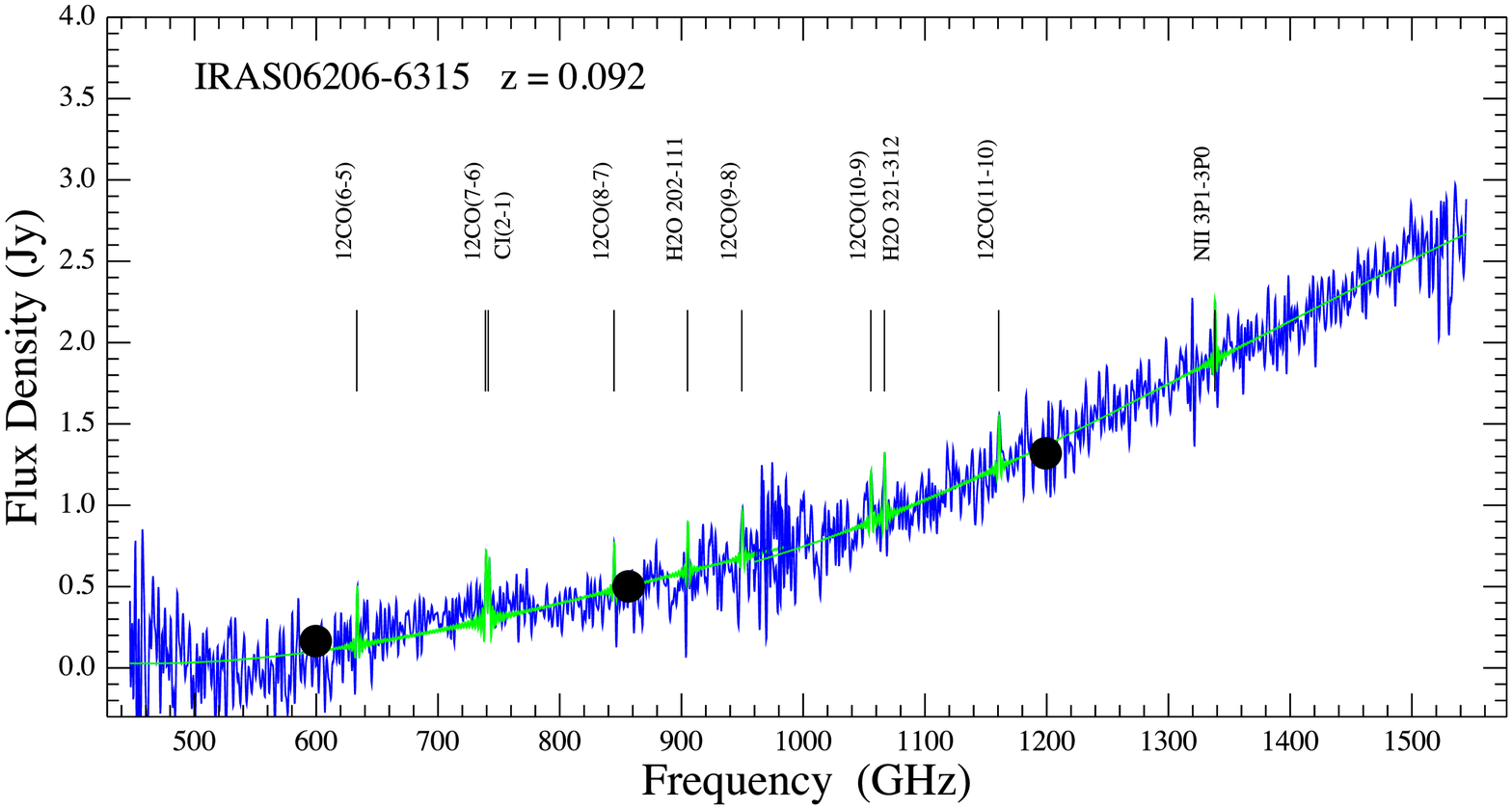}
\includegraphics[width=0.40\columnwidth,angle=0]{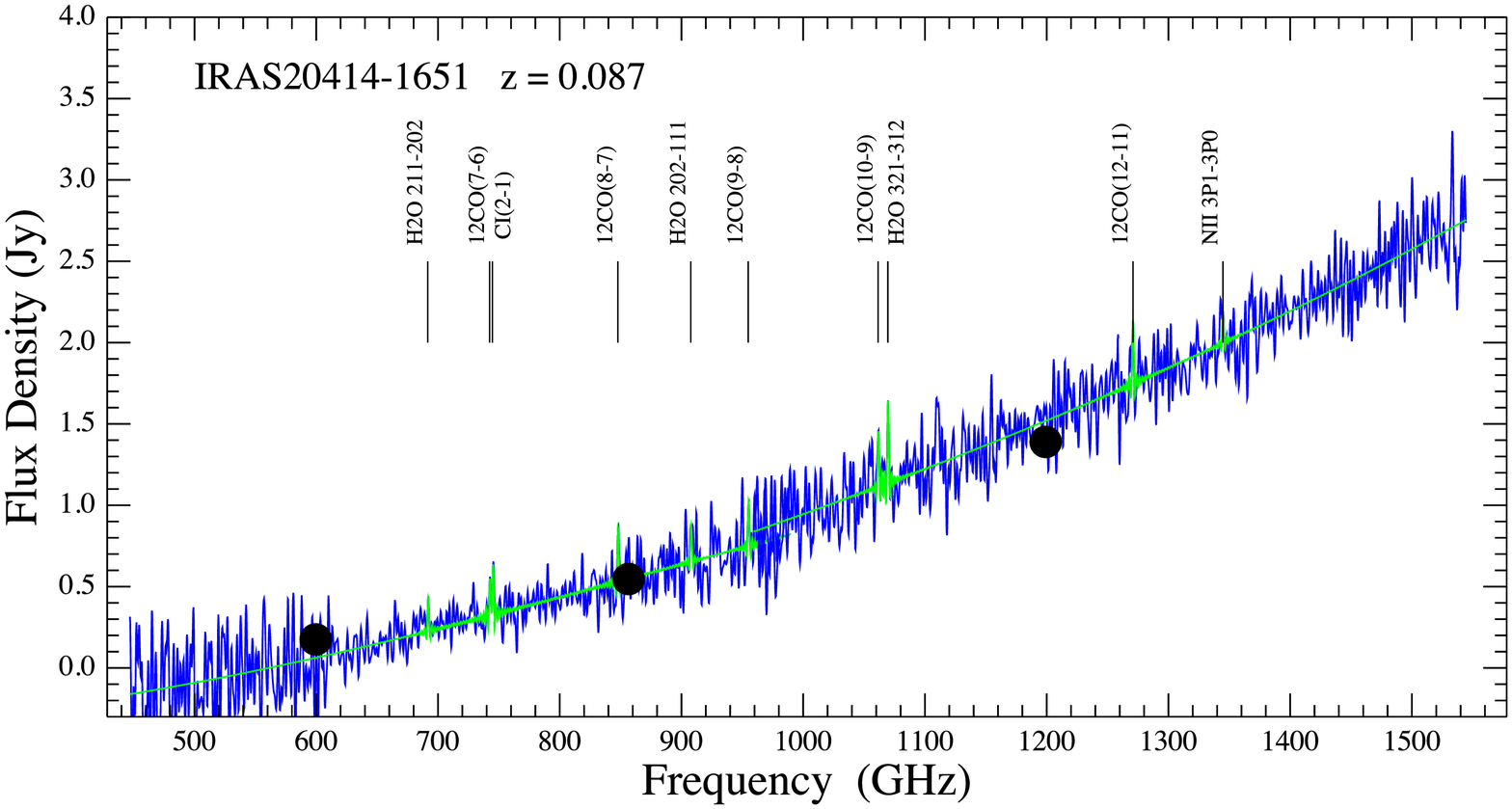}
\caption{Line fitting results for IRAS 12071-0444, IRAS 13451+1232, IRAS 01003-2238, IRAS 11095-0238, IRAS 20087-0308, IRAS 23230-6926, IRAS 08311-2459, IRAS 15462-0450, IRAS 06206-6315, IRAS 20414-1651. The processed spectra (blue) are shown with the model fits overlaid (green). Photometry points are also shown (black dots). Fitted line species are indicated by the vertical bars.}
\label{fig:fittedlines2}
\end{center}
\end{figure}

\begin{figure} 
\begin{center}
\includegraphics[width=0.40\columnwidth,angle=0]{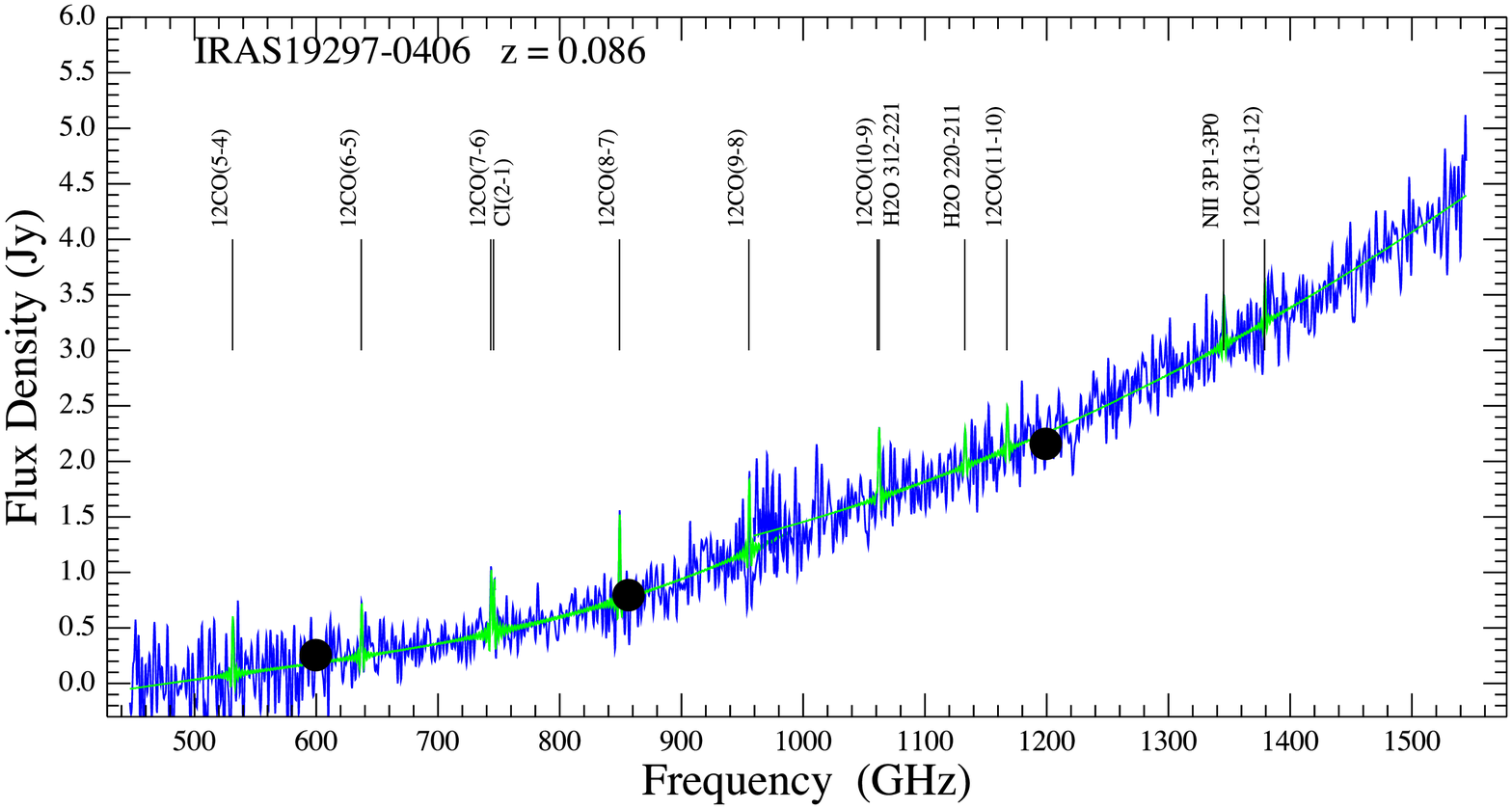}
\includegraphics[width=0.40\columnwidth,angle=0]{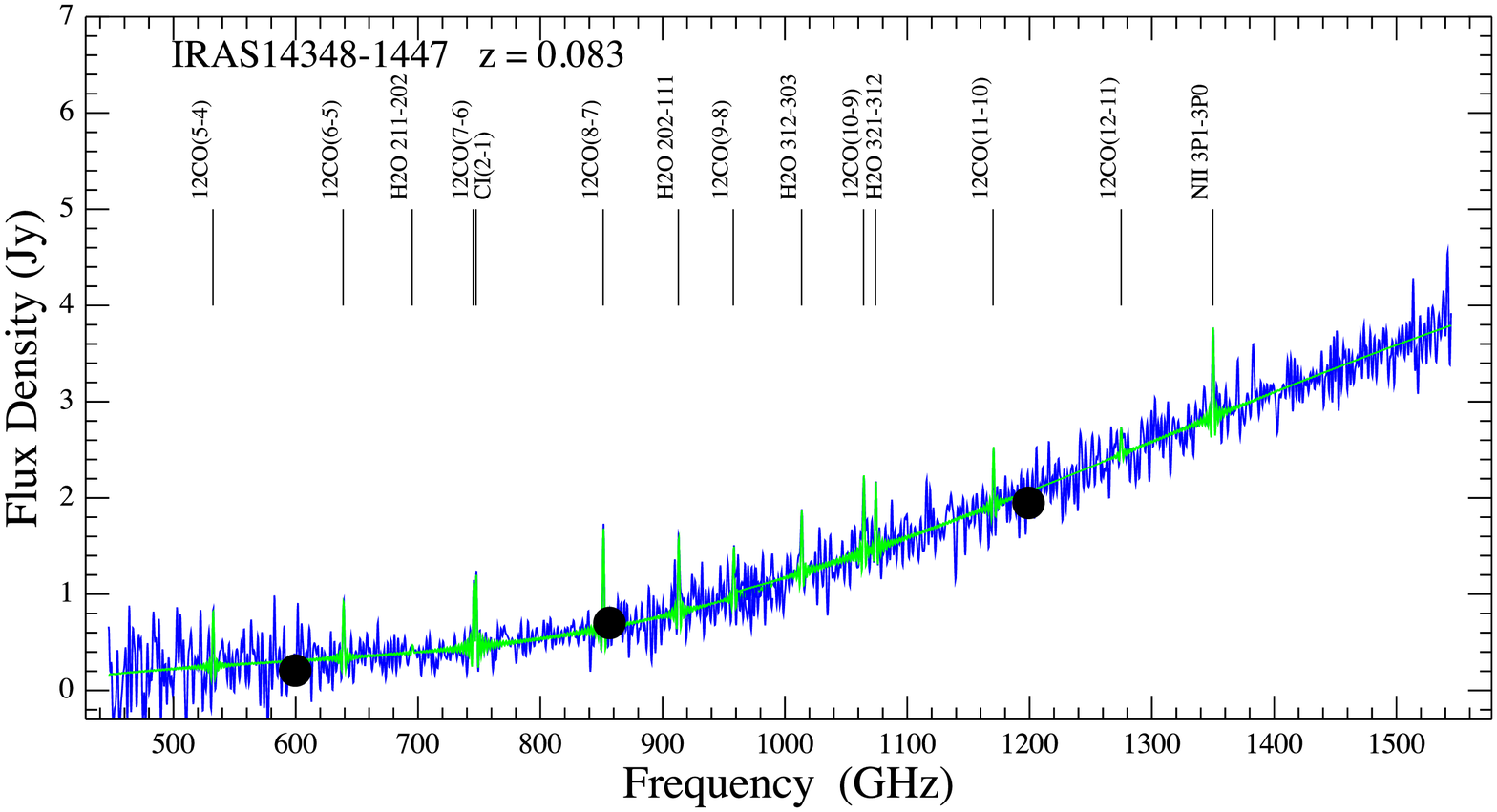}
\includegraphics[width=0.40\columnwidth,angle=0]{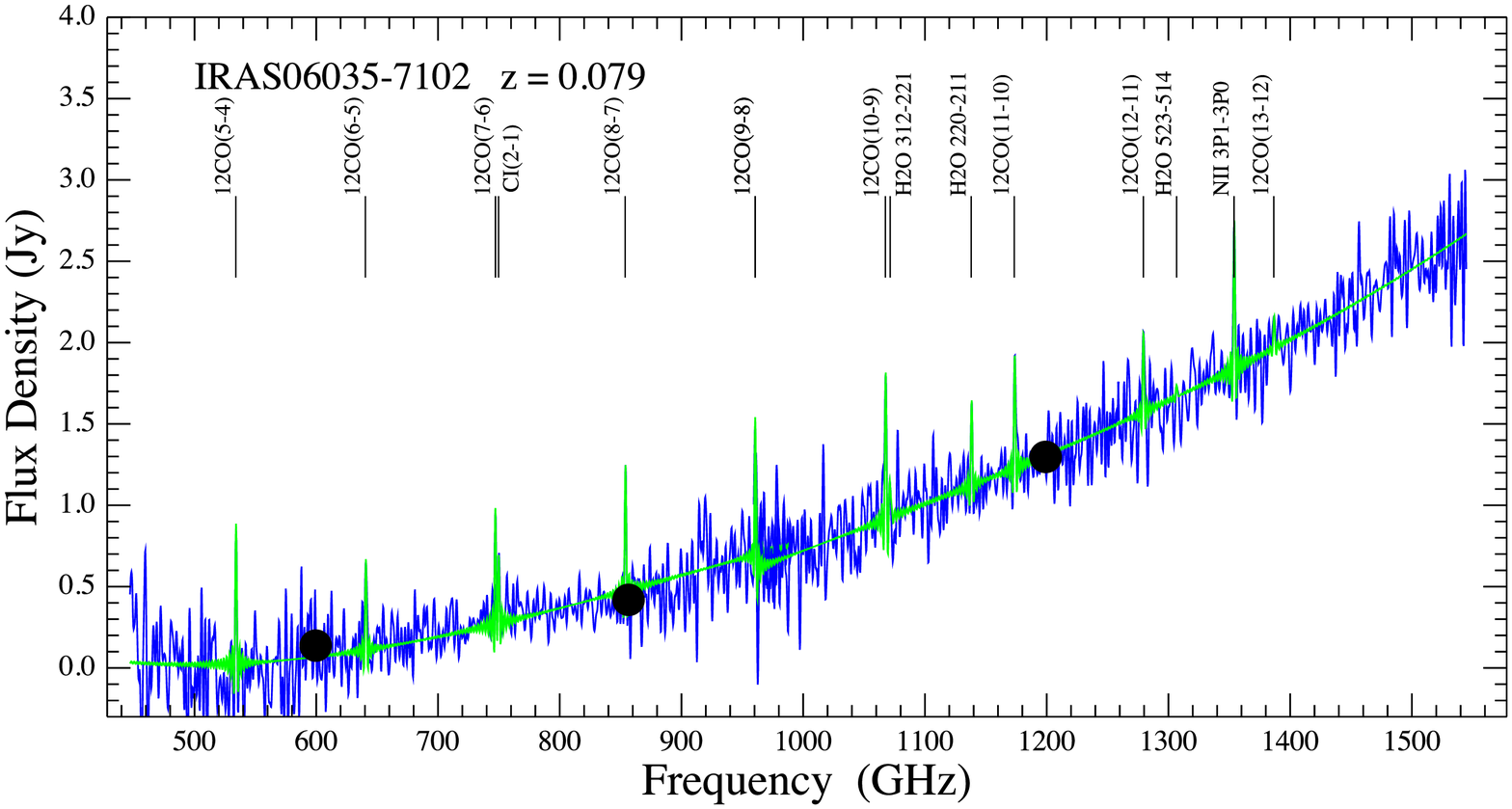}
\includegraphics[width=0.40\columnwidth,angle=0]{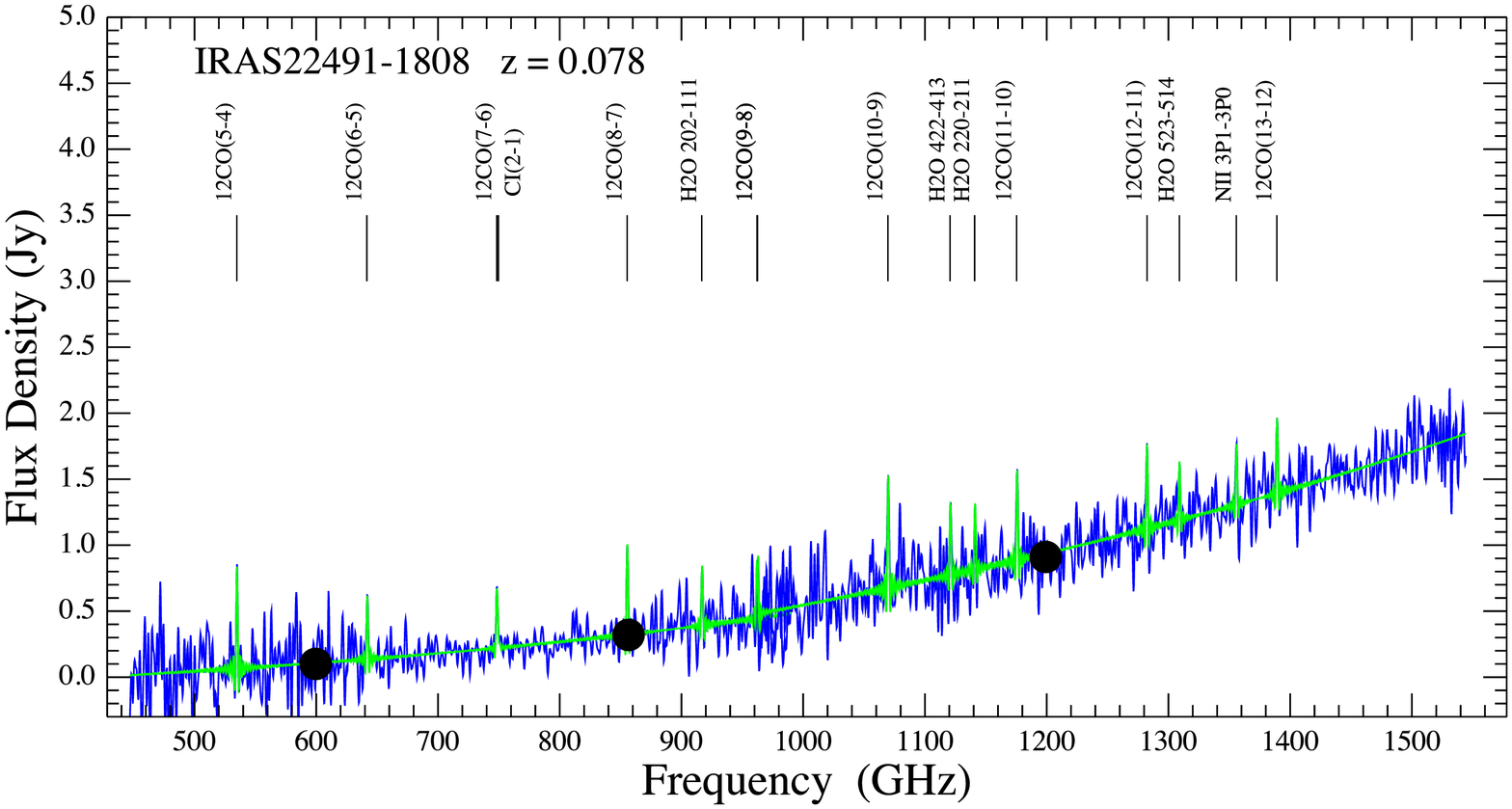}
\includegraphics[width=0.40\columnwidth,angle=0]{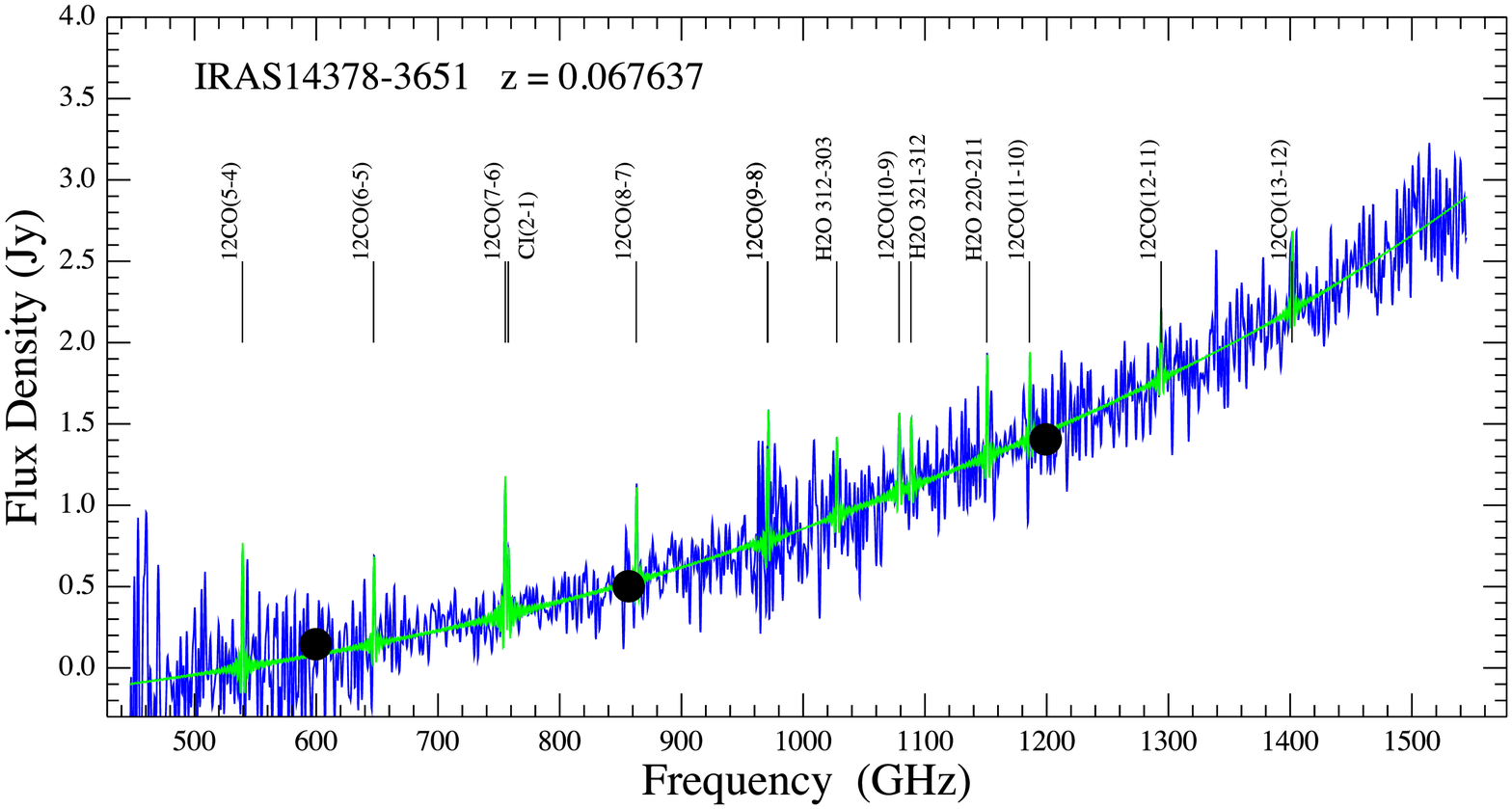}
\includegraphics[width=0.40\columnwidth,angle=0]{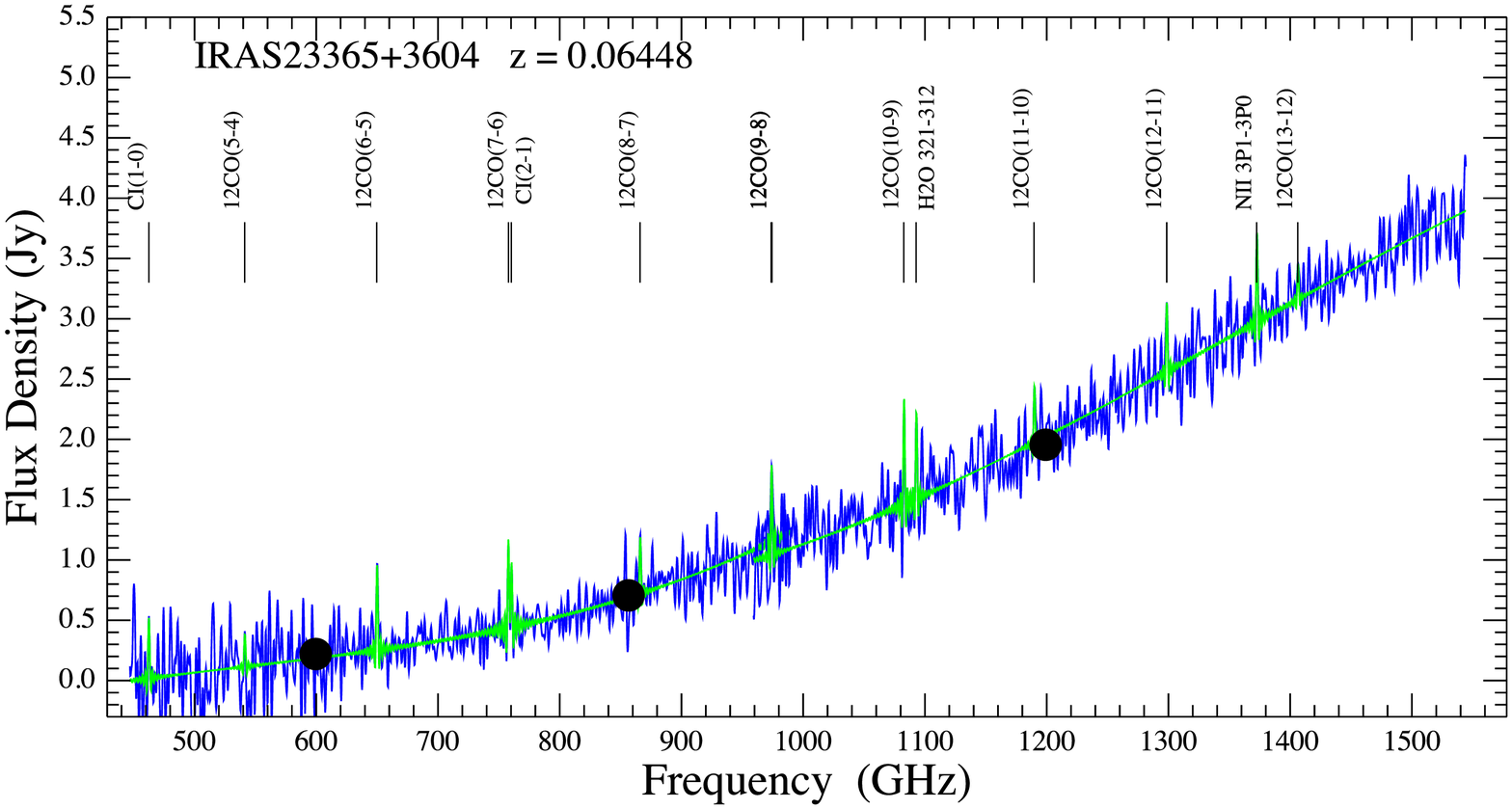}
\includegraphics[width=0.40\columnwidth,angle=0]{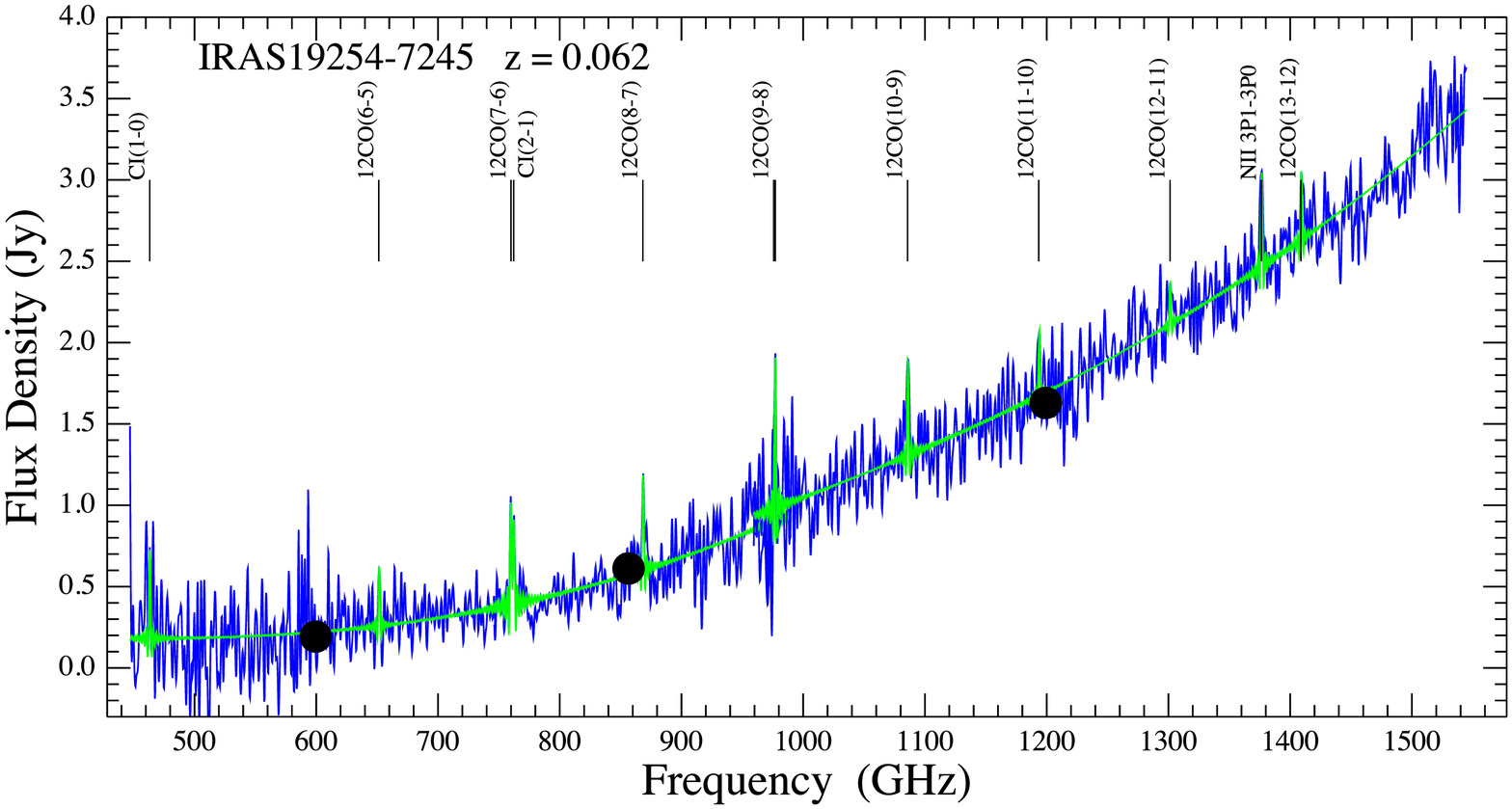}
\includegraphics[width=0.40\columnwidth,angle=0]{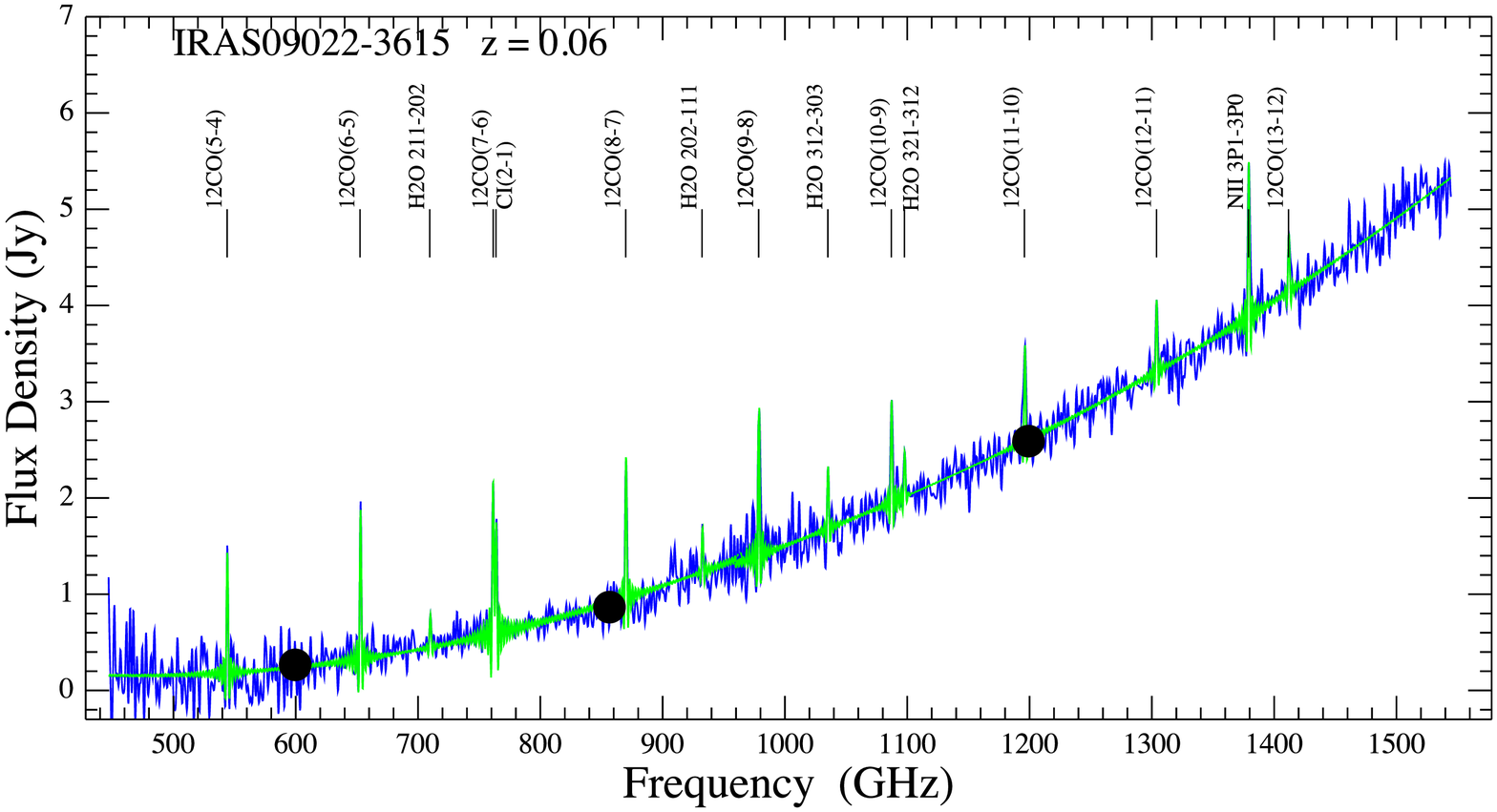}
\includegraphics[width=0.40\columnwidth,angle=0]{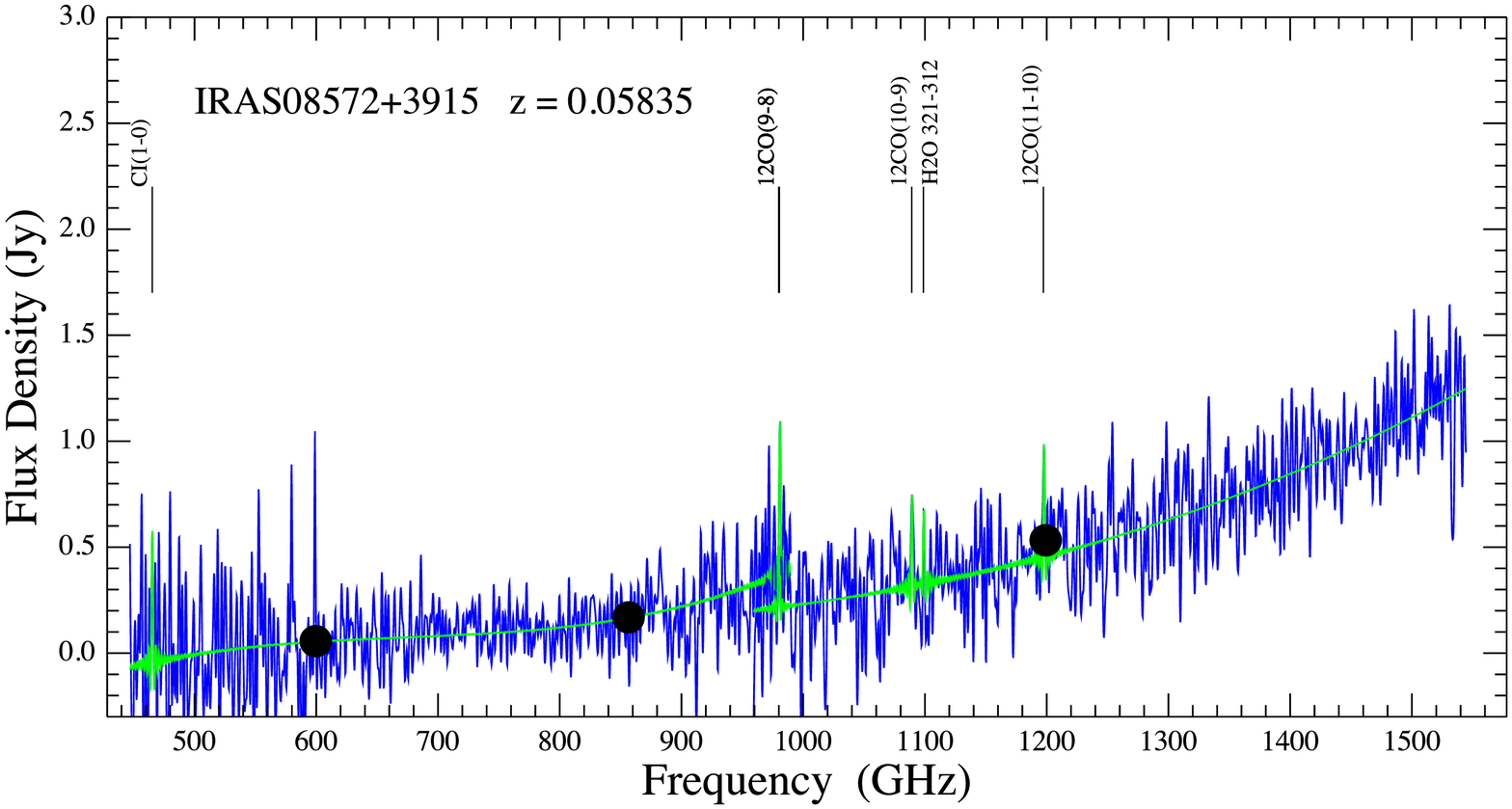}
\includegraphics[width=0.40\columnwidth,angle=0]{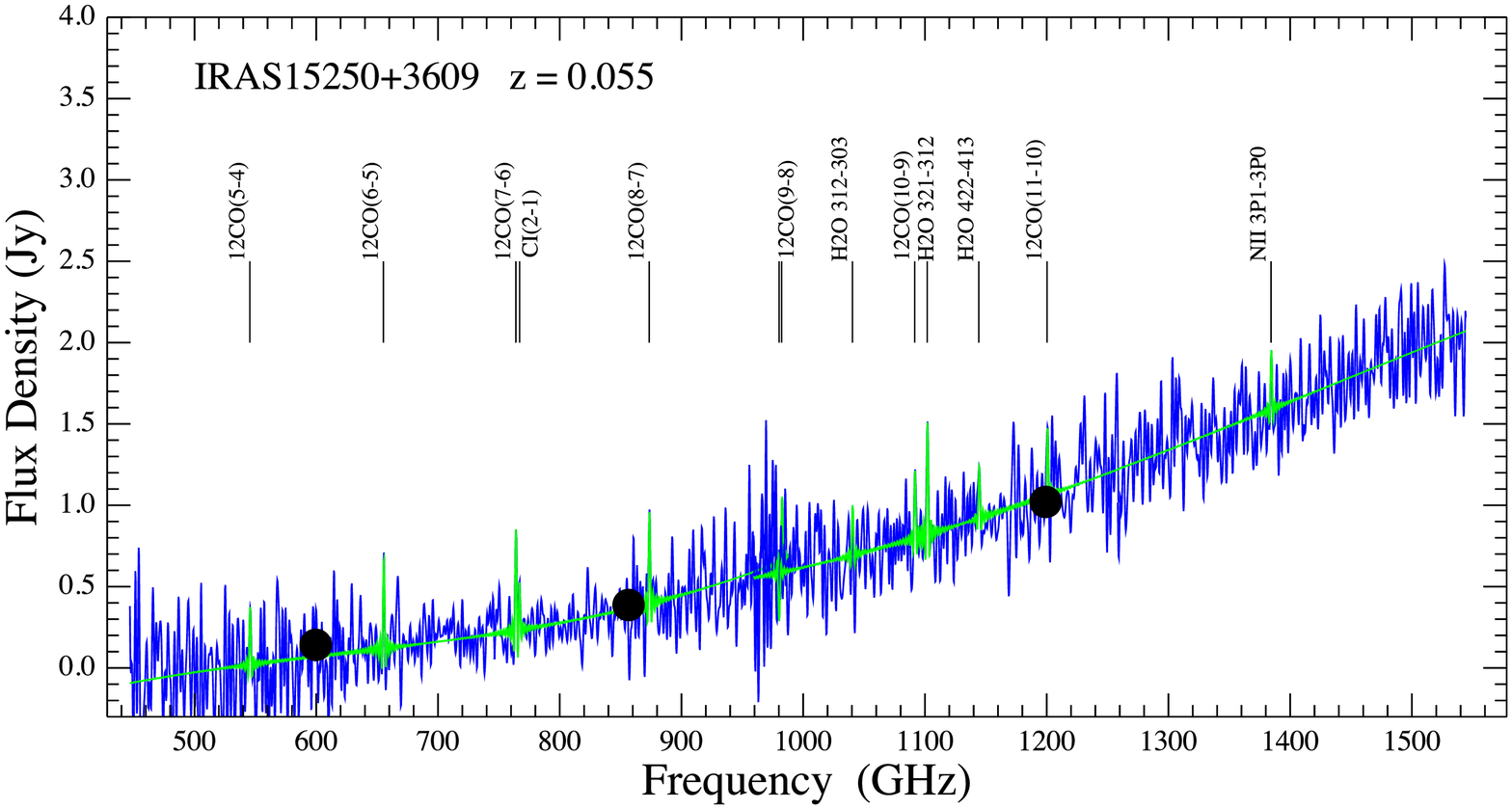}
\caption{Line fitting results for IRAS 19297-0406, IRAS 14348-1447, IRAS 06035-7102, IRAS 22491-1808, IRAS 14378-3651, IRAS 23365+3604, IRAS 19254-7245, IRAS 09022-3615, IRAS 08572+3915, IRAS 15250+3609. The processed spectra (blue) are shown with the model fits overlaid (green). Photometry points are also shown (black dots). Fitted line species are indicated by the vertical bars.}
\label{fig:fittedlines3}
\end{center}
\end{figure}

\begin{figure} 
\begin{center}
\includegraphics[width=0.40\columnwidth,angle=0]{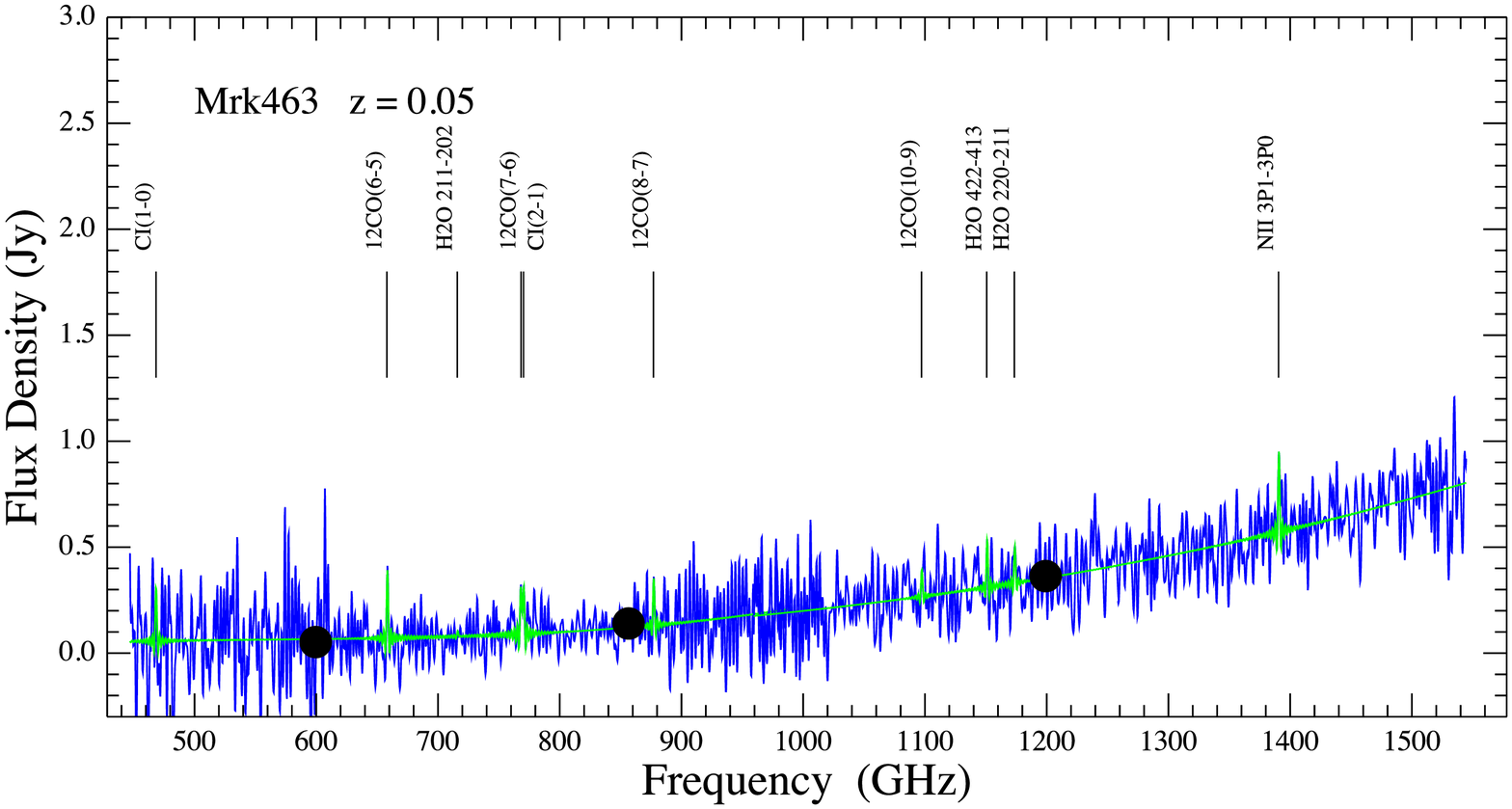}
\includegraphics[width=0.40\columnwidth,angle=0]{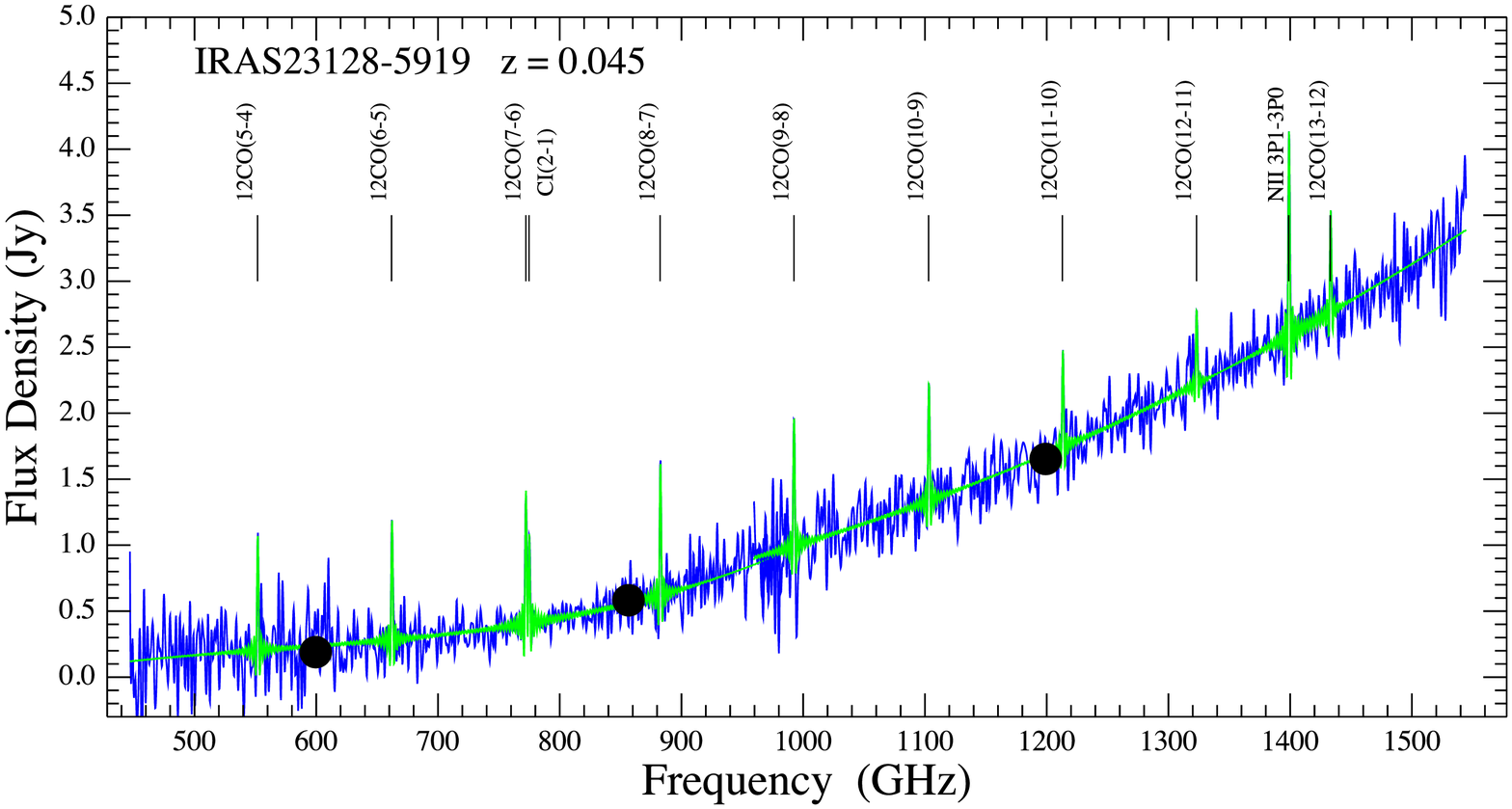}
\includegraphics[width=0.40\columnwidth,angle=0]{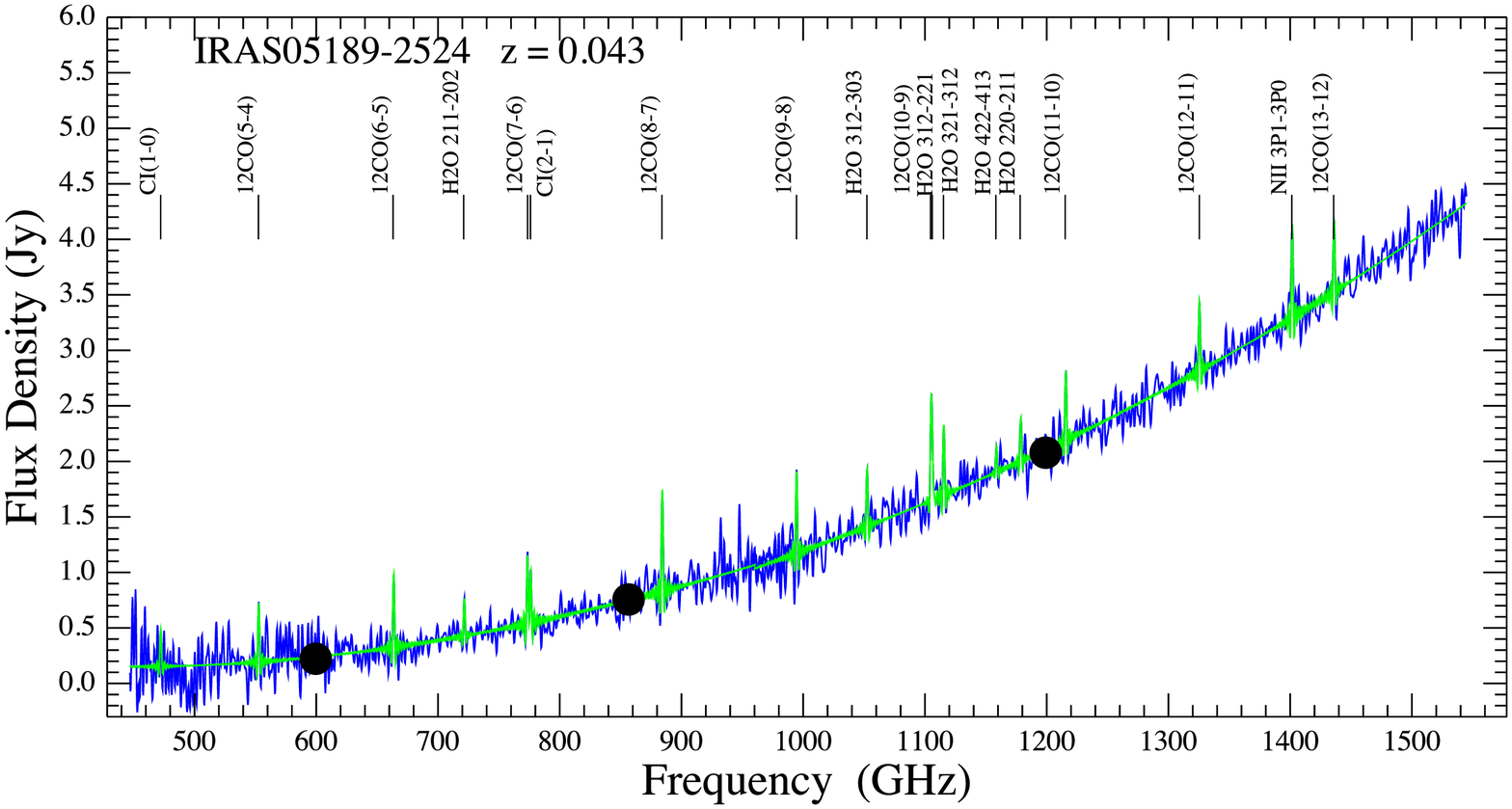}
\includegraphics[width=0.40\columnwidth,angle=0]{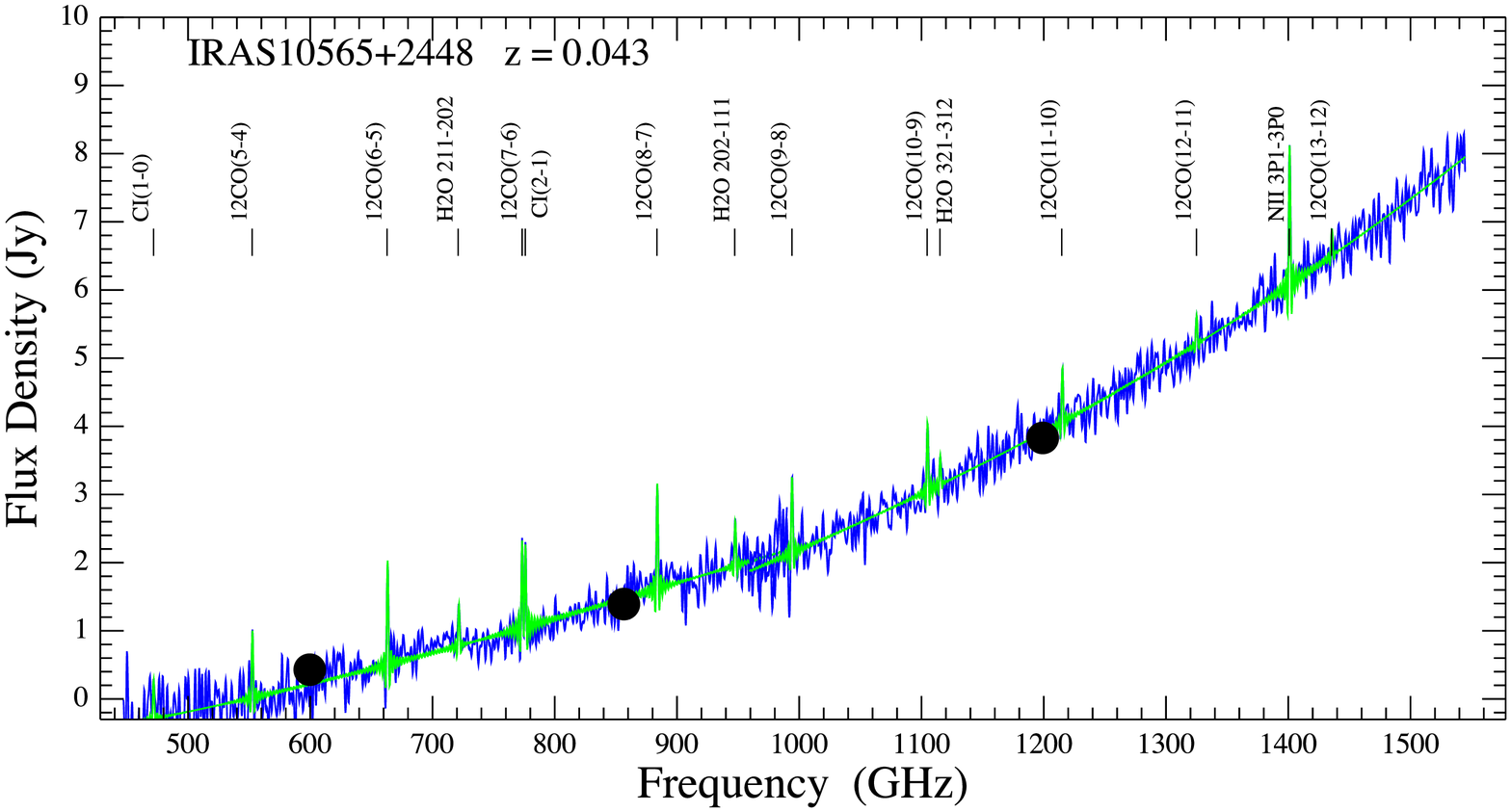}
\includegraphics[width=0.40\columnwidth,angle=0]{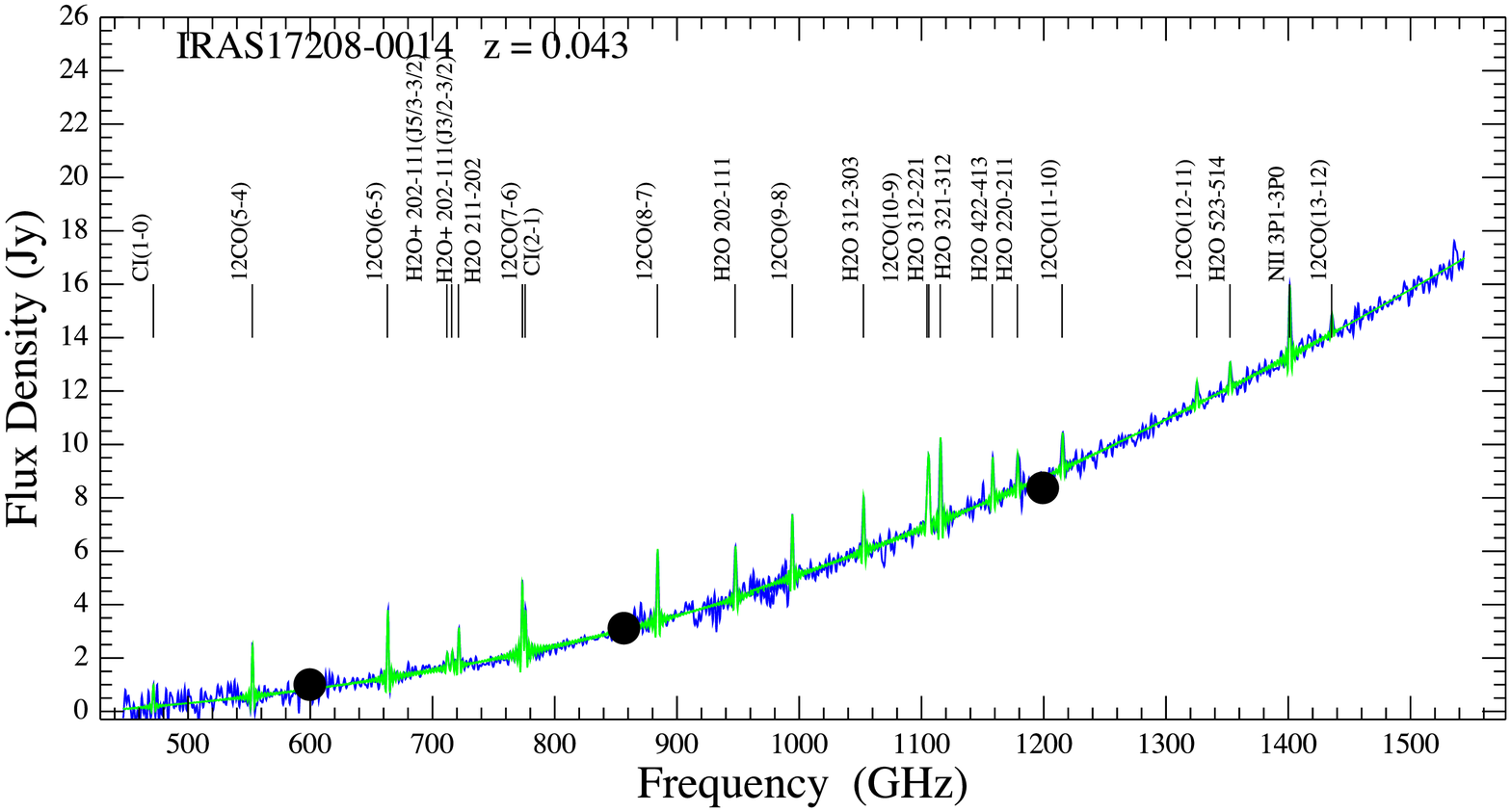}
\includegraphics[width=0.40\columnwidth,angle=0]{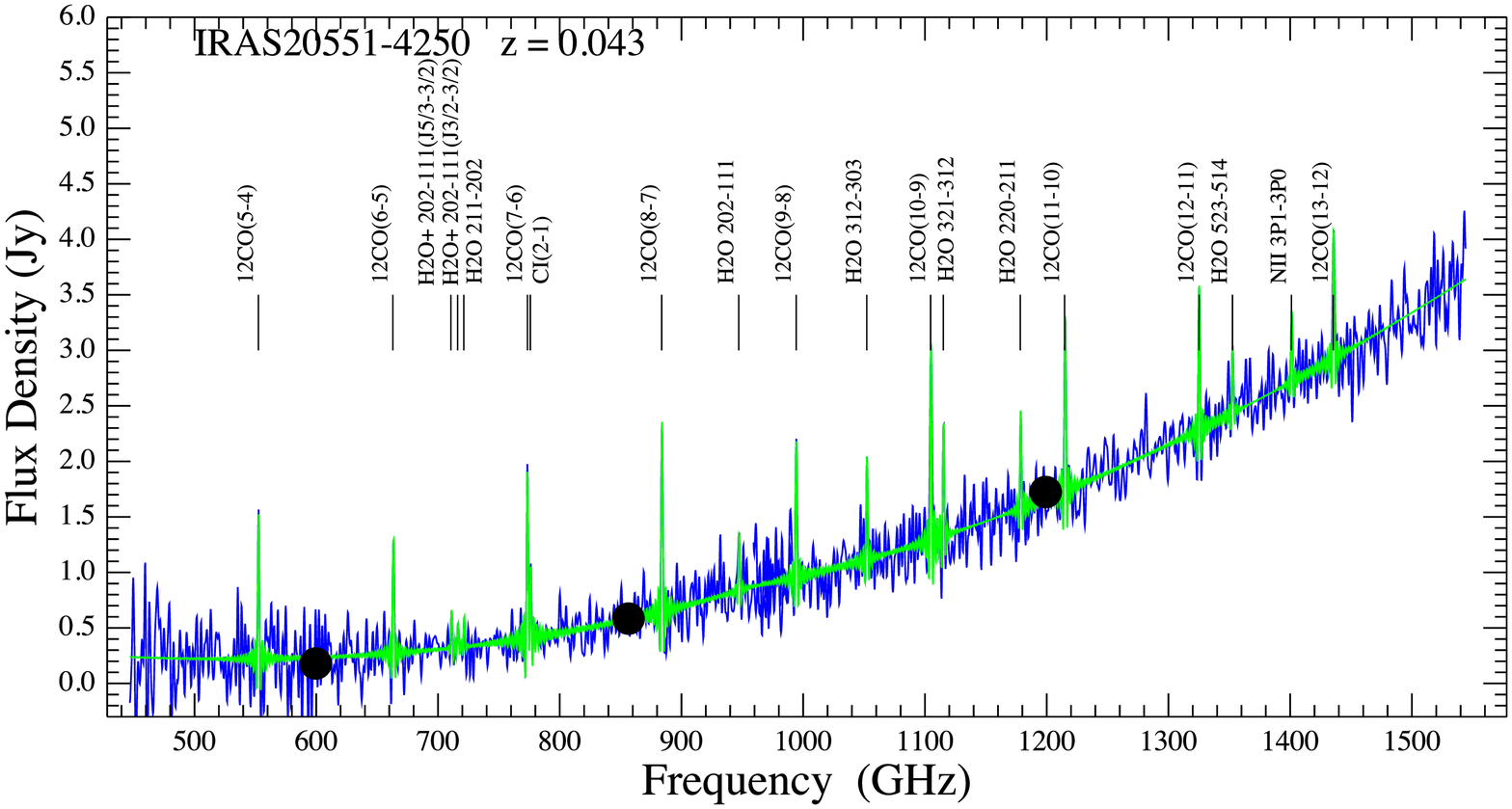}
\includegraphics[width=0.40\columnwidth,angle=0]{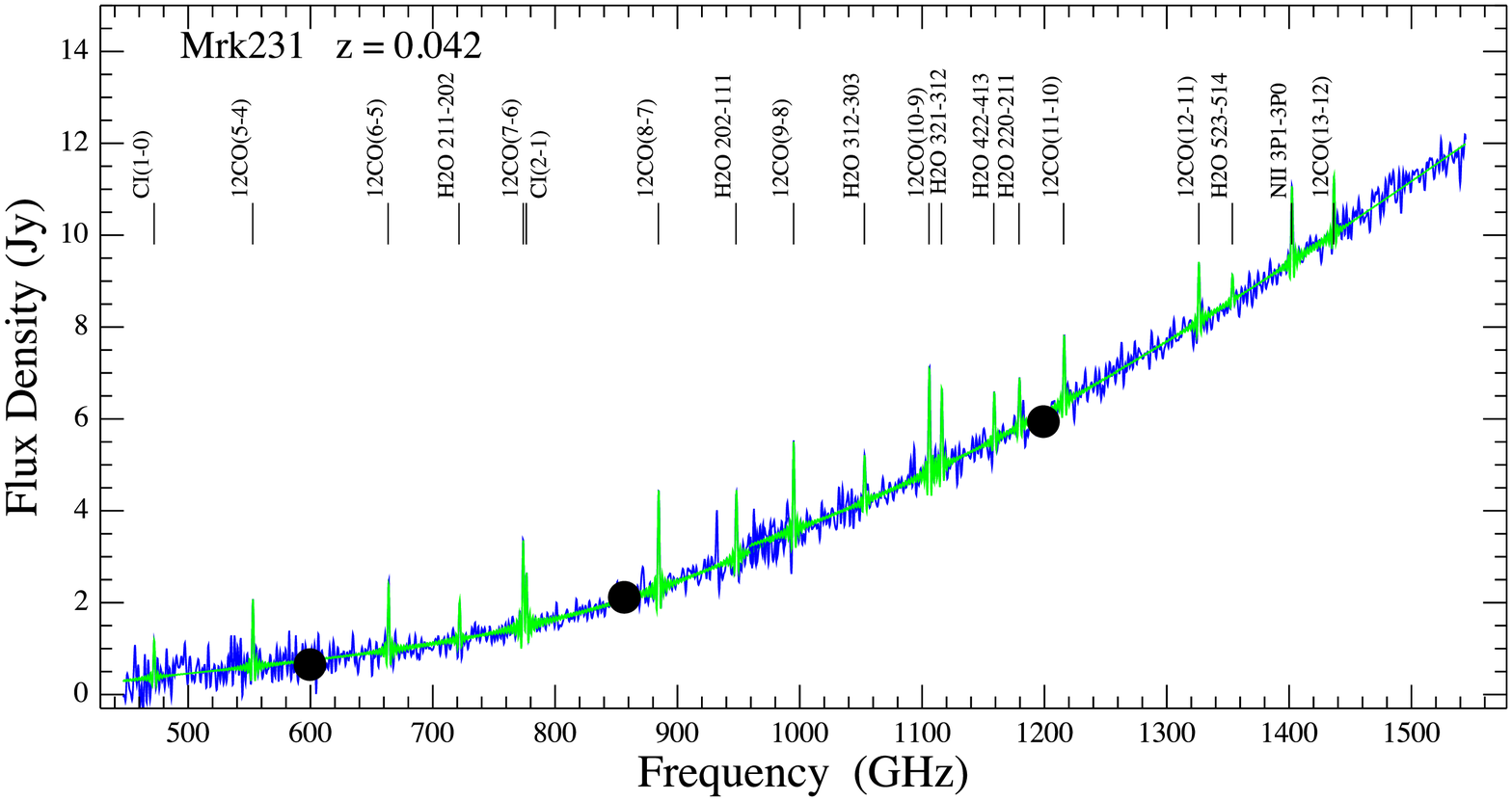}
\includegraphics[width=0.40\columnwidth,angle=0]{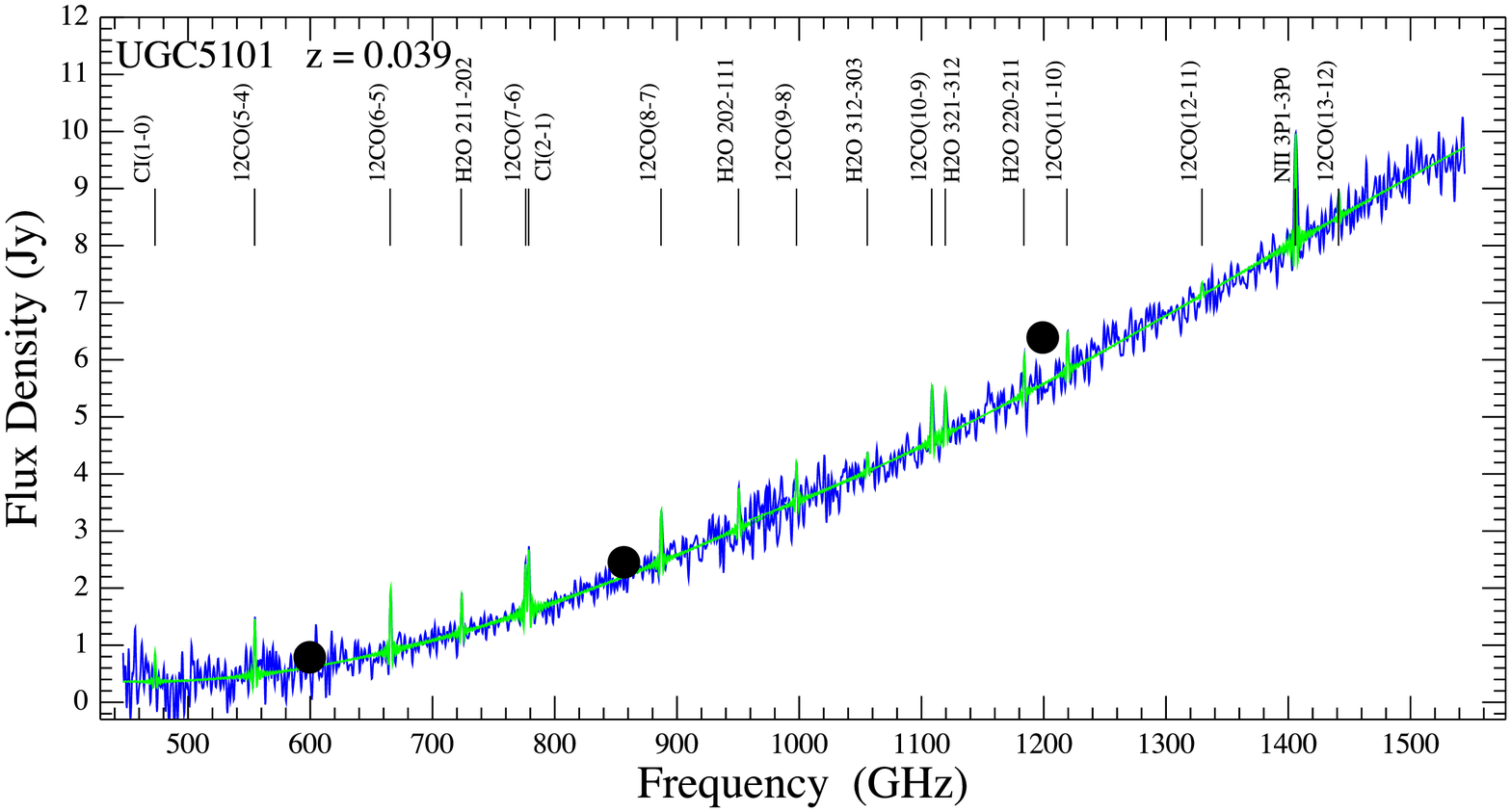}
\includegraphics[width=0.40\columnwidth,angle=0]{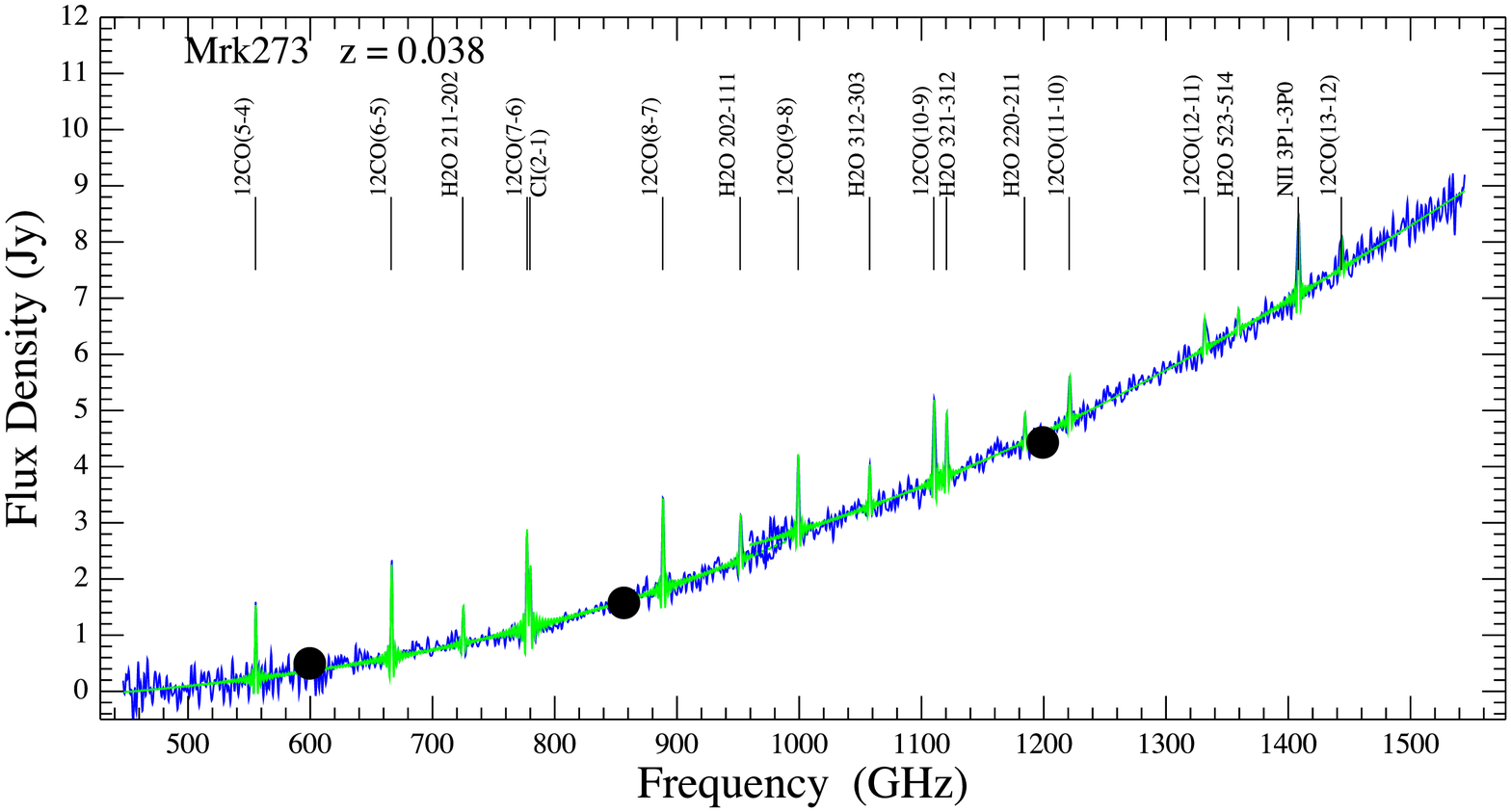}
\includegraphics[width=0.40\columnwidth,angle=0]{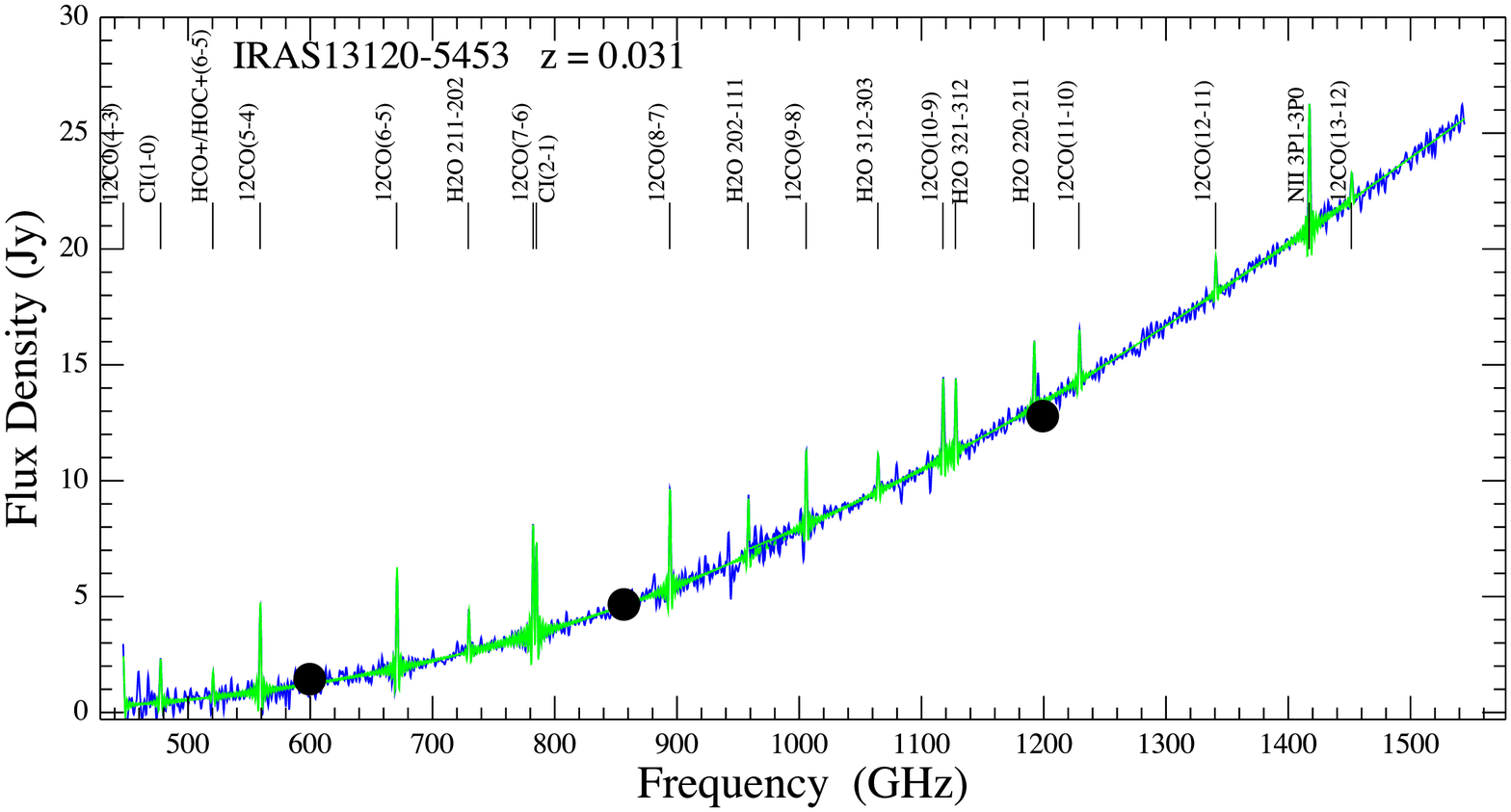}
\caption{Line fitting results for Mrk463, IRAS 23128-5919, IRAS 05189-2524, IRAS 10565+2448, IRAS 17208-0014, IRAS 20551-4250, Mrk231, UGC5101, Mrk273, IRAS 13120-5453. The processed spectra (blue) are shown with the model fits overlaid (green). Photometry points are also shown (black dots). Fitted line species are indicated by the vertical bars.}
\label{fig:fittedlines4}
\end{center}
\end{figure}

\begin{figure} 
\begin{center}
\includegraphics[width=0.40\columnwidth,angle=0]{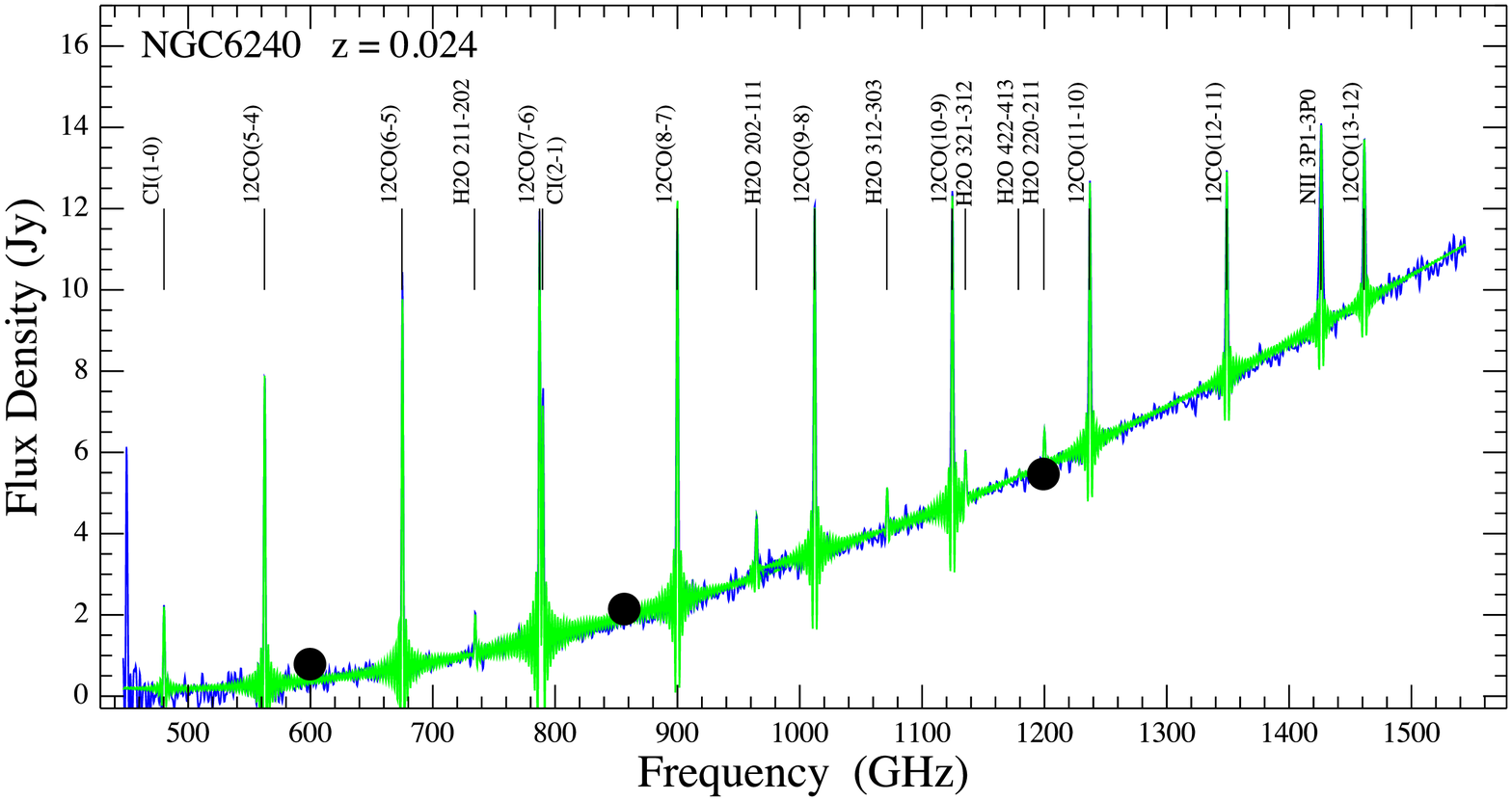}
\includegraphics[width=0.40\columnwidth,angle=0]{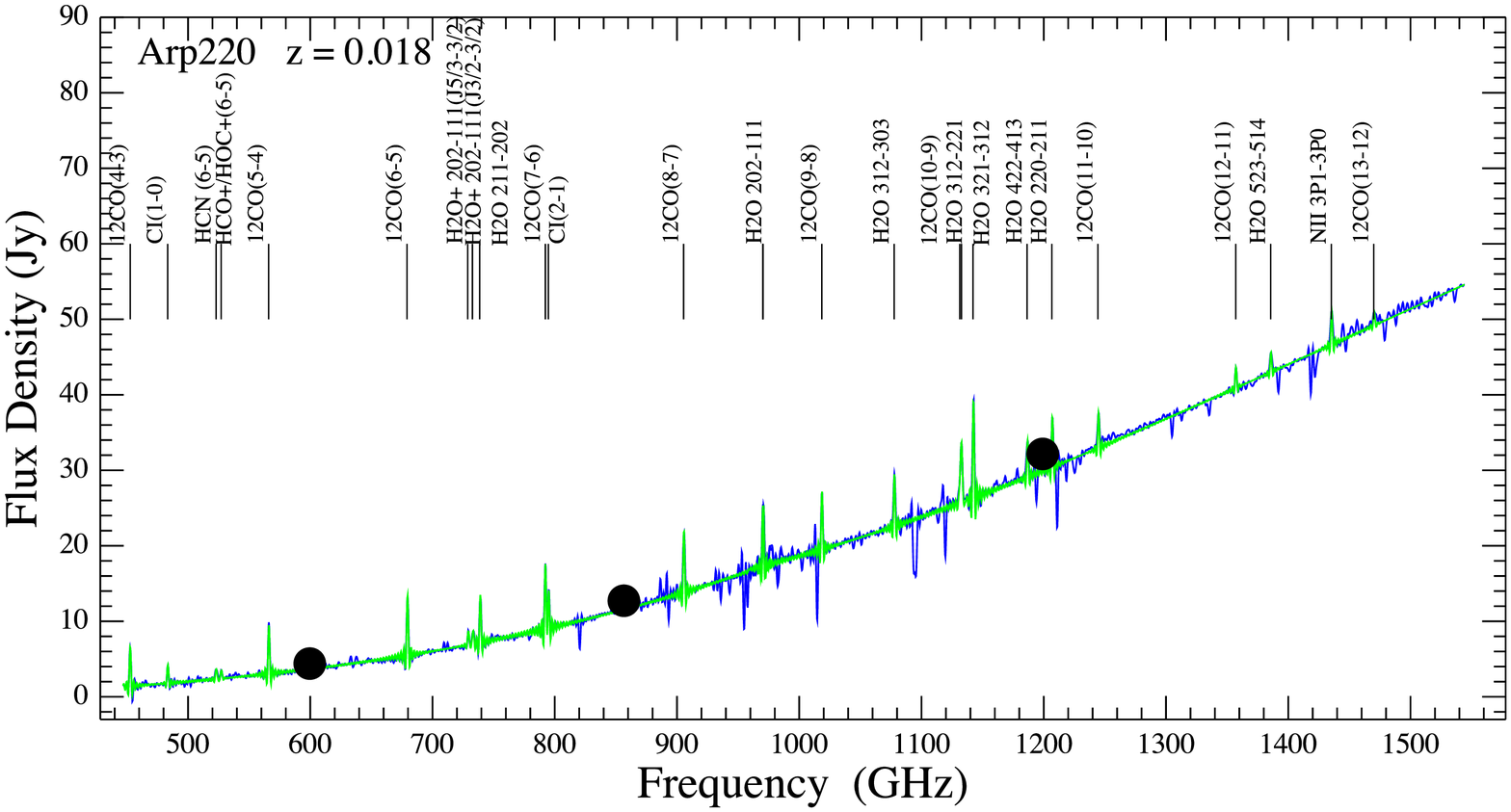}
\caption{Line fitting results for NGC 6240 and Arp220. The processed spectra (blue) are shown with the model fits overlaid (green). Photometry points are also shown (black dots). Fitted line species are indicated by the vertical bars.}
\label{fig:fittedlines5}
\end{center}
\end{figure}

The final spectra for each ULIRG are shown in Figures ~\ref{fig:fittedlines1}, ~\ref{fig:fittedlines2}, ~\ref{fig:fittedlines3}, ~\ref{fig:fittedlines4}, ~\ref{fig:fittedlines5}. The galaxies are tabulated in order of decreasing redshift beginning with IRAS 00397-1312 at $z=0.262$, to Arp 220 at $z=0.018$. The final concatenated spectra for the SSW and SLW bands are plotted as flux density (Jy) as a function of frequency (GHz). Overlaid on the spectra are the SPIRE photometry points and the fitted lines as described in Section~\ref{sec:LineFitting}. 

For our highest redshift object, IRAS 00397-1312 at $z=0.262$, the [CII] line at 157.7$\, \mu$m is redshifted into the SPIRE FTS SSW band pass. For this source a line flux of 28.22$\pm$0.81$\times$10$^{-18}$Wm$^{-2}$ is measured, corresponding to a line luminosity of $10^{9.20\pm0.07}$ L$_{\odot}$, a value comparable with similar ULIRGs at similar redshift (e.g. \citealt{magdis14}). 

CO lines are prominent in the majority of our sources. For two of our sources, IRAS 13120-5453 and Arp 220, the CO(4-3) transition at 461 GHz (649$\, \mu$m) is also detected. Some sources, notably IRAS 07598+6508 (z=0.148), Mrk1014 (z=0.163), IRAS13451+1232 (z=0.122) \& IRAS 08572+3915 (z=0.058) do not show CO lines. These sources are associated with AGN / BAL QSOs \citep{lipari94,boller02,spoon09}. Note that the FTS spectrum of IRAS 08572+3915 has been reported in \citet{efstathiou14} and was modelled with an edge-on AGN torus.

The CO line fluxes for our sample are given in Table \ref{tab:linefluxes}. Comparing our results with published ULIRGs, we find good agreement within the errors for Arp 220 \citep{rangwala11} and Mrk231 \citep{vanderwerf10}. In the case of NGC 6240 we measure substantially lower line fluxes compared to \citet{meijerink13}. The reason for this discrepancy is unclear since it is the same observation, although the data presented in  \citet{meijerink13} was processed with an earlier version of the FTS pipeline (HIPE version 6.0).

In addition, in many sources we detect the two [CI] fine structure lines at 492 GHz (607$\, \mu$m) and 809 GHz (370$\, \mu$m), originating in photodissociation regions in the transition region between the ionised Carbon and molecular CO \citep{kaufman99}. Moreover, the [NII] 3P1-3P0 fine structure line at 1461 GHz (205$\, \mu$m) is detected for the majority of the sample. The [NII] line is indicative of hot HII regions rather than the PDR and can be used to estimate the fraction of [CII] emission that comes from ionized gas, rather than from PDRs. In many cases, the [NII] line is partially resolved and relatively broad (see Figure \ref{fig:SincSincGauss}) and a sinc Gauss profile was used to model the line in many instances where FWHM of $\sim$ 300-400 km$^{-1}$ were measured, Although broad, these widths are still significantly smaller than in systems that unambiguously are shown to have massive outflows (e.g. \citealt{feruglio10}, \citealt{cicone14})

Water is the second most oxygenated molecule (following CO) in the warm interstellar medium of galaxies and has been detected by {\it Herschel} in observations of nearby ULIRGs \citep{gonzalezalfonso10,rangwala11,pereirasantaella13}. The HERUS ULIRGs are also found to be abundant in water lines, in particular the ortho-H$_{2}$O lines, H$_{2}$O 3$_{12}$-3$_{03}$ at 1097GHz (272$\, \mu$m), H$_{2}$O 3$_{12}$-2$_{21}$ at 1153GHz (259$\, \mu$m), H$_{2}$O 3$_{21}$-3$_{12}$ at 1163GHz (257$\, \mu$m), H$_{2}$O 5$_{23}$-5$_{14}$ at 1411GHz (212$\, \mu$m), H$_{2}$O+ 2$_{02}$-1$_{11}$(J5/3-3/2) at 742GHz (404$\, \mu$m), H$_{2}$O+ 2$_{02}$-1$_{11}$(J3/2-3/2) at 742GHz  (402$\, \mu$m) \& H$_{2}$O+ 2$_{11}$-2$_{02}$(J5/2-5/2) at 746GHz  (401$\, \mu$m) and the para-H$_{2}$O lines, H$_{2}$O 2$_{02}$-1$_{11}$ at 988GHz  (303$\, \mu$m), H$_{2}$O 4$_{22}$-4$_{13}$ at 1208 GHz (248$\, \mu$m) \& H$_{2}$O 2$_{20}$-2$_{11}$ at 1229 GHz (243$\, \mu$m). The strength of the water lines can vary considerably from source to source with examples where the water lines have a similar emission strength to the CO lines (IRAS 12071-0444, IRAS 23230-6926, IRAS 06206-6315), and where the water lines, although detected, are significantly weaker than the CO lines (IRAS 09022-3615, IRAS 10565+2448, NGC 6240). A thorough investigation of the water emission in the HERUS sample is beyond the scope of the current work and will be reported on in a forthcoming paper, however we note that our measured fluxes for the water lines in Mrk 231 and Arp 220 are consistent within the errors with the results of \citet{gonzalezalfonso10}, \citet{gonzalezalfonso13}.

Finally, in addition to the above species, we also detect HCO$^{+}$/HOC$^{+}$(6-5) at  537GHz (557$\, \mu$m) in IRAS 13120-5453, and confirm the detection of both HCO$^{+}$/HOC$^{+}$ and HCN at 532GHz (562$\, \mu$m) in Arp 220 \cite{rangwala11}.

In Table ~\ref{tab:H2Olinefluxes} the water line fluxes as measured by the SPIRE FTS for the HERUS local ULIRGs are tabulated in units of 10$^{-18}$Wm$^{-2}$. Line flux estimates were only included for cases where the measured signal to noise of the water lines was $>$3$\sigma$. 

\begin{table*}\tiny
\caption{$^{12}$CO line fluxes as measured by the SPIRE FTS.}
\tabcolsep 4.2pt
\begin{tabular}{@{}llllllllll}
\hline
Transition		    & 5-4     & 6-5     & 7-6     & 8-7     & 9-8     & 10-9     & 11-10     & 12-11     & 13-12            \\
$\nu_{rest}$ (GHz)	& 576.3   & 691.5   & 806.7   & 921.8   & 1036.9  & 1151.9   & 1267.0    & 1382.0    & 1496.9			\\
\hline
Name                & \multicolumn{9}{c}{10$^{-18}$Wm$^{-2}$}		\\
\hline
00397-1312		&	4.05$\pm$0.80	&	2.51$\pm$0.78	&	-	            &	1.91$\pm$0.78	&	2.07$\pm$0.79	&	2.67$\pm$0.79	&	-	            &	-	            &	2.54$\pm$0.80	\\
Mrk1014	        &	3.33$\pm$0.75	&	4.56$\pm$0.75	&	1.96$\pm$0.76	&	-		        &	-	     	    &	-	     	    &	-	      	    &	-		        &	-	    \\
03521+0028		&	-	            &	4.11$\pm$0.78	&	-		        &	3.65$\pm$0.78	&	-		        &	4.46$\pm$0.87	&	2.85$\pm$0.78	&	2.21$\pm$0.78	&	-	    \\
07598+6508		&	- 	            &	3.30$\pm$0.78	&	-		        &	-		        &	-		        &	-	      	    &	-		        &	-		        &	-	    \\
10378+1109		&	-	            &	4.42$\pm$0.76	&	4.92$\pm$0.76	&	4.24$\pm$0.76	&	2.15$\pm$0.76	&	3.02$\pm$0.70	&	1.67$\pm$0.70	&	-		        &	3.59$\pm$0.70	\\
03158+4227 	    &	2.61$\pm$0.94	&	4.69$\pm$0.94	&	4.31$\pm$0.96	&	4.49$\pm$0.94	&	-	      	    &	4.33$\pm$0.91	&	2.90$\pm$0.91	&	-		        &	-	    \\
16090-0139		&	3.68$\pm$0.09	&	2.93$\pm$0.85	&	9.25$\pm$8.94	&	9.74$\pm$0.85	&  13.20$\pm$0.85	&	8.91$\pm$0.85	&	5.88$\pm$0.84	&	4.24$\pm$0.84	&	4.40$\pm$0.84	\\
20100-4156		&	5.92$\pm$0.96	&	-	            &	4.08$\pm$1.00	&	5.63$\pm$0.96	&	-	     	    &	8.10$\pm$0.99	&	3.22$\pm$0.99	&	-	            &	-	    \\
23253-5415		&	- 	            &	2.57$\pm$0.73	&	2.63$\pm$0.75	&	2.99$\pm$0.73	&	3.73$\pm$0.73	&	-	      	    &	1.49$\pm$0.77	&	2.15$\pm$0.77	&	2.06$\pm$0.77	\\
00188-0856		&	-	            &	-	            &	2.39$\pm$0.89	&	3.12$\pm$0.85	&	4.83$\pm$0.85	&	3.45$\pm$0.73	&	-		        &	3.62$\pm$0.73	&	-		    \\
12071-0444		&	3.64$\pm$0.76	&	4.46$\pm$0.76	&	2.35$\pm$0.77	&	3.80$\pm$0.76	&	0.00$\pm$0.76	&	3.69$\pm$0.71	&	2.04$\pm$0.71	&	2.56$\pm$0.71	&	-		    \\
13451+1232		&	3.69$\pm$0.90	&	-	            &	2.41$\pm$0.89	&	-	     	    &	-	     	    &	2.96$\pm$0.82	&	-	            &	-	     	    &	-		    \\
01003-2238		&	-	            &	-	            &	-	 	        &	-		        &	6.61$\pm$0.82	&	0.00$\pm$0.75	&	3.52$\pm$0.75	&	2.42$\pm$0.75	&	2.05$\pm$0.75	\\
11095-0238		&	4.39$\pm$0.82	&	-	            &	3.32$\pm$0.82	&	4.68$\pm$0.82	&	4.10$\pm$0.83	&	4.87$\pm$0.80	&	2.43$\pm$0.80	&	0.00$\pm$0.80	&	3.06$\pm$0.80	\\
20087-0308		&	8.79$\pm$0.90	&	3.96$\pm$0.90	&	7.52$\pm$0.93	&	5.37$\pm$0.90	&	3.70$\pm$0.90	&	6.52$\pm$0.88	&	4.18$\pm$0.88	&	0.00$\pm$0.88	&	4.11$\pm$0.88	\\
23230-6926		&	3.37$\pm$0.83	&	0.83$\pm$0.83	&	3.96$\pm$0.84	&	3.24$\pm$0.83	&	4.70$\pm$0.83	&	5.53$\pm$0.78	&	3.15$\pm$0.78	&	4.54$\pm$0.78	&	3.63$\pm$0.78	\\
08311-2459		&	10.50$\pm$1.00	&	11.50$\pm$1.00	&	8.54$\pm$1.03	&	8.05$\pm$1.00	&	5.19$\pm$1.00	&	7.55$\pm$0.90	&	6.13$\pm$0.92	&	7.86$\pm$0.92	&	2.53$\pm$0.92	\\
15462-0450		&	2.40$\pm$0.81	&	-	       	    &	3.48$\pm$0.82	&	2.07$\pm$0.81	&	-	     	    &	2.91$\pm$0.74	&	-	            &	2.54$\pm$0.74	&	0.00$\pm$0.74	\\
06206-6315		&	-	            &	4.40$\pm$0.94	&	5.26$\pm$0.99	&	3.54$\pm$0.94	&	3.45$\pm$0.94	&	3.91$\pm$0.82	&	3.86$\pm$0.82	&	-	            &	-		    \\
20414-1651		&	-	            &	-         	    &	2.62$\pm$0.97	&	3.96$\pm$0.93	&	3.35$\pm$0.93	&	3.89$\pm$0.91	&	-	            &	4.66$\pm$0.90	&	-		    \\
19297-0406		&	6.23$\pm$1.14	&	5.72$\pm$1.14	&	6.72$\pm$1.19	&	9.19$\pm$1.14	&	7.82$\pm$1.15	&	2.77$\pm$0.13	&	4.64$\pm$0.12	&	-	            &	4.32$\pm$1.16	\\
14348-1447		&	6.90$\pm$1.09	&	7.07$\pm$0.11	&	7.87$\pm$1.14	&  11.20$\pm$1.09	&	6.33$\pm$1.10	&	9.32$\pm$0.11	&	7.23$\pm$0.11	&	3.33$\pm$0.11	&	-		    \\
06035-7102		&	10.20$\pm$1.00	&	6.58$\pm$1.00	&	8.05$\pm$1.02	&	9.15$\pm$1.00	&  10.10$\pm$0.99	&	2.91$\pm$0.96	&	8.40$\pm$0.95	&	5.87$\pm$0.95	&	2.47$\pm$0.95	\\
22491-1808		&	9.14$\pm$0.88	&	5.69$\pm$0.88	&	5.42$\pm$1.13	&	8.20$\pm$0.88	&	5.57$\pm$0.89	&  10.20$\pm$0.96	&	8.07$\pm$0.98	&	7.48$\pm$0.98	&	6.76$\pm$0.99	\\
14378-3651		&	9.03$\pm$1.16	&	6.47$\pm$1.16	&	9.89$\pm$1.20	&	6.80$\pm$1.16	&	6.88$\pm$1.11	&	5.63$\pm$1.10	&	6.41$\pm$1.09	&	5.13$\pm$1.09	&	5.78$\pm$1.09	\\
23365+3604		&	3.20$\pm$1.11	&	8.45$\pm$1.11	&	8.74$\pm$1.16	&	5.56$\pm$1.11	&	6.29$\pm$1.12	&  10.10$\pm$1.11	&	5.55$\pm$1.11	&	6.73$\pm$1.11	&	3.39$\pm$1.11	\\
19254-7245		&	-	            &	4.36$\pm$1.06	&	7.75$\pm$1.11	&	6.97$\pm$1.06	&	7.97$\pm$1.07	&	6.97$\pm$1.00	&	4.75$\pm$1.00	&	3.07$\pm$1.00	&	5.07$\pm$1.00	\\
09022-3615		&	15.00$\pm$1.02	&	18.60$\pm$1.02	&  19.80$\pm$1.07	&  17.80$\pm$1.02	&  17.70$\pm$1.04	&  12.80$\pm$0.98	&  11.90$\pm$0.98	&	8.93$\pm$0.98	&	7.37$\pm$0.98	\\
08572+3915		&	-	            &	-	     	    &	-	      	    &	-               &	8.64$\pm$1.17	&	5.12$\pm$1.10	&	6.33$\pm$1.10	&	-	            &	-	    \\
15250+3609		&	4.17$\pm$1.16	&	6.55$\pm$1.16	&	7.07$\pm$1.17	&	6.64$\pm$1.16	&	4.71$\pm$1.17	&	4.68$\pm$1.17	&	4.96$\pm$1.16	&	-	            &	-	    \\
Mrk463	       	&	-	            &	3.79$\pm$0.80	&	3.04$\pm$0.86	&	2.63$\pm$0.82	&	-	     	    &	1.52$\pm$0.75	&	-	            &	-	            &	-	    \\
23128-5919		&	10.40$\pm$0.94	&	10.80$\pm$0.94	&  15.60$\pm$0.97	&  11.90$\pm$0.94	&  11.70$\pm$0.99	&  10.70$\pm$0.99	&	8.69$\pm$0.96	&	6.72$\pm$0.99	&	9.01$\pm$9.86	\\
05189-2524		&	4.71$\pm$0.93	&	6.44$\pm$0.93	&	7.93$\pm$0.97	&	9.90$\pm$0.93	&	9.95$\pm$0.85	&  12.60$\pm$1.23	&	9.15$\pm$0.85	&	7.57$\pm$0.85	&	7.32$\pm$0.85	\\
10565+2448		&	11.70$\pm$1.37	&	18.30$\pm$1.37	&  14.10$\pm$1.40	&  18.40$\pm$1.37	&  13.40$\pm$1.26	&  11.90$\pm$1.26	&	9.56$\pm$1.25	&	4.92$\pm$1.25	&	4.78$\pm$1.25	\\
17208-0014		&	22.60$\pm$1.43	&	30.40$\pm$1.43	&  31.70$\pm$1.50	&  32.20$\pm$1.43	&  27.80$\pm$1.57	&  28.70$\pm$1.67	&  15.90$\pm$1.56	&	9.99$\pm$1.56	&	8.72$\pm$1.56	\\
20551-4250		&	15.50$\pm$1.18	&	12.40$\pm$1.18	&  17.70$\pm$1.22	&  20.40$\pm$1.18	&  14.60$\pm$1.12	&  20.70$\pm$1.12	&  18.80$\pm$1.12	&  15.60$\pm$1.12	&  13.80$\pm$1.12	\\
Mrk231	        &	17.40$\pm$1.22	&	17.30$\pm$1.22	&  21.80$\pm$1.26	&  25.80$\pm$1.22	&  22.80$\pm$1.21	&  26.50$\pm$1.21	&  17.20$\pm$1.20	&  15.50$\pm$1.21	&  15.00$\pm$1.21	\\
UGC5101	       	&	11.70$\pm$1.25	&	13.30$\pm$1.25	&	9.55$\pm$1.30	&  11.00$\pm$1.25	&	8.19$\pm$1.38	&  12.00$\pm$1.38	&	8.25$\pm$1.37	&	2.46$\pm$1.37	&	5.25$\pm$1.37	\\
13120-5453$^{*}$&	45.30$\pm$1.69	&	51.20$\pm$1.69	&  53.60$\pm$1.74	&  51.00$\pm$1.69	&  38.30$\pm$1.76	&  39.70$\pm$1.76	&  26.10$\pm$1.76	&  17.90$\pm$1.76	&  14.10$\pm$1.76	\\
NGC 6240	    &	94.80$\pm$2.75	&	107.00$\pm$2.75	& 122.00$\pm$2.86	& 121.0$\pm$2.75	& 102.0$\pm$2.82	&  94.20$\pm$2.82	&  81.00$\pm$2.81	&  63.50$\pm$2.82	&  51.70$\pm$2.82	\\
Arp 220$^{**}$  &	76.70$\pm$4.08	&	95.20$\pm$4.07	&  99.00$\pm$4.20	&  97.30$\pm$4.08	&  90.10$\pm$6.82	&  70.10$\pm$7.56	&  55.20$\pm$6.81	&  32.90$\pm$6.81	&  22.10$\pm$6.82	\\
\hline
\multicolumn{10}{l}{* Also: CO(4-3) flux of 25.6$\pm$2.2$\times$10$^{-18}$Wm$^{-2}$.} \\
\multicolumn{10}{l}{** Also: CO(4-3) flux of 62.4$\pm$4.15$\times$10$^{-18}$Wm$^{-2}$.} \\
\end{tabular}
\label{tab:linefluxes}
\end{table*}


\begin{table*}\tiny
\caption{H$_{2}$O line fluxes as measured by the SPIRE FTS.}
\tabcolsep 4.2pt
\begin{tabular}{@{}lllllllll}
\hline
Transition		     & $2_{11}-2_{02}$ & $2_{02}-1_{11}$ & $3_{12}-3_{03}$ & $3_{12}-2_{21}$ & $3_{21}-3_{12}$ & $4_{22}-4_{13}$ & $2_{20}-2_{11}$ & $5_{23}-5_{14}$ \\
$\nu_{rest}$ (GHz)   & 752.0           & 987.9           & 1097.4          & 1153.1          & 1162.9          & 1207.6          & 1228.8          & 1410.6 	    \\
\hline
Name			     &	 \multicolumn{8}{c}{10$^{-18}$Wm$^{-2}$}		\\
\hline
IRAS00397-1312		 &	-	        	& 2.79$\pm$0.78	&	-			&	-	            &	 3.20$\pm$0.79	&	3.20$\pm$0.79	&	-	        	&	-		\\
Mrk1014				 &	-	        	&		-	    & 	-			&	-	            &	-	            &	-	            &	-	        	&	-		\\
IRAS03521+0028		 &	-	        	& 2.58$\pm$0.77	& 3.23$\pm$0.78	&	3.23$\pm$0.78	&	-	            &	-	            &	-	        	&	-		\\
IRAS07598+6508		 &	-	        	&	-        	&	-       	&	-	            &	-	            &	-	            &	-	        	&	-		\\
IRAS10378+1109		 &  2.62$\pm$0.76	&	-        	&	-       	&	-	            &	-	            &	-	            &	-	        	&	-		\\
IRAS03158+4227		 &	-	         	&	-        	& 7.17$\pm$0.95	&	7.17$\pm$0.95	&	-	            &	-	            &	-	        	&	-		\\
IRAS16090-0139		 &  4.43$\pm$0.85	& 3.52$\pm$0.85	& 4.88$\pm$0.86	&	4.88$\pm$0.86	&	-	            &	3.44$\pm$0.84	&	-	        	&	-		\\
IRAS20100-4156		 &	-	         	& 3.59$\pm$0.96	&	-       	&	-	            &	 6.72$\pm$0.99	&	5.69$\pm$0.99	&	4.75$\pm$0.99	&	-		\\
IRAS23253-5415		 &	-	         	&	-        	&	-       	&	-	            &	 5.57$\pm$0.77	&	-	            &	-	        	&	-		\\
IRAS00188-0856		 &	-	         	&	-        	&	-       	&	-	            &	-	            &	-	            &	-	        	&	-		\\
IRAS12071-0444		 &	-	         	&	-        	& 4.63$\pm$0.72	&	4.63$\pm$0.72	&	 2.88$\pm$0.71	&	-	            &	3.02$\pm$0.71	&	3.27$\pm$0.71	\\
IRAS13451+1232		 &	-	         	&	-        	&	-       	&	-	            &	-	            &	-	            &	-	        	&	-		\\
IRAS01003-2238		 &	-	         	&	-        	&	-       	&	-	            &	-	            &	-	            &	-	        	&	-		\\
IRAS11095-0238		 &  4.41$\pm$0.82	&	-        	&	-       	&	-	            &	-	            &	-	            &	-	        	&	-		\\
IRAS20087-0308		 &	-	        	&	-        	&	-       	&	-	            &	 4.10$\pm$0.88	&	-	            &	3.51$\pm$0.88	&	-		\\
IRAS23230-6926		 &	-	        	& 3.44$\pm$0.83	& 4.35$\pm$0.78	&	4.35$\pm$0.78	&	-	            &	-	            &	4.97$\pm$0.78	&	-		\\
IRAS08311-2459		 &	-	        	&	-       	& 3.99$\pm$0.92	&	3.99$\pm$0.92	&	 4.53$\pm$0.92	&	-	            &	-	        	&	-		\\
IRAS15462-0450		 &	-	        	&	-       	&	-        	&	-	            &	-	            &	-	            &	-	        	&	-		\\
IRAS06206-6315		 &	-	        	& 3.73$\pm$0.94	&	-        	&	-	            &	-	            &	-	            &	-	        	&	-		\\
IRAS20414-1651		 &	-	        	&	-       	&	-        	&	-	            &	 5.85$\pm$0.91	&	-	            &	-	        	&	-		\\
IRAS19297-0406		 &	-	        	&	-       	&	-        	&	7.90$\pm$1.29	&	-	            &	-	            &	-	        	&	-		\\
IRAS14348-1447		 &	-	        	& 9.33$\pm$1.09	& 7.68$\pm$1.06	&	-	            &	 8.10$\pm$1.06	&	-	            &	-	        	&	-		\\
IRAS06035-7102		 &	-	        	&	-       	&	-       	&	11.0$\pm$0.96	&	-	            &	-	            &	6.35$\pm$0.95	&	-		\\
IRAS22491-1808		 &	-	        	& 5.35$\pm$0.88	&	-       	&	-	            &	-	            &	6.40$\pm$0.99	&	6.04$\pm$0.99	&	5.22$\pm$0.98	\\
IRAS14378-3651		 &	-	        	&	-       	& 5.84$\pm$1.10	&	-	            &	 4.95$\pm$1.10	&	-	            &	7.45$\pm$1.09	&	-		\\
IRAS23365+3604		 &	-	        	&	-       	&	-       	&	-	            &	 8.24$\pm$1.11	&	-	            &	-	        	&	-		\\
IRAS19254-7245		 &	-	        	&	-       	&	-       	&	-	            &	-	            &	-	            &	-	        	&	-		\\
IRAS09022-3615		 &  4.16$\pm$1.02	& 5.44$\pm$1.03	& 7.82$\pm$0.98	&	5.61$\pm$0.98	&	-	            &	-	            &	-	        	&	-		\\
IRAS08572+3915		 &	-	        	&	-        	&	-       	&	-	            &	-	            &	-	            &	-	        	&	-		\\
IRAS15250+3609		 &	-	        	&	-        	&	-       	&	-	            &	 8.06$\pm$1.17	&	3.87$\pm$1.16	&	-	        	&	-		\\
Mrk463				 &	-	        	&	-        	&	-       	&	-	            &	-	            &	-	            &	-	        	&	-		\\
IRAS23128-5919		 &	-	        	&	-        	&	-       	&	-	            &	-	            &	-	            &	-	        	&	-		\\
IRAS05189-2524		 &  3.74$\pm$0.93	&	-        	& 7.56$\pm$0.85	&	5.17$\pm$1.22	&	 6.29$\pm$0.86	&	3.91$\pm$0.85	&	6.68$\pm$0.86	&	-		\\
IRAS10565+2448		 &  6.88$\pm$1.37	& 7.81$\pm$1.37	&	-         	&	-	            &	-	            &	-	            &	-               &	-		\\
IRAS17208-0014$^{*}$ & 16.40$\pm$1.44	&23.50$\pm$1.44	&24.70$\pm$1.56	&	34.6$\pm$1.67	&	38.30$\pm$1.57	&  19.20$\pm$1.57	&  16.00$\pm$1.56	&  11.40$\pm$1.60	\\
IRAS20551-4250	     &	-	        	& 5.91$\pm$1.18	&10.90$\pm$1.12	&	-	            &	11.70$\pm$1.12	&	-	            &  10.40$\pm$1.12	&	7.19$\pm$1.12	\\
Mrk231				 &  9.76$\pm$1.22	&15.50$\pm$1.22	&11.90$\pm$1.21	&	-	            &	19.60$\pm$1.21	&  12.40$\pm$1.20	&  12.70$\pm$1.20	&	6.70$\pm$1.21	\\
UGC5101				 &  7.66$\pm$1.25	& 8.47$\pm$1.21	& 3.94$\pm$1.37	&	-	            &	 9.35$\pm$1.38	&	-	            &	9.01$\pm$1.37	&	-		\\
Mrk273				 &  7.95$\pm$0.82	& 9.86$\pm$0.83	& 8.88$\pm$0.95	&	-	            &	12.90$\pm$0.95	&	-	            &	5.88$\pm$0.95	&	4.56$\pm$0.95	\\
IRAS13120-5453		 & 20.90$\pm$1.69	&28.80$\pm$1.70	&19.70$\pm$1.76	&	-	            &	36.20$\pm$1.76	&	-	            &  34.10$\pm$1.77	&	-		\\
NGC 6240	 		 & 14.70$\pm$2.75	&14.80$\pm$2.87	&11.00$\pm$2.82	&	-	            &	13.00$\pm$2.80	&	5.42$\pm$2.82	&  12.70$\pm$2.80	&	-		\\
Arp 220$^{*}$		 & 75.10$\pm$4.10	&96.90$\pm$4.11	&80.30$\pm$6.81	&	97.3$\pm$7.58	&  155.00$\pm$6.84	&  58.60$\pm$6.81	&  78.80$\pm$6.81	&	31.50$\pm$6.80	\\
\hline
\multicolumn{9}{l}{* Also: H$_{2}$O$^{+}$ 2$_{02}$-1$_{11(J5/3-3/2)}$ \& H$_{2}$O$^{+}$ 2$_{02}$-1$_{11(J3/2-3/2)}$ fluxes of 7.15$\pm$1.45 \& 6.28$\pm$1.40$\times$10$^{-18}$Wm$^{-2}$ respectively.} \\
\multicolumn{9}{l}{** Also: H$_{2}$O$^{+}$ 2$_{02}$-1$_{11(J5/3-3/2)}$ \& H$_{2}$O$^{+}$ 2$_{02}$-1$_{11(J3/2-3/2)}$ fluxes of 22.0$\pm$4.12 \& 21.2$\pm$4.14$\times$10$^{-18}$Wm$^{-2}$ respectively.} \\
\end{tabular}
\label{tab:H2Olinefluxes}
\end{table*}

\subsection{CO SLED Modelling}\label{sec:COsledmodelling}
\begin{figure} 
\begin{center}
\includegraphics[width=0.85\columnwidth,angle=0]{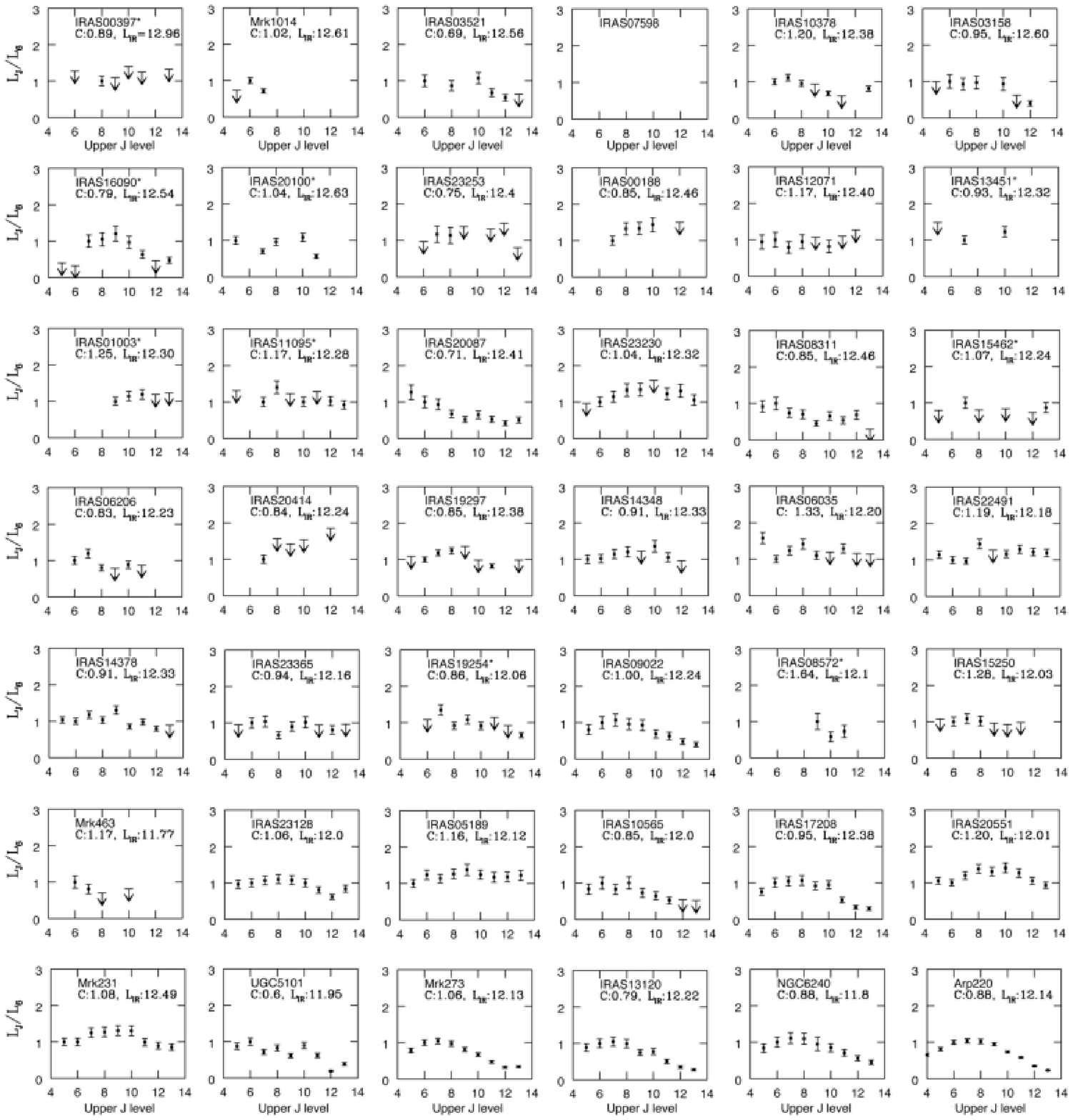}
\caption{CO SLEDs for HERUS galaxies ordered by decreasing redshift as in Figures \ref{fig:fittedlines1}, \ref{fig:fittedlines2}, \ref{fig:fittedlines3}, \ref{fig:fittedlines4}, \ref{fig:fittedlines5}. The line luminosity normalised to the  L$_{CO(6-5)}$ luminosity is plotted against transition level ($J$) for all sources, unless marked with an asterisk then the CO(7-6) or higher transition is used. Also shown is the infrared colour given by the IRAS flux ratio C2=S$_{60}$/S$_{100}$.}
\label{fig:COsled}
\end{center}
\end{figure}

In Figure ~\ref{fig:COsled} we plot the CO spectral line energy distributions (SLED) for HERUS galaxies ordered by decreasing redshift in the same order as Figures \ref{fig:fittedlines1}, \ref{fig:fittedlines2}, \ref{fig:fittedlines3}, \ref{fig:fittedlines4}, \ref{fig:fittedlines5}. For each galaxy the L$_{\rm{J}}$ line luminosity normalised to the  L$_{\rm{CO}(6-5)}$ luminosity is plotted against transition level ($J$). When the L$_{\rm{CO}(6-5)}$ line luminosity is not available, the luminosity is normalised to the  L$_{\rm{CO}(7-6)}$ or that of a higher transition. The L$_{\rm{CO}(6-5)}$ was chosen because the CO(6-5) transition is the most commonly detected line in the HERUS sample. Note that there is no SLED for IRAS 07598+6508 since only a single CO line (CO(6-5)) was detected. 

The ULIRG SLEDs in Figure ~\ref{fig:COsled} can be broadly grouped in three categories: flat, increasing (low to high J) or decreasing. The IR luminosity (reported in each SLED panel) is not a good indicator of the shape of the SLED. For example, IRAS00188-0856 and IRAS 20087-0308 have similar L$_{\rm{IR}}$ but show very different SLEDs; the SLED of the former increases up to J$\sim$10 while the latter peaks at J$\sim$5. 

Since the SLED of a galaxy represents the molecular gas mass distribution across a range of densities and temperatures (provided that the lines do not become optically thick), it is useful to look for possible dependencies of the shape of the SLED on these parameters. First, we investigate the dependence of the shape of the SLED on the FIR colour index C(60$/$100) which is a good proxy for the dust temperature (e.g. \citealt{mrr89}). In addition we look for possible dependencies of the SLED on the
$\alpha$(25/60) colour index which has been widely used to differentiate between cold and warm ULIRGs (e.g. \citealt{sanders88}). Table \ref{tab:IRAScolours} lists the infrared colours for all our sources, both the C(60/100) as well as $\alpha$(25/60). For simplicity, we group FIR colour indices according to their values: we classify C(60$/$100)$>$1.0 as a warm colour index, 0.6$<$C(60$/$100)$<$1.0 as intermediate, and C(60$/$100) $<$ 0.6 as cold. Following this scheme, we find that ULIRGs whose SLED increases with increasing $J$ tend to have warm colour indices, those with decreasing SLEDs (ie SLED peaking at $J \leq$6) have cold colour indices and those displaying flat SLEDS have intermediate colour indices. Thus, the shape of the SLED appears to be influenced by the physical properties (e.g. temperature) of the underlying radiation field, also noted by \citet{lu14} for their sample of LIRG galaxies. Finally, we note that an AGN (i.e. warm ULIRGs) will shift the peak of the SLED to higher $J$ transitions, which may be a contributor to both higher C(60$/$100) colours and warmer SLEDs.

\begin{table}\tiny
\caption{Infrared colours of HERUS ULIRGs derived from {\it IRAS} fluxes at 25$\mu$m, 60$\mu$m and 100$\mu$m, where C1 = S$_{25}$/S$_{60}$ is used to differentiate between cold and hot sources and C2=S$_{60}$/S$_{100}$ can be used as a proxy for the temperature.}
\centering
\begin{tabular}{@{}lll}
\hline
Target          & C1        & C2   \\
                & S$_{25}$/S$_{60}$   & S$_{60}$/S$_{100}$  \\
\hline
IRAS00397-1312	&	0.31	&	0.88	\\
Mrk1014	        &	0.27	&	1.02	\\
3C273	        &	0.42	&	0.76	\\
IRAS03521+0028	&	0.09	&	0.69	\\
IRAS07598+6508	&	0.35	&	0.94	\\
IRAS10378+1109	&	0.15	&	1.20	\\
IRAS03158+4227	&	0.11	&	0.95	\\
IRAS16090-0139	&	0.07	&	0.79	\\
IRAS20100-4156	&	0.08	&	1.04	\\
IRAS23253-5415	&	0.10	&	0.76	\\
IRAS00188-0856	&	0.23	&	0.85	\\
IRAS12071-0444	&	0.21	&	1.18	\\
IRAS13451+1232	&	0.33	&	0.94	\\
IRAS01003-2238	&	0.25	&	1.25	\\
IRAS23230-6926	&	0.09	&	1.04	\\
IRAS11095-0238	&	0.14	&	1.17	\\
IRAS20087-0308	&	0.06	&	0.71	\\
IRAS15462-0450	&	0.15	&	1.07	\\
IRAS08311-2459	&	0.22	&	0.86	\\
IRAS06206-6315	&	0.08	&	0.84	\\
IRAS20414-1651	&	0.09	&	0.89	\\
IRAS19297-0406	&	0.09	&	0.85	\\
IRAS14348-1447	&	0.08	&	0.91	\\
IRAS06035-7102	&	0.12	&	0.87	\\
IRAS22491-1808	&	0.10	&	1.19	\\
IRAS14378-3651	&	0.09	&	0.90	\\
IRAS23365+3604	&	0.11	&	0.94	\\
IRAS19254-7245	&	0.27	&	0.83	\\
P09022-3615	    &	0.10	&	1.00	\\
IRAS08572+3915	&	0.23	&	1.64	\\
IRAS15250+3609	&	0.18	&	1.28	\\
Mrk463	        &	0.74	&	1.17	\\
IRAS23128-5919	&	0.15	&	1.06	\\
IRAS10565+2448	&	0.10	&	0.85	\\
IRAS20551-4250	&	0.15	&	1.20	\\
IRAS05189-2524	&	0.26	&	1.16	\\
IRAS17208-0014	&	0.05	&	0.95	\\
Mrk231      	&	0.25	&	1.09	\\
UGC5101     	&	0.09	&	0.60	\\
Mrk273      	&	0.10	&	1.06	\\
IRAS13120-5453	&	0.07	&	0.79	\\
NGC6240	        &	0.15	&	0.88	\\
Arp220      	&	0.08	&	0.88	\\
\hline
\end{tabular}
\label{tab:IRAScolours}
\end{table}

\section{Discussion}\label{sec:discussion}
\begin{figure} 
\begin{center}
\includegraphics[width=0.75\columnwidth,angle=-90]{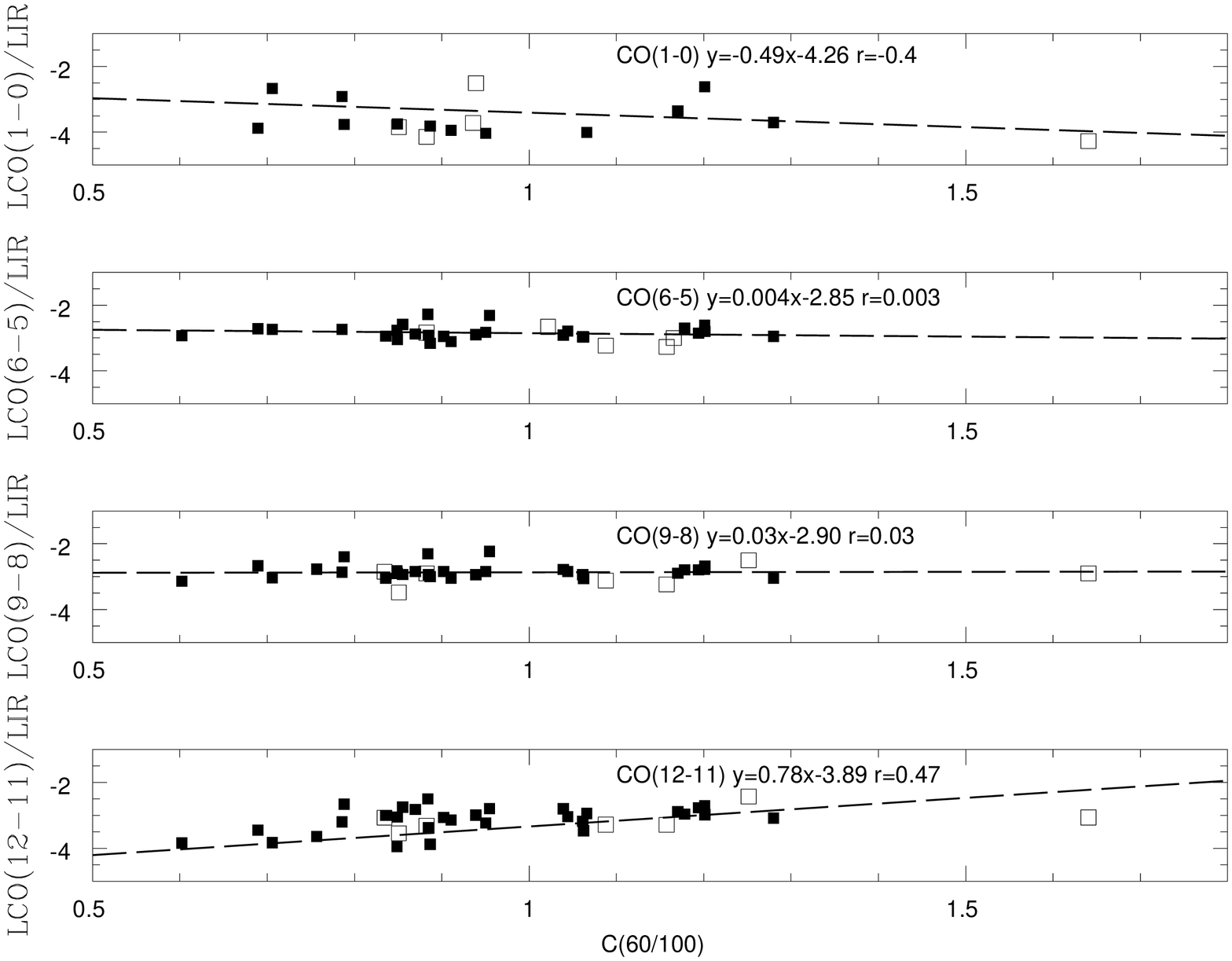}
\caption{$CO$/L$_{FIR}$ line ratios as a function of dust temperature, parameterised as S60/S100 infrared colour (open squares are sources with S25/S60$>$0.2). Parameters for regression analysis and the correlation coefficient are also shown. At low J values, e.g. CO(1-0), there is a poor correlation with L$_{FIR}$. However, between CO(6-5) and CO(9-8) the slope becomes flatter, i.e. independent of temperature with the CO(6-5) or CO(8-7) transitions. From CO(J$>$11) onwards the line strength again deviate from L$_{FIR}$,  possibly due to gas heating due to other sources such as shocks and XDRs.}
\label{fig:COslopesIRAS}
\end{center}
\end{figure}

\begin{figure} 
\begin{center}
\includegraphics[width=0.75\columnwidth,angle=-90]{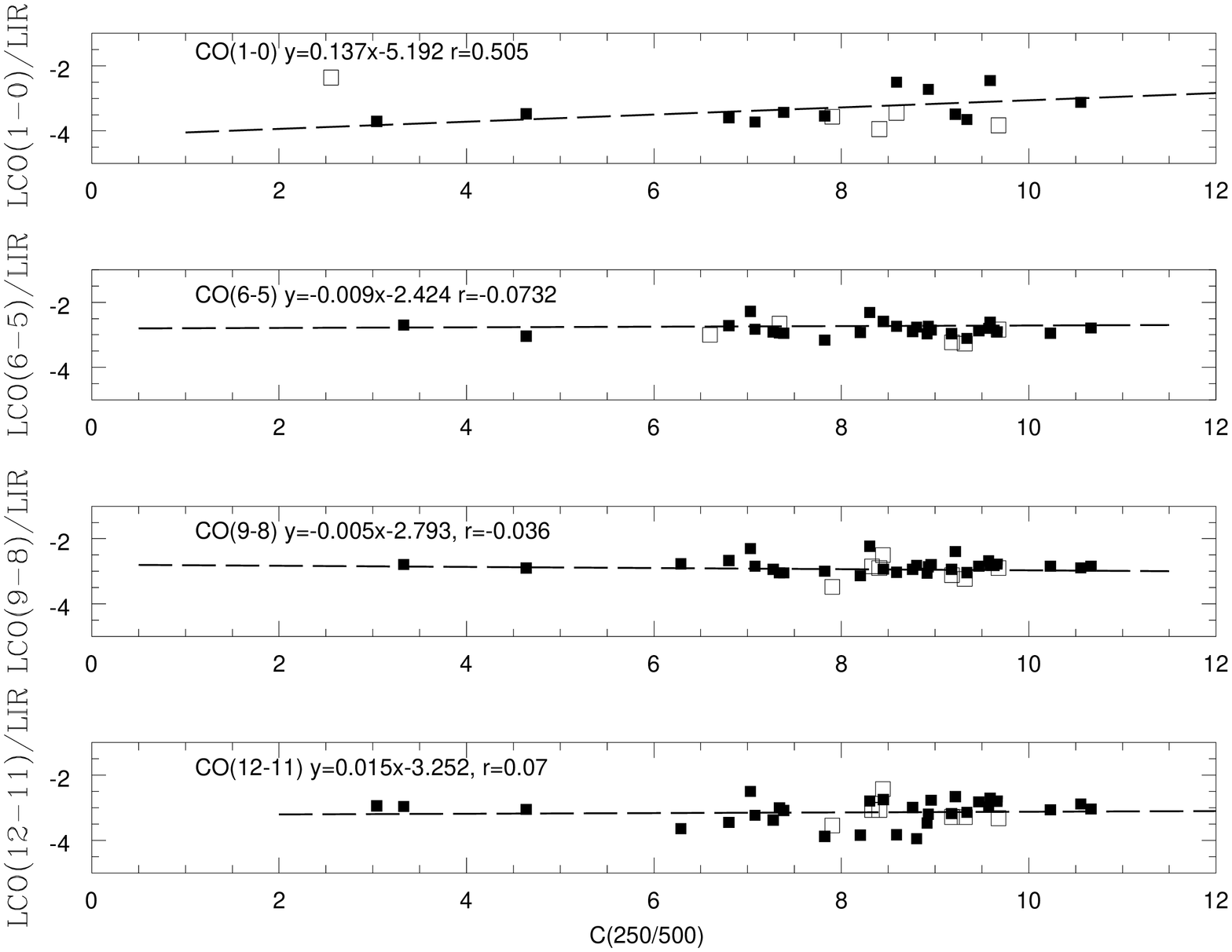}
\caption{The $CO$/L$_{FIR}$ line ratios as a function of dust temperature, parameterised via the SPIRE S$_{250}$/S$_{500}$ colour (open squares are sources with S25/S60$>$0.2).}
\label{fig:COslopesIRASSPIRE}
\end{center}
\end{figure}

\begin{figure} 
\begin{center}
\includegraphics[width=0.75\columnwidth,angle=-90]{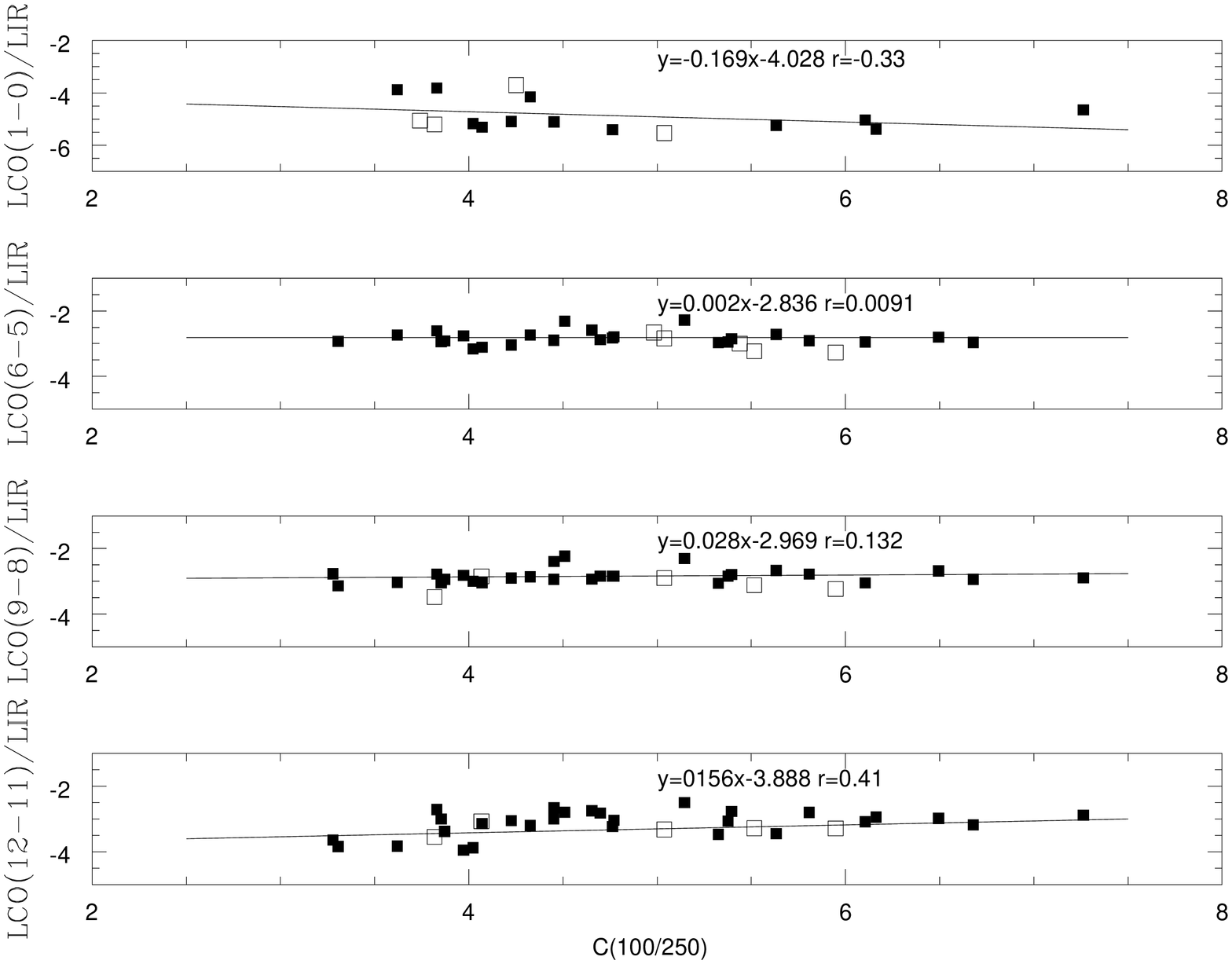}
\caption{The $CO$/L$_{FIR}$ line ratios as a function of dust temperature, parameterised via the S100/S$_{250}$ colour (open squares are sources with S25/S60$>$0.2).}
\label{fig:COslopesSPIRE}
\end{center}
\end{figure}

In order to probe the dependency of the shape of the SLED on the C(60$/$100) index further, in Figure \ref{fig:COslopesIRAS} we plot the  L$_{CO(1-0)}$, L$_{CO(6-5)}$, L$_{CO(9-8)}$ and L$_{CO(13-12)}$ line luminosities normalised by L$_{IR}$ as a function of C(60$/$100) index. We focus on sources with firm detections in all CO transitions. ULIRGs where the fractional contribution of the AGN to the total bolometric output exceeds $\sim$40\% have been denoted with an open square and are not taken into account when estimating the slope of the correlation. The presence of an AGN with significant contribution to the total bolometric output was established based on the presence of [NeV] $\lambda$14.32$\mu$m or a very deep $9.7\mu$m silicate absorption feature (e.g. \citealt{farrah07}). Of the HERUS ULIRGs, six have AGN contributions $>$40\% (\citealt{farrah07}, \citealt{desai07}). The investigation of the SLED dependence on broadband colours is extended to SPIRE wavelengths in Figures \ref{fig:COslopesIRASSPIRE} \& \ref{fig:COslopesSPIRE} which again show the L$_{CO}$  line luminosity normalised by L$_{IR}$ this time as a function of IRAS/SPIRE 100$\mu$m$/$250$\mu$m and SPIRE 250$\mu$m$/$500$\mu$m colours respectively.

Figure \ref{fig:COslopesIRAS} shows that as the C(60$/$100) index increases, the CO gas gets warmer. This is demonstrated by the inversion in the slopes of the linear fits to the data as a function of $J$, changing from -0.54 for the CO(1-0), through -0.2 for CO(6-5) and $+$0.031 for CO(9-8), to $+$0.63 for CO(12-11). The Pearson correlation coefficients confirm this: $r$ is negative for CO(1--0), close to zero for CO(6-5) and CO(9-8) and positive for CO(12-11). The near-zero value (r=0.03) computed for the mid-J transitions implies that the ratios L$_{CO(6-5)}/$L$_{IR}$ and L$_{CO(9-8)}/$L$_{IR}$ are constant over the values of the C(60$/$100) explored here. The scatter in the relations is also noteworthy. For the moderate J lines (6-5 and 9-8),  the CO/L$_{IR}$ ratio has approximately the same value, with little scatter, for a wide range in S60/S100 ratio. For the 1-0 and 12-11 lines however the CO/L$_{IR}$ ratio changes, and the scatter is greater. The same pattern is seen for the S100/S$_{250}$ ratio (Figure \ref{fig:COslopesIRASSPIRE}). Notably though, the S$_{250}$/S$_{500}$ ratio shows a different behavior; a positive correlation with L$_{CO(1-0)}/$L$_{IR}$, and approximately flat relations with the other CO line ratios (Figure \ref{fig:COslopesIRAS}). 

We checked these results by repeating the analysis for the C(60$/$100) indices using PACS data. The PACS indices were estimated from continua measurements around the [OI]$\lambda$63$\mu$m and [NII]$\lambda$122$\mu$m lines. However, the PACS observations of the HERUS ULIRGs include only a sub-sample of the SPIRE-FTS sample. Nonetheless, it is reassuring that we find similar slopes for the CO(6-5), CO(9-8) and CO(12-11) correlations. The CO(1-0) data show a larger scatter which we attribute to smaller number statistics from the smaller sample.

In what follows we outline a simple scenario that matches, at least quantitatively, these results. For starbursts in local ULIRGs, the total IR luminosity is a proxy for the total number of hot young stars in the starburst, largely independently of how those stars are distributed. The S$_{60}$/S$_{100}$ (and to a certain extent the S$_{100}$/S$_{250}$ ratio) ratio on the other hand traces the broad-scale geometry of the starburst - how compact it is given the number of stars it has - with a higher ratio implying greater compactness. This can be justified in terms of the correlation found between the equivalent width of OH(65) and C(60/100) in \citet{gonzalezalfonso15}.

The S$_{250}$/S$_{500}$ ratio on the other hand traces cold, but still star formation heated dust, whose properties are more decoupled from the CO gas reservoir than the hotter dust. These ratios may also provide an indication of the age and star formation history of the starburst, therefore determining the relative amounts of hot, warm, and cold gas.

Turning to the CO lines, the baseline assumption is that the CO(1-0) line is tracing the bulk of the cold CO gas reservoir, at some distance from star forming regions. The lines from approximately CO(6-5) through CO(9-8) on the other hand trace warmer CO in the outer parts of PDRs, or the nearby ISM. Finally, CO lines from approximately CO(12-11) and up trace hot CO in the inner PDRs or HII regions. Their luminosities are thus affected by small scale `microphysics' of the starburst, giving rise to a larger scatter with FIR color since this measures the geometry on larger scales of the star forming regions. A consequence of this scenario is that CO SLEDs of star-forming ULIRGs could be matched with three components;  a cold gas component, responsible for the low-J (J$<$4) transitions, a `warm' gas component giving rise to the mid-J transitions and a `hot' gas component which is responsible for J$>$9 transitions. This component will likely also have a contribution from AGN-heated gas. 

This simple picture is however complicated by more energetic processes that are likely to be present in the majority of ULIRGs. Mechanical heating (supernova-driven turbulence, shocks) are also likely to be present and play a major role in powering CO transitions with J$>$6 \citep[e.g.][]{greve14}.  The detection of water transitions in particular, immediately suggests that shocks are likely to play a role. For example, although not a ULIRG, the CO-SLED of NGC 1266 has been modeled with a combination of low velocity C-shocks and PDRs \citep{pell13}.

\begin{figure} 
\begin{center}
\includegraphics[width=0.75\columnwidth,angle=0]{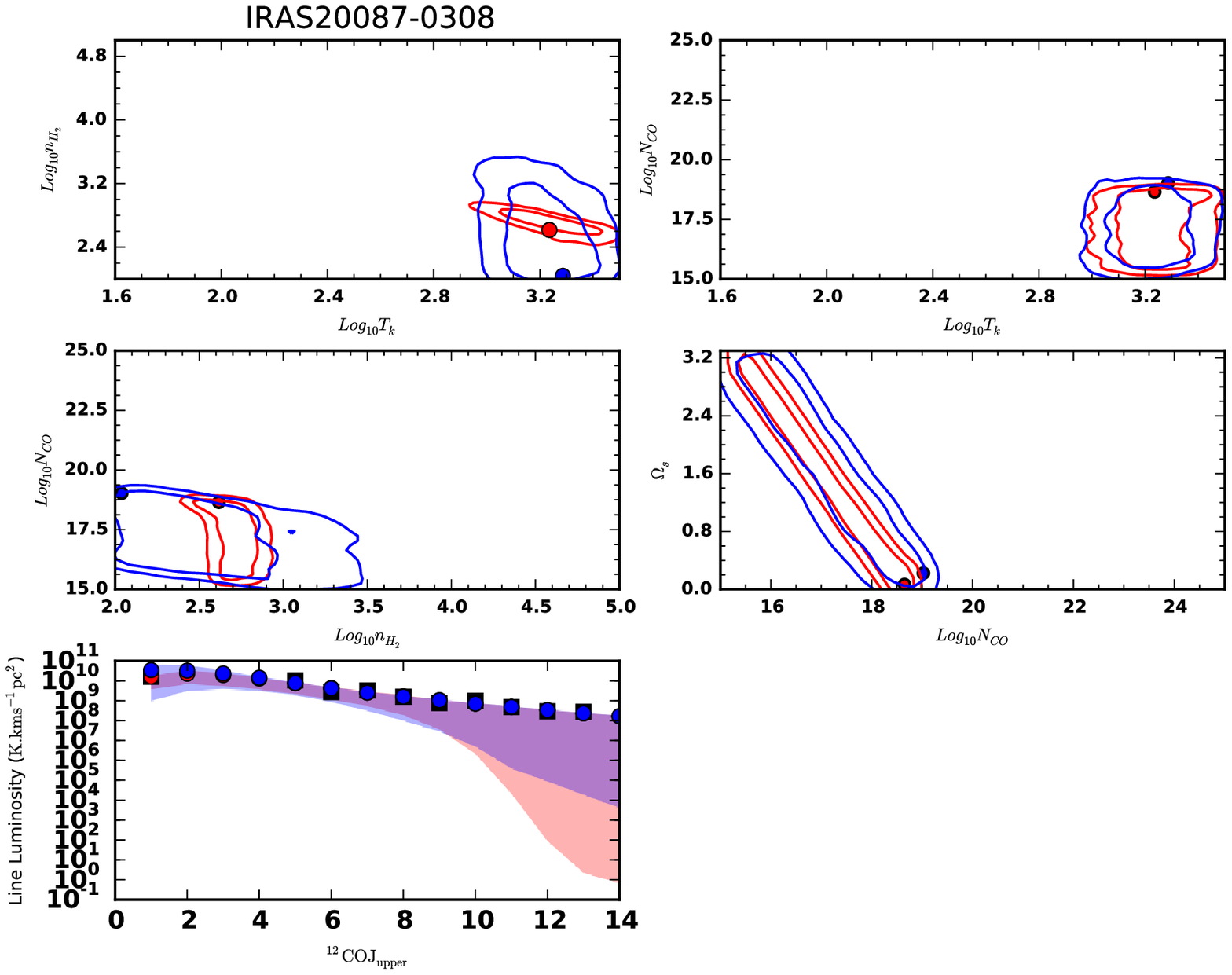}
\caption{Example maximum likelihood {\itshape single component} fit to IRAS 20087-0308. We select this object as it shows any evidence for shock-heated gas in its optical spectra. The upper four panels show the parameter constraints achieved with only the SPIRE data (blue), and with the SPIRE and ground-based data (red). The lower panel shows the CO SLED, where the black squares are the observed data (with error bars omitted for clarity), and the blue and circles are the predicted CO fluxes using just SPIRE, and SPIRE plus ground based data, respectively. The single component model adequately reproduces the whole SLED.}
\label{fig:modela}
\end{center}
\end{figure} 

\begin{figure} 
\begin{center}
\includegraphics[width=0.75\columnwidth,angle=0]{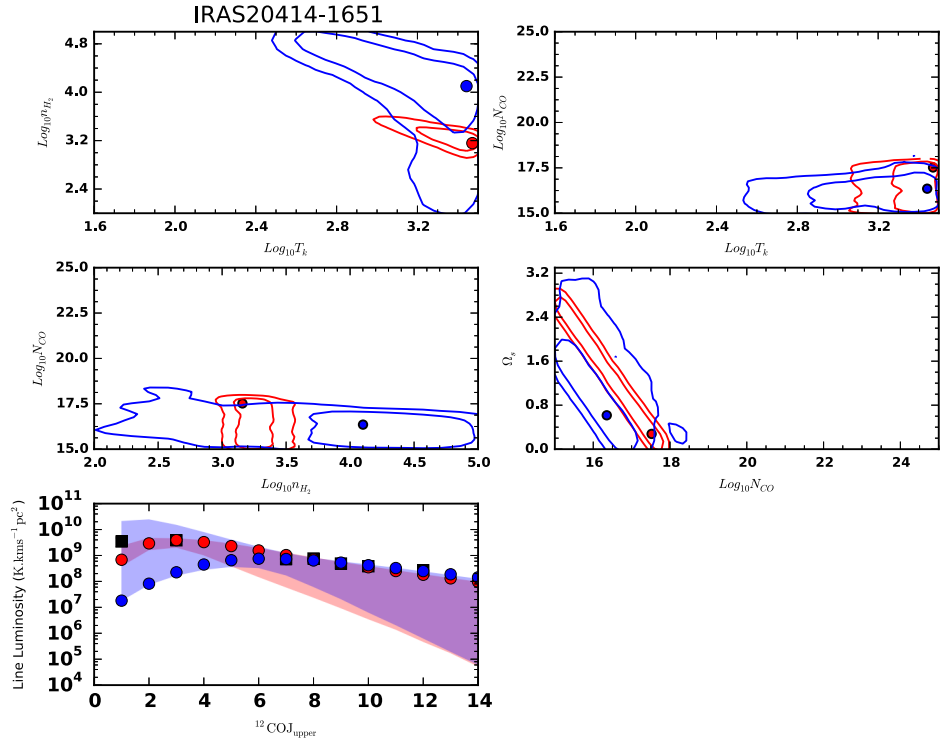}
\caption{Example maximum likelihood {\itshape single component} fit to IRAS 20414-1651. See the caption of Figure \ref{fig:modela} for modelling details a a key to the plotted data. Here, unlike the fit to IRAS 20087-0308, a single component fit does not adequately reproduce the whole CO SLED, being unable to account for the lowest transition.}
\label{fig:modelb}
\end{center}
\end{figure} 

 \begin{figure} 
 \begin{center}
 \includegraphics[width=0.75\columnwidth,angle=-00]{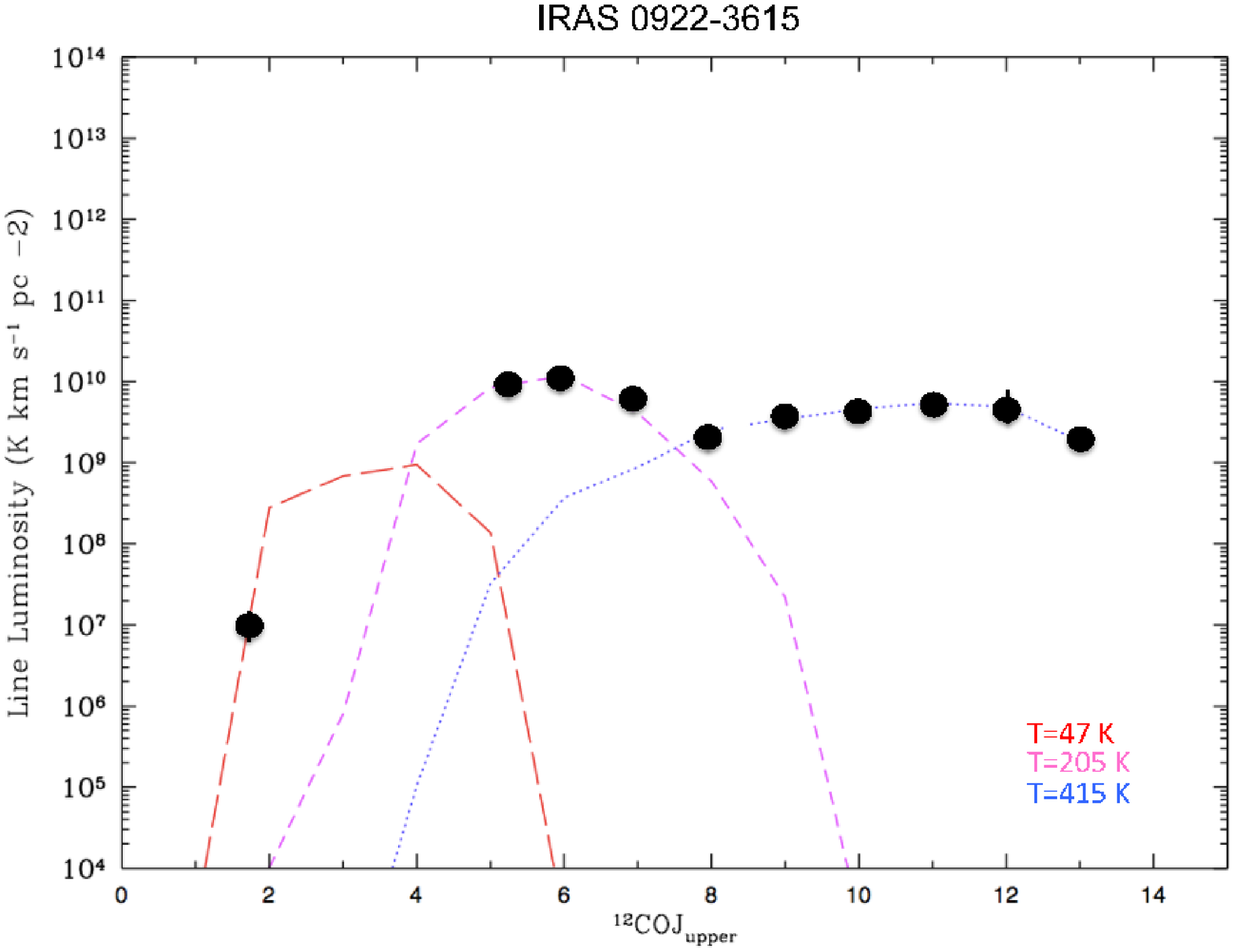}
 \caption{Example fit for a 3 component temperature model generated by the Radex non-LTE radiative transfer code \citep{vandertak07} for IRAS 0922-3615. The CO SLED is fit by a cold (47K) component for the low-J transitions, warm (205K) component for the mid-J transitions and a hot (415K) component for the high-J transitions. The reduced $\chi^2$ for this fit is below unity.}
 \label{fig:radex}
 \end{center}
 \end{figure} 

We defer a rigorous analysis of this possibility to Hurley et al. in preparation, but here show three illustrative examples. Using RADEX we searched a large grid of temperatures (T$_{kin}$), densities (n(H$_{2}$), column densities (N$_{CO}$), line widths, and source sizes, for {\itshape single component} fits. The parameter space is searched using the nested sampling routine MULTINEST \citep{feroz08} which uses Bayesian evidence to select the best model and samples multiple nodes and$/$or degeneracies using posterior distributions. The details of the modelling can be found in \citet{rigopoulou13}. From these fits, we can extract maximum likelihood contours for key physical parameters. Adopting such a process allows the effects of combining Herschel and ground-based CO data to be examined. Figures ~\ref{fig:modela} and \ref{fig:modelb} show examples of such an extraction, for two ULIRGs, IRAS 20087-0308 and  IRAS 20414-1651, respectively, for which we have obtained ground-based data. In the case of IRAS 20087-0308, a single component model can reproduce the entire CO SLED, as has been found for a small number of systems in previous studies \citep{kamenet14,mashi15,rosen15}. In the case of IRAS 20414-1651 however a single component model is clearly inadequate, as it fails to reproduce the CO(1-0) emission. Conversely, in Figure \ref{fig:radex} we show an example of a ULIRG where three components can adequately reproduce the entire CO SLED.

\section{Conclusions}\label{sec:conclusions}
A CO atlas from the HERUS programme has been presented for our flux limited (S$_{60\mu m}>$1.8Jy) sample of 43 local ULIRGS observed with the {\it Herschel} SPIRE FTS instrument, with complementary SPIRE photometry at 250, 350 and 500$\mu$m. Post-pipeline processing was employed in order to produce high quality spectra between 194 - 671$\mu$m. Our spectra were analysed using HIPE v11 although custom made routines were used to correct for the effects of `Cooler burps'. Our conclusions are:

(1) The CO ladder from the J=4 to the J=13 transitions is clearly seen in multiple detections for more than half our sample. In addition, atomic Carbon (607$\mu$m \& 370$\mu$m) and the ionised Nitrogen (205$\mu$m) fine structure lines are detected for many sources, along with the detection of various water lines. The important ionised Carbon cooling line at 197$\mu$m was detected in our highest redshift source IRAS00397-1312 at z=0.262 where the line is redshifted into the SPIRE FTS observational bands.

We find that ULIRG CO-SLEDS do not correlate with L$_{FIR}$. However, a comparison of the SLEDS with their far-infrared colours reveals a correlation between the 60$\mu$m/100$\mu$m colour index and the slope of the CO SLED with ULIRGs exhibiting an increasing slope with J transition having warmer far-infrared colours, and ULIRGs exhibiting a decreasing slope with J ($>$6) transition having cooler far-infrared colours.  This infers that all ULIRG SLEDs require at least three different gas components, in accordance with the trends seen when examining the variation of individual line luminosities with changes in the C(60$/$100) colour index. Mid-j transitions originate in warm gas (T$\sim$140-260 K). Since the ratio L$_{mid-J}/$L$_{IR}$ remains constant over a wide range of C(60$/$100), we suggest that the gas originating in these mid-J transitions is directly linked to on-going star-formation. It is likely that the hot gas component in ULIRGs is also associated with shocks or processes directly linked to the presence of AGN (XDRs) although we cannot probe this further based on the current datasets.

\acknowledgments
SPIRE has been developed by a consortium of institutes led by Cardiff Univ. (UK) and including: Univ. Lethbridge (Canada); NAOC (China); CEA, LAM (France); IFSI, Univ. Padua (Italy); IAC (Spain); Stockholm Observatory (Sweden); Imperial College London, RAL, UCL-MSSL, UKATC, Univ. Sussex (UK); and Caltech, JPL, NHSC, Univ. Colorado (USA). This development has been supported by national funding agencies: CSA (Canada); NAOC (China); CEA, CNES, CNRS (France); ASI (Italy);MCINN (Spain); SNSB (Sweden); STFC, UKSA (UK); and NASA (USA). HIPE is a joint development by the Herschel Science Ground Segment Consortium, consisting of ESA, the NASA Herschel Science Center, and the HIFI, PACS and SPIRE consortia. J.A. gratefully acknowledges support from the Science and Technology Foundation (FCT, Portugal) through the research grant PTDC/FIS-AST/2194/2012 and UID/FIS/04434/2013.

{\it Facilities:}  \facility{Herschel}.


\begin{thebibliography}{}

\bibitem[\protect\citeauthoryear{Armus et al.}{2007}]{armus07}
Armus L., Charmandaris V., Bernard-Salas J., et al., 2007, ApJ, 656, 148 

\bibitem[\protect\citeauthoryear{Bethermin et al.}{2014}]{bethermin14}
B{\'e}thermin M., Daddi E., Magdis, G. et al., 2014, arXiv:1409.5796 

\bibitem[\protect\citeauthoryear{Boller et al.}{2002}]{boller02}
Boller T., Gallo L.C., Lutz D.,  Sturm E., 2002, MNRAS, 336, 1143

\bibitem[\protect\citeauthoryear{Cicone et al.}{2014}]{cicone14}
Cicone C., Maiolino R., Sturm E. et al., 2014, A\&A, 562, 21

\bibitem[\protect\citeauthoryear{Clements et al.}{1996}]{clements96}
Clements D. L., Sutherland W. J., SaundersW., Efstathiou G. P., McMahon R. G., Maddox S., Lawrence A., Rowan-Robinson M., 1996, MNRAS, 279, 459

\bibitem[\protect\citeauthoryear{Desai et al.}{2007}]{desai07} 
Desai V., Armus L., Spoon H.W.W. et al., 2007, ApJ, 669, 810 

\bibitem[\protect\citeauthoryear{Dowell et al.}{2010}]{dowell10} 
Dowell C., Pohlen M., Pearson C.P. et al., 2010, Proc. SPIE 7731, 36 

\bibitem[\protect\citeauthoryear{Downes et al.}{1993}]{downes93} 
Downes D., Solomon P.M., \& Radford S.J.E., 1993, ApJ, 414, L13 

\bibitem[\protect\citeauthoryear{Efstathiou et al.}{2014}]{efstathiou14}
Efstathiou A., Pearson C.,  Farrah D. et al., 2014, MNRAS, 437, 16

\bibitem[\protect\citeauthoryear{Elbaz et al.}{2011}]{elbaz11}
Elbaz D., Dickenson M.,  Hwang H.S. et al., 2011, A\&A, 533, 119

\bibitem[\protect\citeauthoryear{Farrah et al.}{2001}]{farrah01}
Farrah D., Rowan-Robinson M., Oliver S. et al.,  2001,MNRAS, 326, 1333

\bibitem[\protect\citeauthoryear{Farrah et al.}{2007}]{farrah07}
Farrah, D., Bernard-Salas J., Spoon H.W.W.  et al., 2007, ApJ, 667, 149 

\bibitem[\protect\citeauthoryear{Farrah et al.}{2008}]{farrah08}
Farrah D., Lonsdale C.J., Weedman D.W., et al., 2008, ApJ, 677, 957 

\bibitem[\protect\citeauthoryear{Farrah et al.}{2013}]{farrah13}
Farrah D.,  Lebouteiller V., Spoon H.W.W. et al., 2013, ApJ, 776, 38

\bibitem[\protect\citeauthoryear{Feroz \& Hobson}{2008}]{feroz08}
Ferorz F., Hobson M.P., 2008, MNRAS, 384, 449

\bibitem[\protect\citeauthoryear{Feruglio et al.}{2010}]{feruglio10}
Feruglio C., Maiolino R., Piconcelli E., Menci N., Aussel H., Lamastra A., Fiore F., 2010, A\&A, 518, 155

\bibitem[\protect\citeauthoryear{Fischer et al.}{2010}]{fischer10}
Fischer J., Sturm E., Gonzalez-Alfonso E. et al., 2010, A\&A, 518, L41 

\bibitem[\protect\citeauthoryear{Fulton et al.}{2010}]{fulton10}
Fulton T., Baluteau J-P., Bendo G. et al., 2010, Proc.SPIE, 7731, 34

\bibitem[Fulton et al.(2016)]{fulton14} 
Fulton, T., Naylor, D.~A., Polehampton, E.~T., et al.\ 2016, \mnras, 458, 1977 

\bibitem[\protect\citeauthoryear{Gonz{\'a}lez-Alfonso  et al.}{2010}]{gonzalezalfonso10}
Gonz{\'a}lez-Alfonso, E., Fischer J., Isaak K. et al., 2010, A\&A, 518, L43

\bibitem[\protect\citeauthoryear{Gonz{\'a}lez-Alfonso  et al.}{2013}]{gonzalezalfonso13}
Gonz{\'a}lez-Alfonso, E., Fischer J., Bruderer S. et al., 2013, A\&A, 550, A25 

\bibitem[\protect\citeauthoryear{Gonz{\'a}lez-Alfonso  et al.}{2014}]{gonzalezalfonso14}
Gonz{\'a}lez-Alfonso, E., Fischer J., Aalto S., Falstad N., 2014, A\&A, 567, A91

\bibitem[\protect\citeauthoryear{Gonz{\'a}lez-Alfonso  et al.}{2015}]{gonzalezalfonso15}
Gonz{\'a}lez-Alfonso, E., Fischer J., Sturm E. et al., 2015, ApJ, 800, 69

\bibitem[\protect\citeauthoryear{Greve et al.}{2014}]{greve14}
Greve T.R., Leonidaki  I., Xilouris E.M. et al., 2014, ApJ, 794, 142 

\bibitem[\protect\citeauthoryear{Griffin et al.}{2010}]{griffin10}
Griffin, M.~J., Abergel, A., Abreu, A., et al.\ 2010, \aap, 518, L3 

\bibitem[\protect\citeauthoryear{Hailey-Dunsheath et al.}{2010a}]{haileydunsheath10a}
Hailey-Dunsheath S., Sturm E., Fischer J. et al., 2012, ApJ, 755, 57 

\bibitem[\protect\citeauthoryear{Hailey-Dunsheath et al.}{2010b}]{haileydunsheath10b}
Hailey-Dunsheath S., Nikola T., Stacey G.J. et al., 2010, ApJ, 714, L162 

\bibitem[\protect\citeauthoryear{Hopwood et al.}{2015}]{hopwood15}
Hopwood R., polehampton E.T., Valtchanov I.  et al., 2015, MNRAS, 449, 2274

\bibitem[\protect\citeauthoryear{Houck et al.}{2004}]{houck04}
Houck J.R., Roellig T.L., Van Cleve J. et al. 2004, ApJSS, 154, 18 

\bibitem[Kamenetzky et al.(2014)]{kamenet14} 
Kamenetzky, J., Rangwala, N., Glenn, J., Maloney, P.~R., \& Conley, A.\ 2014, \apj, 795, 174 

\bibitem[\protect\citeauthoryear{Kamenetsky et al.}{2016}]{kamenet16}
Kamenetzky, J., Rangwala, N., Glenn, J., Maloney, P.~R., \& Conley, A.\ 2016, \apj, 829, 93 

\bibitem[\protect\citeauthoryear{Kaufman et al.}{1999}]{kaufman99}
Kaufman M.J., Wolfire M.G., Hollenbach D.J., Luhman M.L., 1999, ApJ, 527, 795

\bibitem[\protect\citeauthoryear{Kim \& Sanders}{1998}]{kim98}
Kim D.C., Sanders D.B., 1998, ApJS, 119, 41

\bibitem[\protect\citeauthoryear{Le Floc'h et al}{2005}]{lefloch05}
Le Floc'h E., Papovich C., Dole H., et al., 2005, ApJ, 632, 169 

\bibitem[\protect\citeauthoryear{Lipari}{1994}]{lipari94}
Lipari, S., 1994, ApJ, 436, 102 

\bibitem[\protect\citeauthoryear{Lonsdale et al.}{2006}]{lfs06}
Lonsdale, C.~J., Farrah, D., \& Smith, H.~E., Astrophysics Update 2, Springer Praxis Books. ISBN 978-3-540-30312-1. Praxis Publishing Ltd, Chichester, UK, 2006, p. 285

\bibitem[\protect\citeauthoryear{Lu et al}{2014}]{lu14}
Lu N., Zhao Y., Xu C.K. et al., 2014, ApJ, 787, L23

\bibitem[\protect\citeauthoryear{Luhman et al}{1998}]{luhman98}
Luhman M.L., Satyapal S., Fischer J. et al., 1998, ApJ, 504, L11 

\bibitem[\protect\citeauthoryear{Luhman et al}{2003}]{luhman03}
Luhman M.L., Satyapal S., Fischer J. et al., 2003, ApJ, 594, 758 

\bibitem[\protect\citeauthoryear{Magdis et al. }{2011}]{magdis11}
Magdis G.E., Daddi E., Elbaz D. et al.\ 2011, ApJ, 740, L15 

\bibitem[\protect\citeauthoryear{Magdis et al. }{2014}]{magdis14}
Magdis, G.~E., Rigopoulou, D., Hopwood, R., et al.\ 2014, \apj, 796, 63 

\bibitem[Mashian et al.(2015)]{mashi15} 
Mashian, N., Sturm, E., Sternberg, A., et al.\ 2015, \apj, 802, 81 

\bibitem[\protect\citeauthoryear{Meijerink  \& Spaans}{2005}]{meijerink05}
Meijerink R., Spaans M., 2005, A\&A, 436, 397

\bibitem[\protect\citeauthoryear{Meijerink et al. }{2013}]{meijerink13}
Meijerink, R., Kristensen, L.~E., Wei{\ss}, A., et al.\ 2013, \apjl, 762, L16 

\bibitem[\protect\citeauthoryear{Melnick \& Mirabel}{1990}]{melnick90}
Melnick J., \& Mirabel I.F., 1990, A\&A, 231, L19 

\bibitem[\protect\citeauthoryear{Murphy et al. }{2011}]{murphy11}
Murphy E.J., Chary R.-R., Dickinson M. et al., 2011, ApJ, 732, 126  

\bibitem[\protect\citeauthoryear{Ott et al.}{2010}]{ott10} 
Ott S., 2010, ASP Conference Series, 434, 139

\bibitem[\protect\citeauthoryear{Pearson et al.}{2014}]{pearson14}
Pearson C., Lim T., North C. et al., 2014, Experimental Astronomy, 37, 175

\bibitem[Pellegrini et al.(2013)]{pell13} 
Pellegrini, E.~W., Smith, J.~D., Wolfire, M.~G., et al.\ 2013, \apjl, 779, L19 

\bibitem[\protect\citeauthoryear{Pereira-Santaella et al.}{2013}]{pereirasantaella13}
Pereira-Santaella M., Spinoglio L., Busquet G. et al., 2013, ApJ, 768, 55

\bibitem[\protect\citeauthoryear{Pilbratt et al.}{2010}]{pilbratt10}
Pilbratt, G.~L., Riedinger, J.~R., Passvogel, T., et al.\ 2010, \aap, 518, L1 

\bibitem[\protect\citeauthoryear{Poglitsch et al.}{2010}]{poglitsch10}
Poglitsch, A., Waelkens, C., Geis, N., et al.\ 2010, \aap, 518, L2 

\bibitem[\protect\citeauthoryear{Pope et al.}{2006}]{pope06}
Pope A., Scott D., Dickinson M. et al. 2006, MNRAS, 370, 1185

\bibitem[\protect\citeauthoryear{Rangwala et al. }{2011}]{rangwala11}
Rangwala N., Maloney P.R., Glenn J. et al. 2011, ApJ, 743, 94

\bibitem[Rigopoulou et al.(1999)]{rigo99} 
Rigopoulou, D., Spoon, H.~W.~W., Genzel, R., et al.\ 1999, \aj, 118, 2625 

\bibitem[\protect\citeauthoryear{Rigopoulou et al. }{2013}]{rigopoulou13}
Rigopoulou D., Hurley P.D., Swinyard B.M. et al., 2013, MNRAS, 434, 2051

\bibitem[Rigopoulou et al.(2014)]{rigo14} 
Rigopoulou, D., Hopwood, R., Magdis, G.~E., et al.\ 2014, \apjl, 781, L15 

\bibitem[Rosenberg et al.(2015)]{rosen15} 
Rosenberg, M.~J.~F., van der Werf, P.~P., Aalto, S., et al.\ 2015, \apj, 801, 72 

\bibitem[\protect\citeauthoryear{Rowan-Robinson \& Crawford }{1989}]{mrr89}
Rowan-Robinson M., Crawford P. 1989, MNRAS, 238, 523

\bibitem[\protect\citeauthoryear{Sanders et al. }{1988}] {sanders88} 
Sanders D.B., Soifer B.T. Elias J.H., Madore B.F., Matthews K., Neugebauer G., Scoville N. Z., 1988, ApJ, 325, 74

\bibitem[\protect\citeauthoryear{Sanders et al. }{1991}] {sanders91}
Sanders D.B., Scoville N.Z., Soifer B.T., 1991, \apj, 370, 158

\bibitem[\protect\citeauthoryear{Sanders \& Mirabel}{1996}] {sanders96} 
Sanders D.B., Mirabel I.F., 1996, ARAA, 34, 725

\bibitem[\protect\citeauthoryear{Sanders et al. }{2003}] {sanders03} 
Sanders D.B., Mazzarella J.M., Kim D.C. Surace J.A., Soifer B.T, 2003, \aj, 126, 1607

\bibitem[\protect\citeauthoryear{Saunders et al. }{2000}] {saunders00} 
Saunders, W., Sutherland, W.~J., Maddox, S.~J., et al.\ 2000, \mnras, 317, 55  

\bibitem[\protect\citeauthoryear{Savage \& Oliver}{2007}] {savage07} 
Savage R., Oliver S., 2007, ApJ, 661, 1339

\bibitem[\protect\citeauthoryear{Soifer et al. }{1986}] {soifer86} 
Soifer B.T., Sanders D.B. Neugebauer G., Danielson G.E., Lonsdale C.J., Madore B.F., Persson S.E., 1986, ApJ, 303, L41

\bibitem[\protect\citeauthoryear{Soifer et al. }{2000}] {soifer00} 
Soifer, B. T.; Neugebauer, G.; Matthews, K. et al., 2000, AJ, 119, 509

\bibitem[\protect\citeauthoryear{Solomon et al. }{1997}] {solomon97} 
Solomon P. M., Downes D., Radford S. J. E., Barrett J. W., 1997, ApJ, 478, 144

\bibitem[\protect\citeauthoryear{Spoon et al. }{2009}] {spoon09} 
Spoon H.W.W., Armus  L., Marshall J.A. et al., 2009, ApJ, 693, 1223 

\bibitem[\protect\citeauthoryear{Spoon et al. }{2013}] {spoon13} 
Spoon, H.W.W., Farrah D., Lebouteiller V. et al., 2013, ApJ, 775, 127

\bibitem[\protect\citeauthoryear{Stacey et al. }{2010}] {stacey10} 
Stacey G.~J., Hailey-Dunsheath S., Ferkinhoff C. et al., 2010, ApJ, 724, 957

\bibitem[\protect\citeauthoryear{Sturm et al. }{2011}] {sturm11} 
Sturm E., Poglitsch A. Contursi A. et al., 2011, ApJ, 733, L16 

\bibitem[\protect\citeauthoryear{Swinyard et al. }{2010}] {swinyard10} 
Swinyard B., Ade P., Baluteau J.P. et al., 2010, A\&A, 518, 4 

\bibitem[\protect\citeauthoryear{Swinyard et al. }{2014}] {swinyard14} 
Swinyard B., Polehampton E., Hopwood R. et al., 2014, MNRAS, 440, 3658 

\bibitem[\protect\citeauthoryear{Symeonidis et al. }{2013}] {symeonidis13} 
Symeonidis M., Vaccari M., Berta S. et al., 2013, MNRAS, 431, 2317 

\bibitem[\protect\citeauthoryear{van der Tak et al. }{2007}] {vandertak07} 
van der Tak F.F.S., Black J.H., Schoier F.L., Jansen D.J., van Dishoeck E.F., 2007, A\&A, 468, 627

\bibitem[\protect\citeauthoryear{van der Werf et al. }{2010}] {vanderwerf10} 
van der Werf P.P., Isaak K.G., Meijerink R. et al. 2010, A\&A, 518, L42

\bibitem[Werner et al.(2004)]{wern04} 
Werner, M.~W., Roellig, T.~L., Low, F.~J., et al.\ 2004, \apjs, 154, 1 

\bibitem[\protect\citeauthoryear{Wolfire et al.}{2010}] {wolfire10} 
Wolfire M.G.,  Hollenbach D.,  McKee C.F., 2010, ApJ, 716, 1191

\bibitem[\protect\citeauthoryear{Wu et al.}{2013}] {wu13} 
Wu R., Polehampton E., Etxaluze M. et al., 2013, A\&A, 556, 116

\end{thebibliography}
\end{document}